\documentclass[a4paper,fleqn,usenatbib]{mnras}

\usepackage{newtxtext,newtxmath}

\usepackage[T1]{fontenc}
\usepackage{ae,aecompl}
\usepackage{pdflscape}
\usepackage{threeparttable}
 \usepackage{color}

\usepackage{graphicx}	
\usepackage{amsmath}	
\usepackage{amssymb}	

\title[CN, HCN, HNC observations of proto-BDs]{Chemical tracers in proto-brown dwarfs: CN, HCN, and HNC observations}

\author[Riaz et al.]{
Riaz, B.,$^{1, 2}$\thanks{E-mail: briaz@usm.lmu.de}
Thi, W.-F.,$^{1}$
Caselli, P.$^{1}$
\\
$^{1}$  Max-Planck-Institut f\"{u}r Extraterrestrische Physik, Giessenbachstrasse 1, D-85748 Garching, Germany \\
$^{2}$ Universit\"{a}ts-Sternwarte M\"{u}nchen, Ludwig Maximilians Universit\"{a}t, Scheinerstra$\beta$e 1, D-81679 M\"{u}nchen, Germany \\
}

\date{Accepted XXX. Received YYY; in original form ZZZ}

\pubyear{2018}

\begin{document}
\label{firstpage}
\pagerange{\pageref{firstpage}--\pageref{lastpage}}
\maketitle

\begin{abstract}

We present results from a study of nitrogen chemistry in Class 0/I proto-brown dwarfs (proto-BDs). We have used the IRAM 30 m telescope to observe the CN (2-1), HCN (3-2), and HNC (3-2) lines in 7 proto-BDs. All proto-BDs show a large CN/HCN abundance ratio of $>$20, and a HNC/HCN abundance ratio close to or larger than unity. The enhanced CN/HCN ratios can be explained by high UV flux originating from an active accretion zone in the proto-BDs. The larger than unity HNC/HCN ratio for the proto-BDs is likely caused by a combination of low temperature and high density. Both CN and HNC show a flat distribution with CO, indicating that these species can survive in regions where CO is depleted. We have investigated the correlations in the molecular abundances of these species for the proto-BDs with Class 0/I protostars. We find tentative trends of CN (HCN) abundances being about an order of magnitude higher (lower) in the proto-BDs compared to protostars. HNC for the proto-BDs shows a nearly constant abundance unlike the large spread of $\sim$2 orders of magnitude seen for the protostars. Also notable is a rise in the HNC/HCN abundance ratio for the lowest luminosity objects, suggesting that this ratio is higher under low-temperature environments. None of the relatively evolved Class Flat/Class II brown dwarfs in our sample show emission in HNC. The HNC molecule can be considered as an efficient tracer to search and identify early stage sub-stellar mass objects.

\end{abstract}




\section{Introduction}

The relative abundance of CN, HCN, and HNC varies significantly among different astrophysical environments. The CN/HCN abundance ratio has been typically used as a probe of photon-dominated regions (PDRs). The ratio is found to be the largest near the source of the UV radiation field, and the intensity and density ratios decrease with increasing distance and optical depth from the UV source. In well-known PDRs such as Orion Bar or NGC 7023 regions, the CN/HCN ratio decreases from $\sim$8--10 close to the UV source to $\sim$1 at a distance of a few arcminutes from it (e.g., Hogerheijde et al. 1995; Schilke et al. 1992; Fuente et al. 1993; Cuadrado et al. 2016). The CN/HCN ratio can be further enhanced if the source of the UV flux is from the accretion zone in a protostar or protostellar disk, where the stellar FUV flux can be 10$^{2}$--10$^{4}$ times higher than the interstellar UV radiation field (ISRF) (e.g., Bergin et al. 2003; Chapillon et al. 2012). This results in the photo-dissociation of HCN by the Ly-$\alpha$ radiation from the star to yield CN. Some prime examples are the actively accreting T Tauri disk DM Tau and Herbig Ae stars MWC 480 and HD 163296 for which CN/HCN ratios of $>$10 have been measured (e.g., Dutrey et al. 1997; Thi et al. 2004; Chapillon et al. 2012). 

The two isomeric forms of HCN and HNC are common tracers of dense gas in molecular clouds. The HNC/HCN abundance ratio is observed to be $\approx$1 in cold and dense interstellar clouds, where the gas is shielded from the UV radiation, while in regions illuminated by UV photons, HNC becomes $\sim$5--10 times less abundant than HCN in both diffuse interstellar clouds and in PDRs (e.g., Schilke et al. 1992; Fuente et al. 1993; Liszt \& Lucas 2001; Godard et al. 2010). HNC/HCN ratios larger than unity have been reported for dark cloud cores, with an average value of $\sim$2 measured for more than 20 cores and the largest ratio being $\sim$5 for the L1498 core (e.g., Churchwell et al. 1984; Hirota et al. 1998). The HNC abundances are thus generally higher than HCN in cold, dense environments, in contrast to being significantly depleted in high-temperature ($>$20-40 K) regions. The temperature dependence of the HNC/HCN ratio possibly reflects isomerization mechanisms of HNC to HCN (e.g., Hirota et al. 1998) and suggests a chemical differentiation between the two isomers, with HNC having either a higher formation rate or lower destruction rate than HCN (e.g., Chenel et al. 2016). 

We have conducted the first substantial molecular line survey to study the chemistry of the envelopes in early-stage Class 0/I sub-stellar mass objects (M$_{total} \leq$ 0.08 M$_{\sun}$; L$_{bol} \leq$ 0.1 L$_{\sun}$), also termed as `proto-brown dwarfs' (proto-BDs). The aim of this exploratory study is to identify the best chemical tracers to probe the chemical composition of brown dwarfs during their early evolutionary stages. This survey will look into the analogies in the chemical properties between the proto-BDs and Class 0/I protostellar objects, in particular, to investigate if the molecular abundances of some of the commonly detected species in protostars show a different trend for bolometric luminosities below the sub-stellar limit. Our deep molecular line survey has revealed emission in 13 molecular species in proto-BD cores. This paper presents the results from the CN, HCN, and HNC observations. Section~\ref{obs} presents details on the sample and observations. In Sect.~\ref{analysis}, we present an analysis of the line profiles and the derivation of the line parameters, column densities, and the molecular abundances. Section~\ref{discussion} presents a discussion on the CN, HCN, and HNC abundance ratios in proto-BDs, and the trends in their molecular abundances in comparison with pre- and protostellar objects.


\section{Targets, Observations and Data Reduction}
\label{obs}

\subsection{Sample}

\begin{table*}
\centering
\caption{Sample}
\label{sample}
\begin{threeparttable}
\begin{tabular}{lllp{0.8cm}p{0.8cm}lp{1.2cm}} 
\hline
Object & RA (J2000) & Dec (J2000) & L$_{bol}$ (L$_{\sun}$)\tnote{a} & M$_{total}^{d+g}$ (M$_{Jup}$) & Classification\tnote{b} & Region \\
\hline
\multicolumn{7}{c}{Stage 0+I} \\
SSTc2d J182854.9+001833 (J182854) & 18h28m54.90s &  00d18m32.68s &  0.05 & 48$\pm$11 & Stage 0+I, Stage I, Class 0/I & Serpens \\ 
SSTc2d J182844.8+005126 (J182844) & 18h28m44.78s  & 00d51m25.79s &  0.04 & 70$\pm$20 & Stage 0+I, Stage 0, Class 0/I & Serpens \\ 
SSTc2d J183002.1+011359 (J183002) &  18h30m02.09s  & 01d13m58.98s &  0.09 & 74$\pm$18 & Stage 0+I, Stage 0, Class 0/I & Serpens \\ 
SSTc2d J182959.4+011041 (J182959) & 18h29m59.38s  & 01d10m41.08s &  0.008 & 62$\pm$20 & Stage 0+I, Stage II, Class 0/I & Serpens \\ 
SSTc2d J163143.8-245525 (J163143) & 16h31m43.75s & -24d55m24.61s &  0.16 & 65$\pm$14 &  Stage 0+I, Stage I, Class Flat & Ophiuchus \\ 
SSTc2d J182953.0+003607 (J182953) & 18h29m53.05s  & 00d36m06.72s &  0.15 & $<$25 		& Stage 0+I, Stage 0, Class Flat & Serpens \\ 
SSTc2d J163136.8-240420 (J163136) & 16h31m36.77s & -24d04m19.77s &  0.09 & 56$\pm$28 &  Stage 0+I, Stage I, Class Flat & Ophiuchus \\ 
 \multicolumn{7}{c}{Stage I-T/II} \\
SSTc2d J182940.2+001513 (J182940) & 18h29m40.20s &  00d15m13.11s &  0.074 & 28$\pm$12 & Stage II, Stage I-T, Class 0/I & Serpens \\ 
SSTc2d J182927.4+003850 (J182927) & 18h29m27.35s &  00d38m49.75s &  0.012 & $<$12 	& Stage II, Stage I-T, Class 0/I & Serpens \\ 
SSTc2d J182952.1+003644 (J182952) & 18h29m52.06s &  00d36m43.63s &  0.016 & $<$28 	& Stage II, Stage II, Class Flat & Serpens \\ 
\hline
\end{tabular}
\begin{tablenotes}
\item[a] Errors on L$_{bol}$ are estimated to be $\sim$30\%. 
\item[b] The first, second, and third values are using the classification criteria based on the integrated intensity in the HCO$^{+}$ (3-2) line, the physical characteristics, and the SED slope, respectively.
\end{tablenotes}
\end{threeparttable}
\end{table*}

Our sample consists of 10 proto-BD candidates, 8 of which were identified in Serpens and 2 in the Ophiuchus region. Table~\ref{sample} lists the properties for the targets in the sample. We applied the same target selection and identification criteria as described in Riaz et al. (2015; 2016), which is based on a cross-correlation of infrared data from the UKIDSS and/or 2MASS, Spitzer, and Herschel archives, followed by deep sub-millimeter 850$\mu$m continuum observations obtained with the JCMT/SCUBA-2 bolocam. Out of the 10 targets, 7 are detected at a $>$5-$\sigma$ level in the 850$\mu$m sub-millimeter continuum observations, while the rest are detected at a marginal ($\sim$2-$\sigma$) level. We have derived the total (dust+gas) mass, M$^{d+g}_{total}$, for the proto-BD candidates using their 850 $\mu$m flux density or the $\sim$2-$\sigma$ upper limits for the marginal detections, assuming a dust temperature of 10 K, a gas to dust mass ratio of 100, and a dust mass opacity coefficient at 850 $\mu$m of 0.0175 cm$^{2}$ gm$^{-1}$ (Ossenkopf \& Henning 1994). The masses thus derived are listed in Table~\ref{sample}, and are found to be in the range of $\sim$0.01--0.1 M$_{\sun}$. The bolometric luminosity, L$_{bol}$, for the targets as measured from integrating the observed infrared to sub-millimeter spectral energy distribution (SED) is in the range of $\sim$0.008--0.1 L$_{\sun}$. We assumed a distance to the Serpens region of 436$\pm$9 pc (Ortiz-Leon et al. 2017; Dzib et al. 2010) based on astrometric observations, and 140$\pm$6 pc to Ophiuchus (Mamajek 2008; Schlafly et al. 2014). We note that since these objects are still accreting, the total mass may not represent the final mass for these systems. However, based on the accretion models by Baraffe et al. (2012), the L$_{bol}$ for these candidates is below the luminosity threshold considered between very low-mass stars and brown dwarfs ($<$0.2 L$_{\sun}$), and considering the sub-stellar mass reservoir in the (envelope+disk) for these systems, the final mass is also expected to stay within the sub-stellar limit.

We have determined the evolutionary stage of the candidates using the Stage 0+I/II criteria based on the strength in the HCO$^{+}$ (3-2) line emission, and the Stage 0/I/I-T/II classification criteria based on the physical characteristics of the system. A detailed description of these criteria is presented in Riaz et al. (2016). The Stage thus determined has been compared to the Class 0/I/Flat classification based on the 2-24 $\mu$m slope of the SED. Stage I-T and Class Flat are considered to be intermediate between the Stage I and Stage II, or Class I and Class II evolutionary phases. As listed in Table~\ref{sample}, we find a good match between the Stage of the system based on the physical characteristics and the strength in the molecular line emission, whereas the classification based on the SED slope does not relate well to this evolutionary stage. This is particularly true for the Class Flat cases, three of which are found to be Stage 0/I objects. The physical structure from modelling of these objects indicates an edge-on inclination, which could result in a flatter spectral slope in the SED compared to genuine Stage 0/I systems (e.g., Whitney et al. 2003; Riaz et al. 2016). Thus, 7 proto-BD candidates in our sample are both in the early Stage 0/I evolutionary phase and sub-stellar in nature, and can be considered as bona fide proto-BDs, whereas the sources J182940, J182927, and J182952 are sub-stellar objects that are in the intermediate Stage I-T/Stage II phase (Table~\ref{sample}). A detailed analysis on the physical and chemical structure of the targets based on SED and line modelling will be presented in a forthcoming paper.


\subsection{Molecular Line Observations and Data Reduction}
\label{observations}

We observed the 10 targets with the IRAM 30m telescope using the EMIR heterodyne receiver in May, 2016, and December, 2017. We observed the CN (2-1), HCN (3-2), and HNC (3-2) lines in the E230 band (Table~\ref{line-obs}), using the FTS backend in the wide mode with a spectral resolution of 200 kHz ($\sim$0.3 km s$^{-1}$ at 226 and 267 GHz). Follow-up observations in the isotopologues of HN$^{13}$C (3-2), H$^{13}$CN (3-2), $^{13}$CN (2-1), and HC$^{15}$N (3-2) lines were obtained for the proto-BD J182844 that shows the strongest emission in both HCN (3-2) and CN (2-1). Due to time limitations, these isotopologues could not be observed for the whole sample. The observations were taken in the frequency switching mode with a frequency throw of approximately 7 MHz. The source integration times ranged from 3 to 4 hours per source per tuning reaching a typical RMS (on T$_{A}^{*}$ scale) of $\sim$0.02--0.05 K. The telescope absolute RMS pointing accuracy is better than 3$\arcsec$ (Greve et al. 1996). All observations were taken under good weather conditions (0.08 $<$$\tau$ $<$0.12; PWV $<$2.5 mm). The HCN (3-2) spectrum for the proto-BD J183002 shows instrumental effects and was removed from further analysis.


The spectra were calibrated at the telescope onto the antenna temperature T$_{A}^{*}$ scale using the standard chopper wheel method. The absolute calibration accuracy for the EMIR receiver is around 10\% (Carter et al. 2012). The telescope intensity scale was converted into the main beam brightness temperature (T$_{mb}$) using standard beam efficiencies that vary between 59\% at 210 GHz to 49\% at 280 GHz. The half power beam width of the telescope beam varies between 11$\arcsec$ at 210 GHz and 9$\arcsec$ at 260 GHz. The spectral reduction was conducted using the CLASS software (Hily-Blant et al. 2005) of the GILDAS facility\footnote{http://www.iram.fr/IRAMFR/GILDAS}. The standard data reduction process consisted of averaging multiple observations for each transition line, extracting a subset around the line rest frequency, folding the spectrum to correct for the frequency switching effects, and finally a low-order polynomial baseline was subtracted for each spectrum. In case of a line detection, a single or multiple Gaussians were fitted, as described further in Sect.~\ref{analysis}.

\section{Data Analysis and Results}
\label{analysis}

\subsection{Line Profiles and Parameters}

We have measured the parameters of the line center, line width, the peak and integrated intensities for all spectra (Figs.~\ref{hcn-figs};~\ref{cn-figs};~\ref{hnc-figs}). A single-peaked fit was used for typical Gaussian shaped profiles. For all spectra that show non-Gaussian line shapes with multiple peaks, we have used the IDL routine $MPFIT$ in an attempt to decompose the observed profile into multiple Gaussians. The line fits are shown in Figs.~~\ref{hcn-figs};~\ref{cn-figs};~\ref{hnc-figs}, and the parameters derived from the Gaussian fits are listed in Tables~\ref{hcn};~\ref{cn};~\ref{hnc}. Also listed is the signal-to-noise ratio (SNR) in each line detection. We estimate uncertainties of $\sim$15\%-20\% for the integrated intensity, $\sim$0.1 km s$^{-1}$ in V$_{lsr}$, $\sim$10\% on T$_{mb}$ and $\Delta$v. We have opted to fit multiple Gaussians to the HCN and CN hyperfine lines to measure the line parameters instead of using the CLASS hyperfine fitting routine as it failed to provide a good-quality fit to the ``anomalies'' observed in the hyperfine structure, as discussed in Sect.~\ref{class}. In case of a non-detection, a 3-$\sigma$ upper limit for HNC (3-2) was calculated by integrating over the velocity range of $\pm$2 km s$^{-1}$ from the cloud systemic velocity of $\sim$8 km s$^{-1}$ for Serpens (e.g., Burleigh et al. 2013) and $\sim$4.4 km s$^{-1}$ for Ophiuchus (e.g., White et al. 2015). The 3-$\sigma$ upper limits for HCN (3-2) and CN (2-1) was estimated by integrating over the velocity range of the brightest hyperfine component, which is $J$=3-2, $F \rightarrow F^{\prime}$=4 $\rightarrow$ 3 for HCN (3-2), and $J \rightarrow J^{\prime}$=5/2 $\rightarrow$ 3/2, $F \rightarrow F^{\prime}$=7/2 $\rightarrow$ 5/2 for CN (2-1). The spectra for J182953 and J182952 are shifted from the cloud systemic velocity by $\sim$2 km s$^{-1}$ (Figs.~\ref{cn-figs};~\ref{hnc-figs}). This suggests that the observed emission for these objects is tracing molecular gas at a different bulk velocity than the others, which is possibly due to differences in the dynamical processes such as collapse versus co-rotation or expansion that can lead to slight Doppler shifts of the gas components (e.g., Evans 1999). Both Serpens and Ophiuchus are very large molecular cloud complexes, with significant velocity gradients across, therefore all of the objects in the whole cloud cannot be associated with a single cloud systemic velocity.

\begin{table*}
\centering
\caption{IRAM Molecular line observations}
\label{line-obs}
\begin{threeparttable}
\begin{tabular}{lllllllll} 
\hline
Molecule & Line & Frequency (GHz) 	& $E_{j}$/k (K) & $A_{jk}$ (s$^{-1}$) & $n_{crit}^{thin}$(10 K) (cm$^{-3}$)   	\\
\hline
CN \tnote{a}		& 2-1 		& 226.8748 		& 16.34 	& 1.1E-4 	& 3E+6		\\
HCN	\tnote{a}		& 3-2 		& 265.8865 		& 25.52 	& 8.4E-4 	& 1E+7		\\
HNC				& 3-2 		& 271.9811 		& 26.11 	& 9.3E-4 	& 5E+6		\\
HC$_{3}$N		& 24-23		& 218.3247		& 120.50 	& 8.3E-4 	& 5E+6 		\\
HC$_{3}$N		& 25-24 		& 227.4189 		& 130.98	& 9.3E-4 	& 5E+6		\\
HC$_{3}$N		& 27-26 		& 245.6063 		& 153.24	& 1.3E-3 	& 5E+6		\\
\hline
\end{tabular}
\begin{tablenotes}
  \item[a] Hyperfine splitting observed. The frequency is listed for the brightest hyperfine component, which is $J$=3-2, $F$=4-3 for HCN (3-2) and $J$=5/2-3/2, $F$=7/2-5/2 for CN (2-1). 
\end{tablenotes}
\end{threeparttable}
\end{table*}


\subsubsection{HCN and CN hyperfine lines}
\label{hcn-lines}

\begin{figure*}
  \centering       
     \includegraphics[width=2.8in]{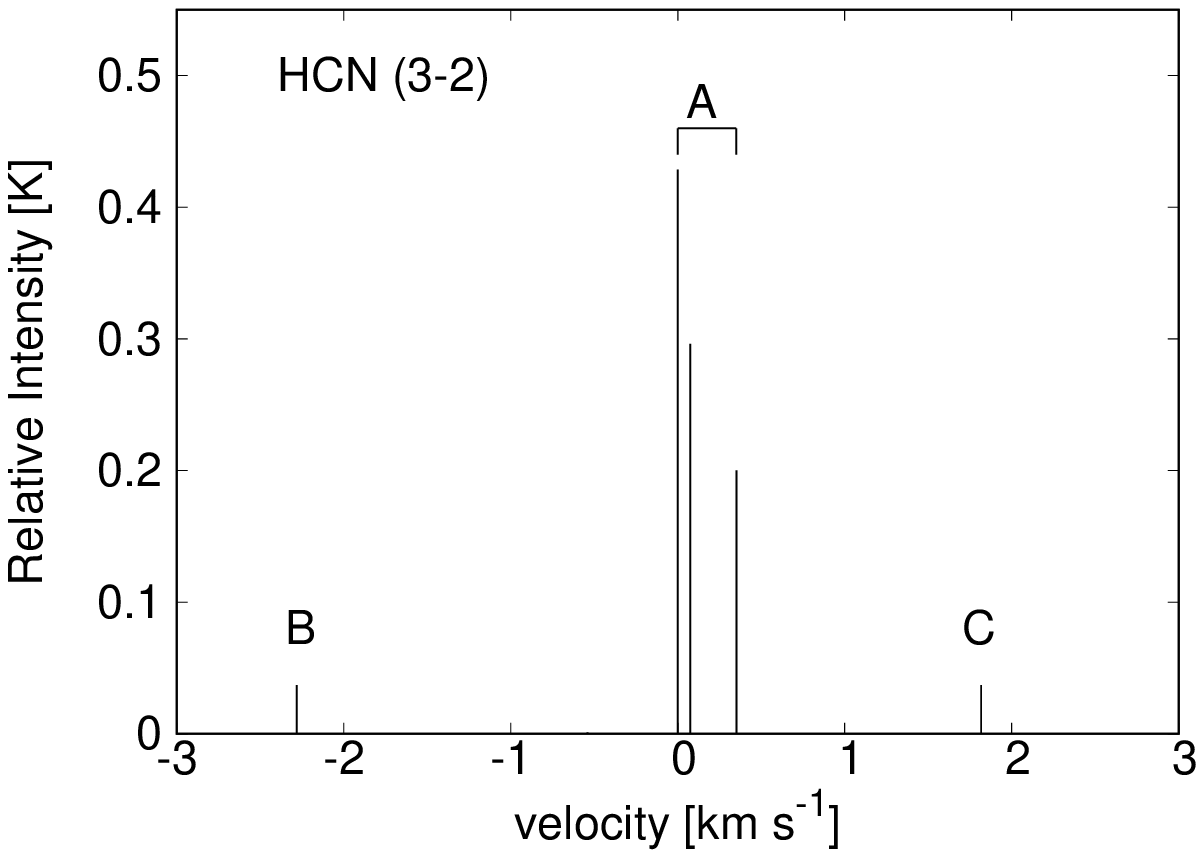}         \\ \vspace{0.1in}
     \includegraphics[width=2.8in]{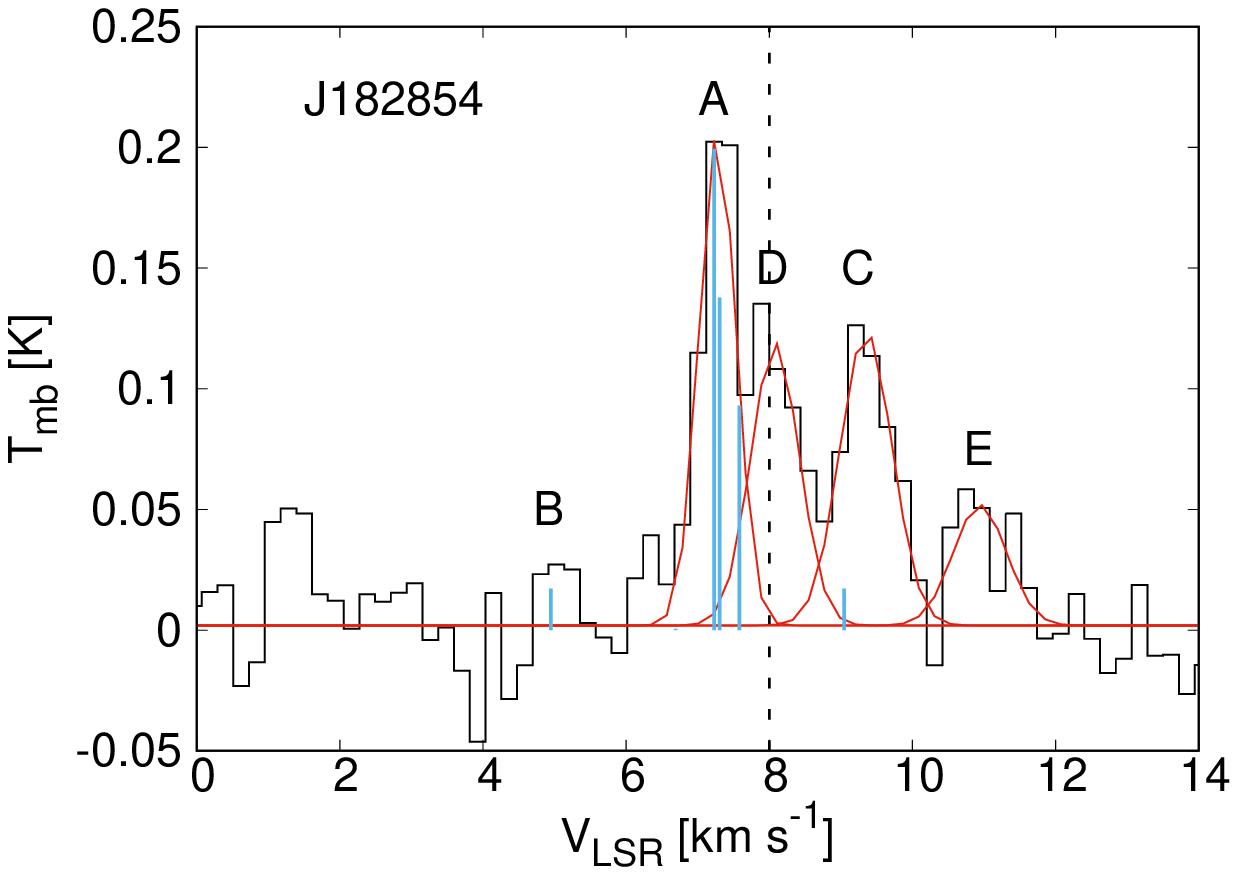}
     \includegraphics[width=2.8in]{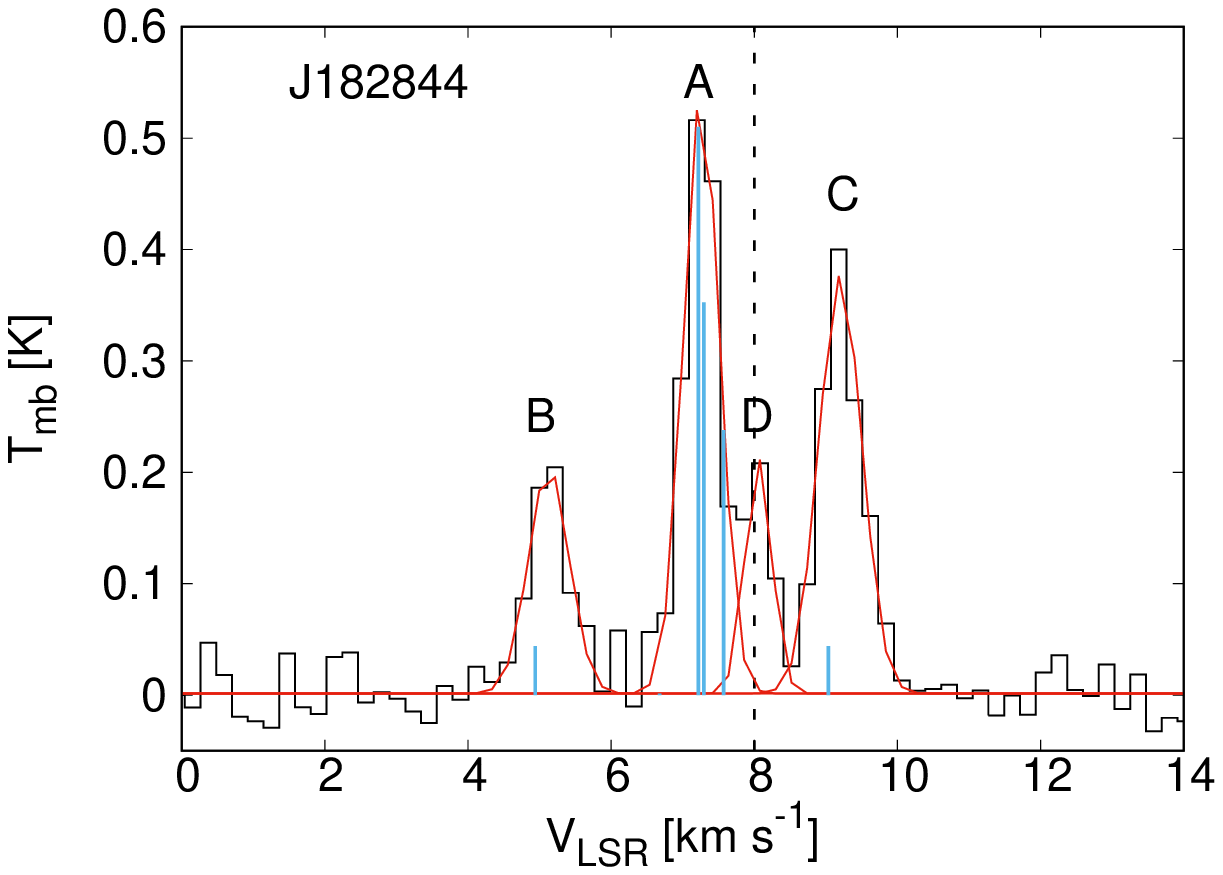}
     \caption{The observed HCN (3-2) spectra (black) with the Gaussian fits (red) for the proto-BDs. Top panel shows the hyperfine structure expected under optically thin LTE conditions with the three main hyperfine lines labelled. The hyperfine component `A' is a composite of four spectrally unresolved components. Blue lines in the observed spectra mark the hyperfine components shown in the top panel, and have been scaled to match the peak intensity in the `A' component of the observed spectrum. Black dashed line marks the cloud systemic velocity of $\sim$8 km/s and $\sim$4.4 km/s in Serpens and Ophiuchus, respectively. }
     \label{hcn-figs}     
  \end{figure*}

There is detection in HCN (3-2) for the proto-BDs J182854 and J182844, and the spectra show hyperfine splitting. Figure~\ref{hcn-figs} shows the HCN (3-2) spectra; also over plotted are the relative intensities of the hyperfine components. The hyperfine structure expected under optically thin LTE conditions is plotted in Fig.~\ref{hcn-figs} (top panel), and the frequencies and the relative intensities of the hyperfine components are listed in Table~\ref{hcn-hfs}. The HCN (3-2) hyperfine structure data is from Laughnane et al. (2012). HCN (3-2) has six hyperfine components. The central bright feature (labelled `A'; Fig.~\ref{hcn-figs}) is a composite of four components ($F \rightarrow F^{\prime}$ = 2 $\rightarrow$ 1; 3 $\rightarrow$ 2; 4 $\rightarrow$ 3; 2 $\rightarrow$ 3) that are not spectrally resolved, and the spectrum thus shows a strong central component and two satellite components (labelled `B' and `C'). Under optically thin LTE conditions, the relative weightings of the HCN (3-2) hyperfine lines are 1:25:1 or 0.037:0.926:0.037. The satellite hyperfine components `B' and `C' are thus expected to be of equal intensities. However, HCN molecular line surveys of both low-mass dense cores and massive star-forming turbulent cores find the `C' component to be ``anomalous'', such that its intensity is always found to be higher than the `B' component (e.g., Loughane et al. 2012).

The term ``anomalous'' implies different hyperfine line ratios than expected due to various effects, as discussed below. The relative intensities of the hyperfine components can be used as a measure of the extent of anomalies seen in the observed hyperfine structure. Under optically thin conditions, the ratio of the intensity of the satellite components relative to the main component (B/A and C/A) would be equal to 0.04, assuming that the level populations of the hyperfine states are in LTE and hence proportional to the statistical weights. Under optically thick conditions, the lines would saturate and reach equal intensities, or a line intensity ratio equal to 1. The intermediate ratios between 0.04 and 1 would then indicate anomalous sources (Loughane et al. 2012).

We see clear anomalies in the HCN (3-2) spectra for the proto-BDs (Fig.~\ref{hcn-figs}). The `B' component in the HCN spectrum for J182854 is suppressed, even undetected, while the emission in `C' is boosted to intensities nearly comparable to the central component (Fig.~\ref{hcn-figs}). J182844 shows much stronger emission in `B' and `C' than expected, and the component `C' is a factor of $\sim$2 stronger than `B'. Table~\ref{hcn-ratios} lists the integrated intensity ratios of the outlying to the central hyperfine component, B/A and C/A, and the ratio of the satellite components, B/C. The B/A and C/A line ratios are $\sim$0.3-0.4, intermediate between the expected ratio of 0.04 but smaller than the saturation limit of unity. The ratio of the satellite components, B/C is $\sim$0.4-0.5, less than the expected ratio of 1, and indicates that these hyperfine components show anomalous intensities. Note that the central component `A' has another feature (labelled `D') that lies adjacent to it close to the cloud systemic velocity ($\sim$8 km s$^{-1}$) (Fig.~\ref{hcn-figs}). The features `D' and `E' are unassociated with any of the hyperfine components, and are likely the contamination from the foreground/background cloud material. The combined profile of `A'+`D' gives the appearance of a ``self-absorbed'' blue-asymmetric line with a dip close to the cloud systemic velocity. The possible causes for the self-absorbed shape of the profile, which is typically indicative of infall activity, are discussed in Sect.~\ref{overlap}.



The CN (2-1) hyperfine lines are detected in J182844, J182959, J183002, J182953, and J163143, with weak detection in J182952 (Fig.~\ref{cn-figs}). CN (2-1) has 19 hyperfine components, 11 of which have relative intensities of $<$0.01 and are too weak to be detected. The hyperfine structure of the 9 brightest components under optically thin LTE conditions is plotted in the top panels in Fig.~\ref{cn-figs}, and the frequencies and the normalized intensities of the hyperfine components are listed in Table~\ref{cn-hfs}. The CN (2-1) hyperfine structure data is from Hily-Blant et al. (2010). The 9 brightest components are detected in two proto-BDs, J182844 and J182959, while the rest mainly show emission in the central `C', `D', and `E' components that are blended together into a broad profile centered close to the cloud systemic velocity. In particular, the `A' and `B' hyperfine components, which are the faintest with the lowest relative intensities (Table~\ref{cn-hfs}), are only detected in the proto-BDs J182844 and J182959 that show the brightest emission in CN, but are undetected in the remaining objects. This can be expected from flux-limited observations with the same on-source integration time for all targets. The high-velocity hyperfine components are spectrally resolved, and are weaker in strength in J182959, J183002, and J182953 compared to J182844.

There are anomalies seen in the observed CN spectra when compared with the expected hyperfine structure shown in the top panels of Fig.~\ref{cn-figs}. The anomalies are not similar for all sources or even for the individual hyperfine components. For instance, the `A' component is as strong as `D' in J182844, and the high-velocity `F', `G', `I' components are nearly the same intensity as `H' in J182844 and J182959. Table~\ref{cn-ratios} lists the the ratio of the integrated intensity of each hyperfine component relative to the sum of the integrated intensities for the `C'+`D'+`E' components. Since the `C', `D', and `E' components are blended, we have considered the sum of the intensities in these components. The ratios are listed for J182844, J182959, and J183002, which are the only objects that show emission in the weaker hyperfine components. Table~\ref{cn-ratios} also lists the line ratios expected under LTE optically thin conditions, which have been calculated from the normalized intensities listed in Table~\ref{cn-hfs}. The line ratios for `F', `G', `H' are the closest to the expected LTE value, suggesting that these components are the least anomalous ones. The `A', `B', and `I' components appear the most anomalous possibly due to the effects of high opacity as their line ratios are $\sim$10 times higher than the optically thin case.

Hyperfine anomalies are, in all cases, related to non equal excitation temperature among the hyperfine lines of a given multiplet. Yet, there can be several causes rather than just a non-LTE opacity effect (e.g., Guilloteau \& Baudry 1981, Walmsley et al 1982; Loughane et al. 2012; Magalhaes et al 2018). Radiative transfer effects of absorption or scattering of the photons emitted from the high-density core in the less dense outer envelope could also produce anomalies in the expected relative strengths of the hyperfine components. A line overlap in a higher rotational level transition could cause a small change in the line strength and disturb the expected line shape. It is difficult to discern the effects of a large optical depth from radiative transfer, line trapping effects, and extra velocity components along the line of sight, which may produce higher or lower relative line intensity ratios of the hyperfine components than the expected values.



 \begin{figure*}
   \centering       
     \includegraphics[width=2.8in]{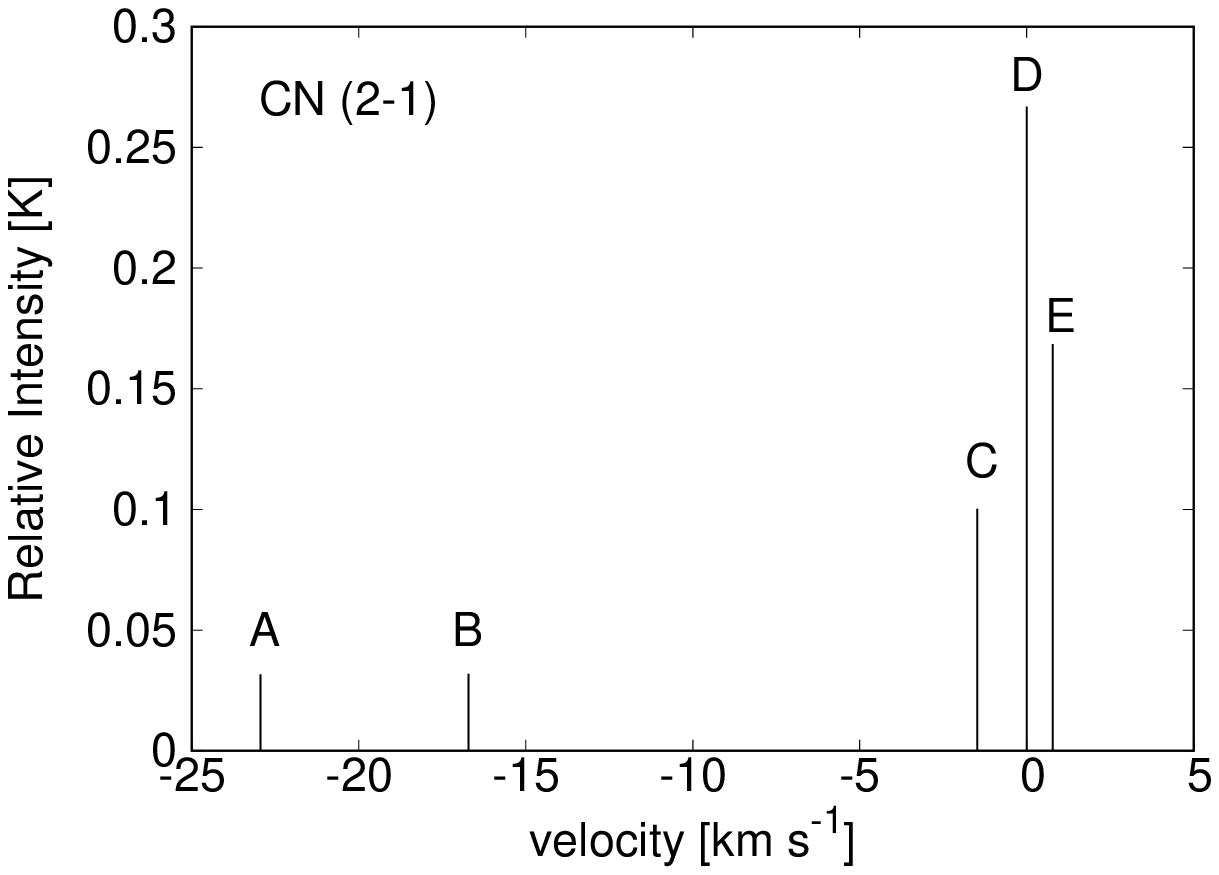}      
     \includegraphics[width=2.8in]{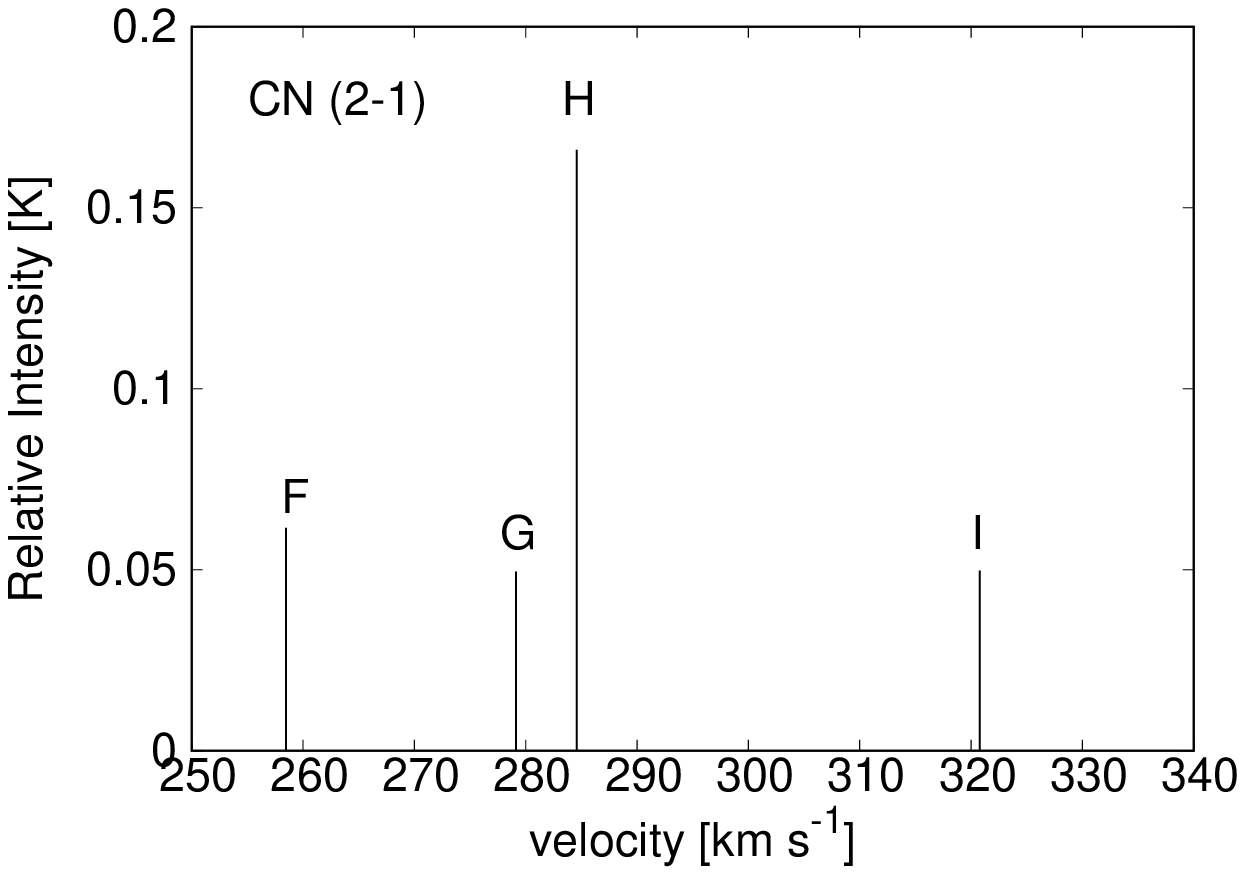}       \\                
     \includegraphics[width=2.8in]{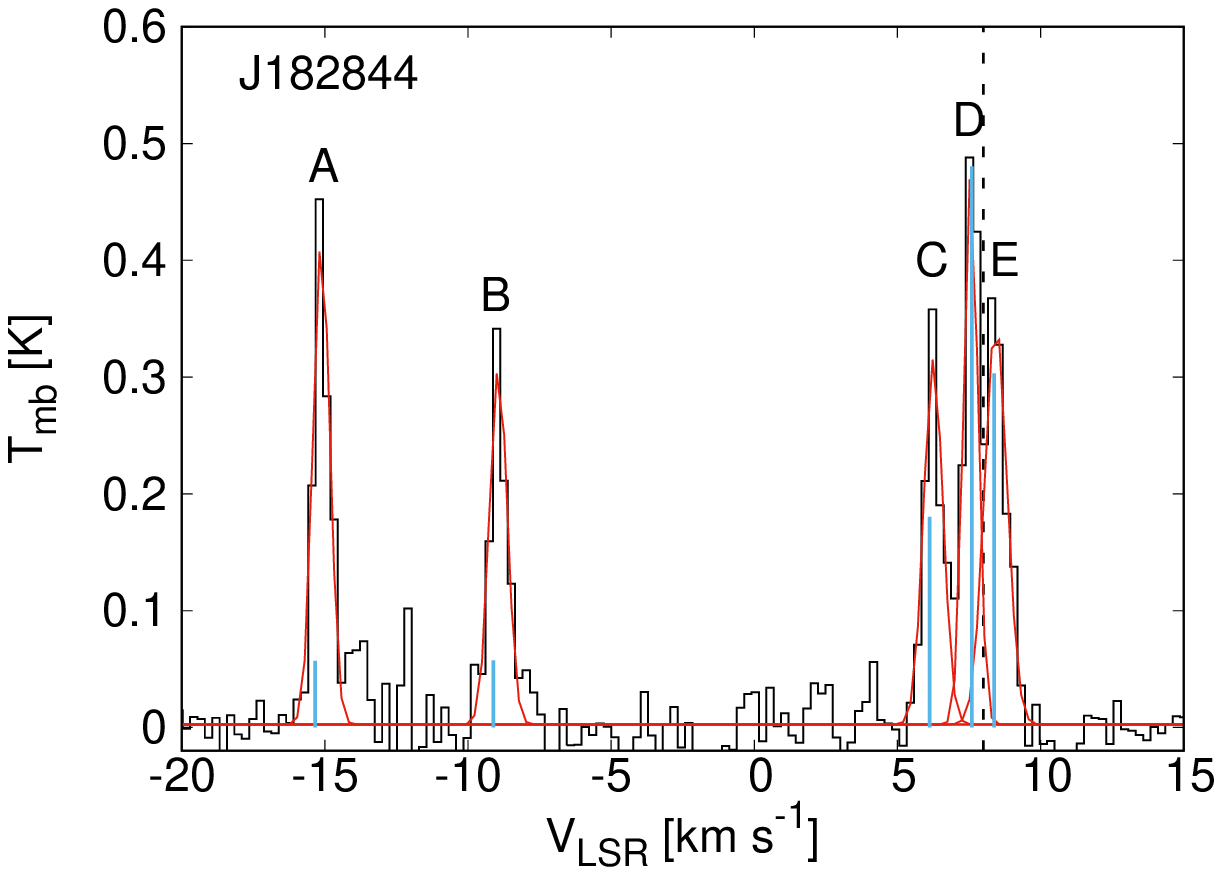}      
     \includegraphics[width=2.8in]{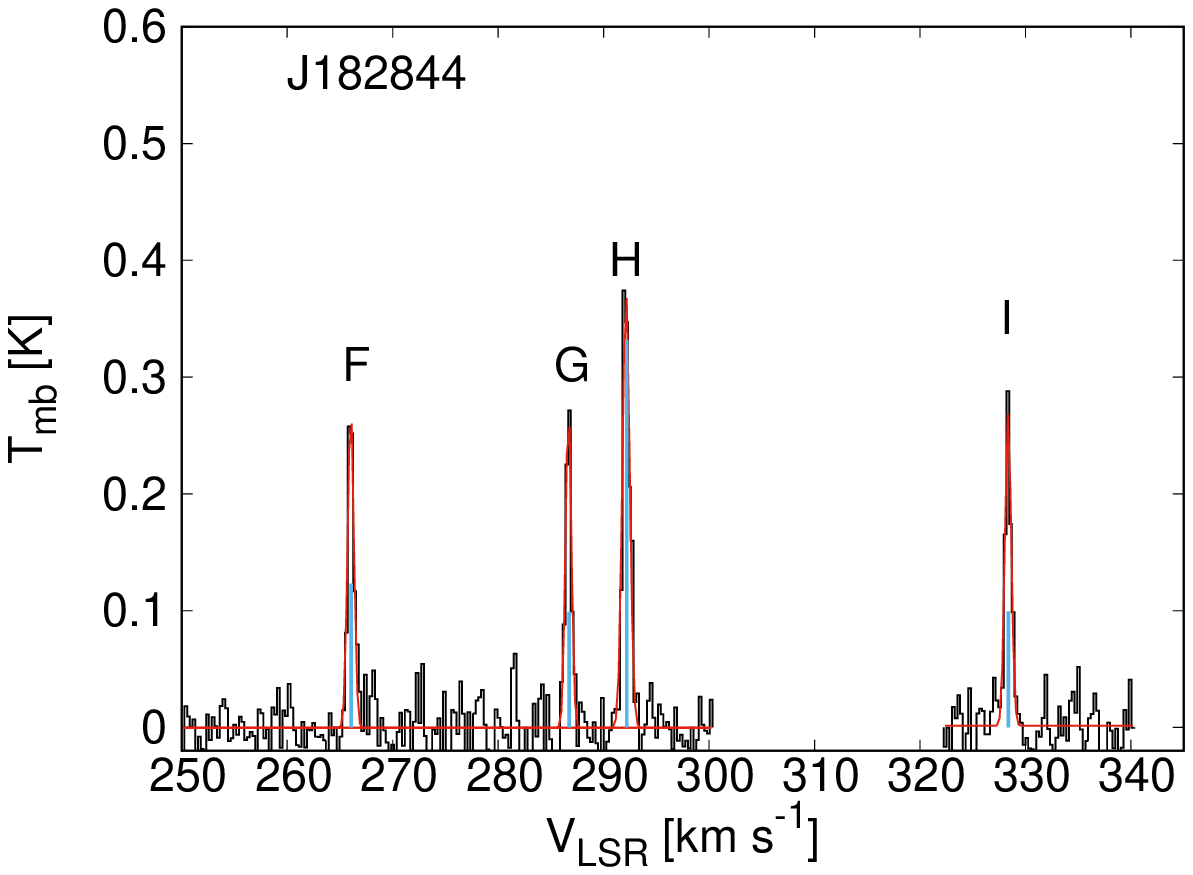}       \\     
     \includegraphics[width=2.8in]{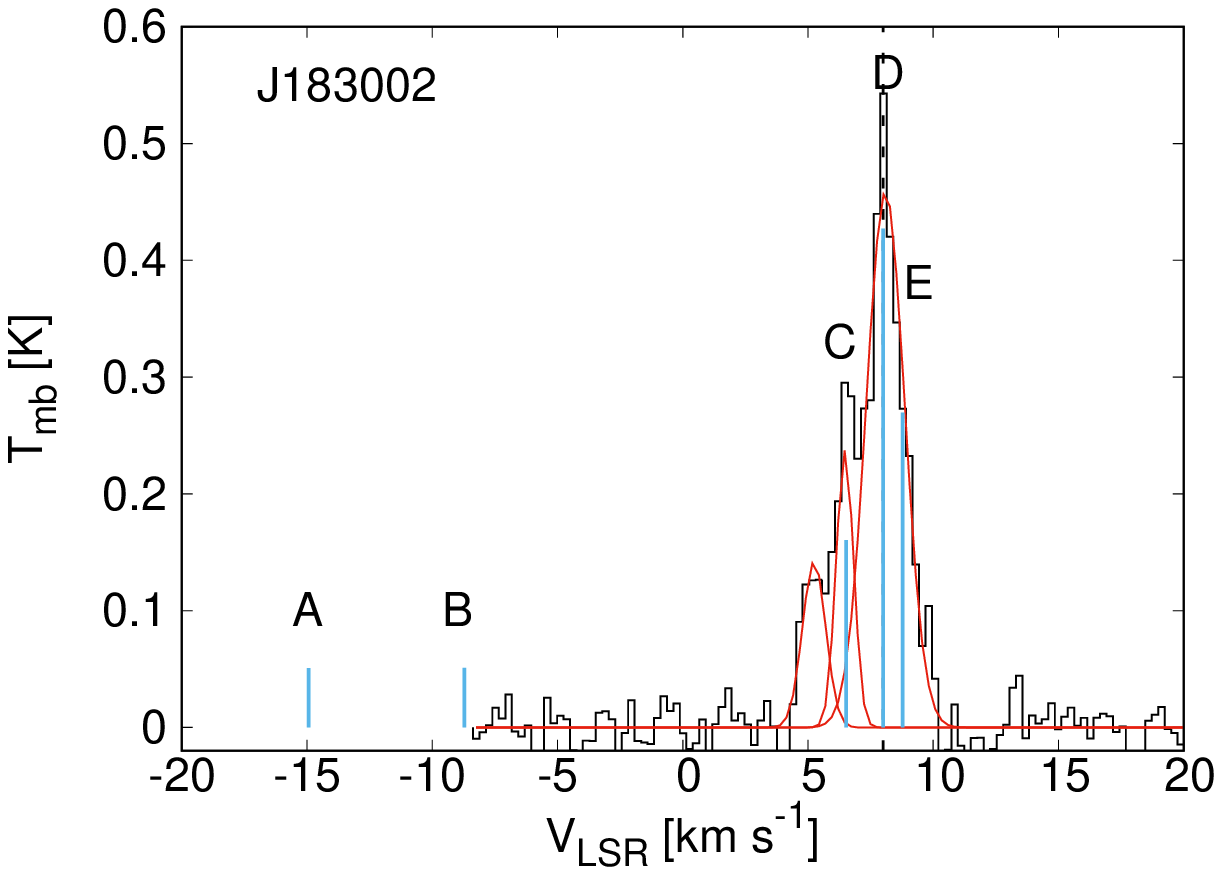}      
     \includegraphics[width=2.8in]{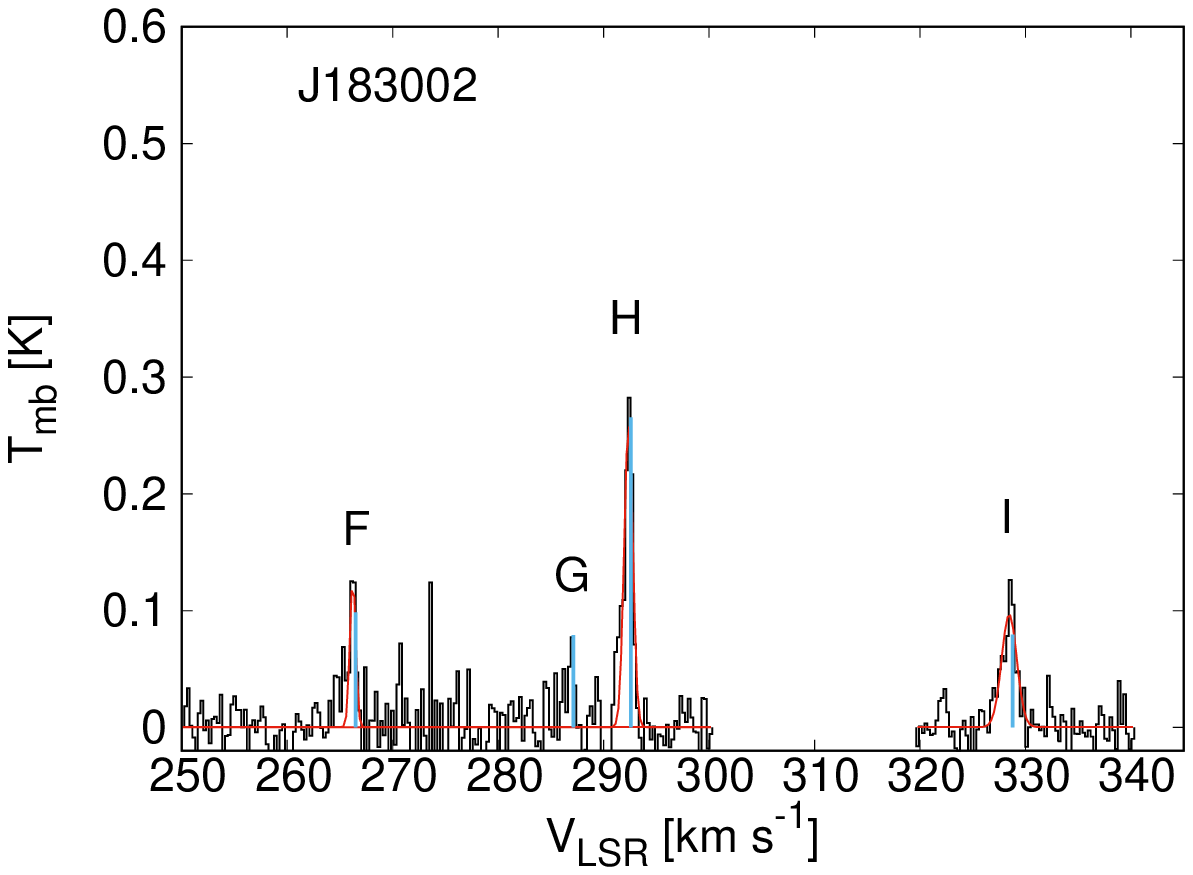}       \\       
     \includegraphics[width=2.8in]{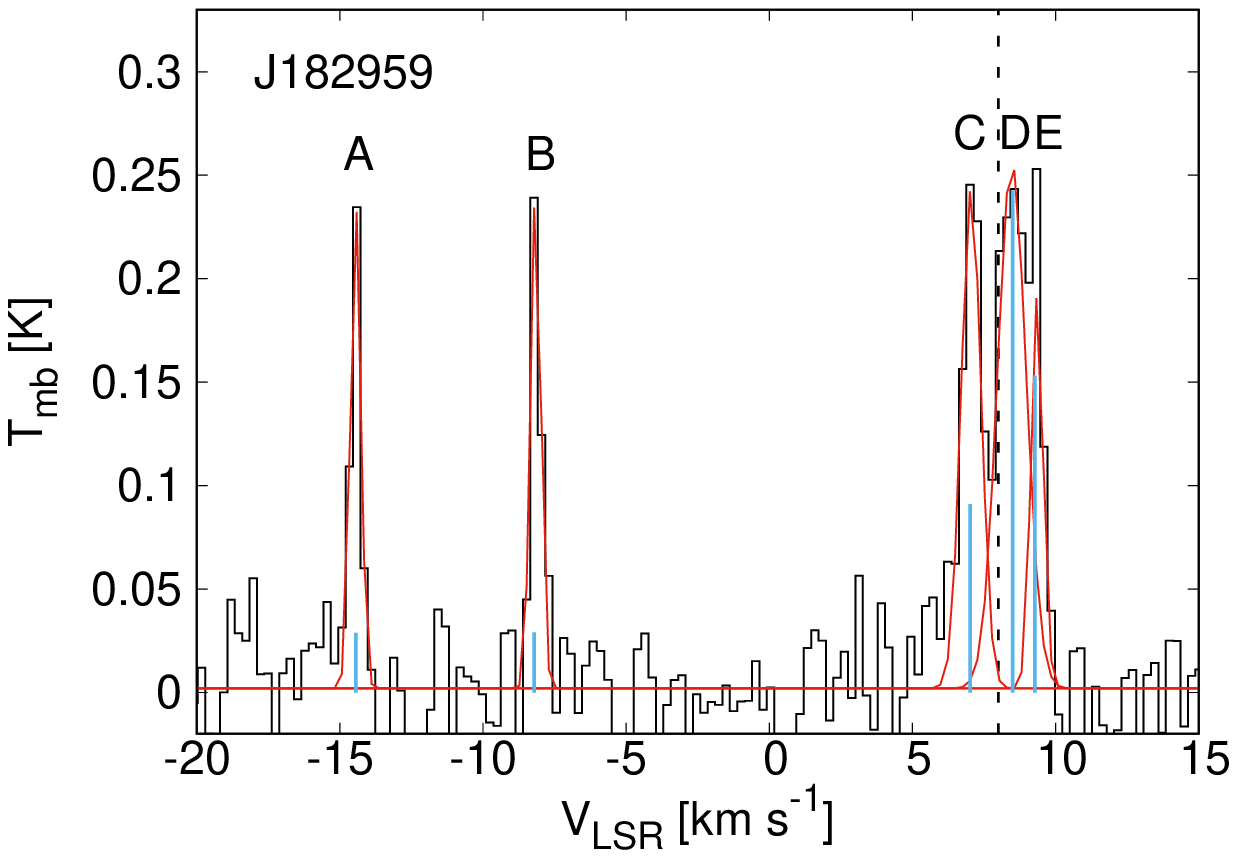}  
     \includegraphics[width=2.8in]{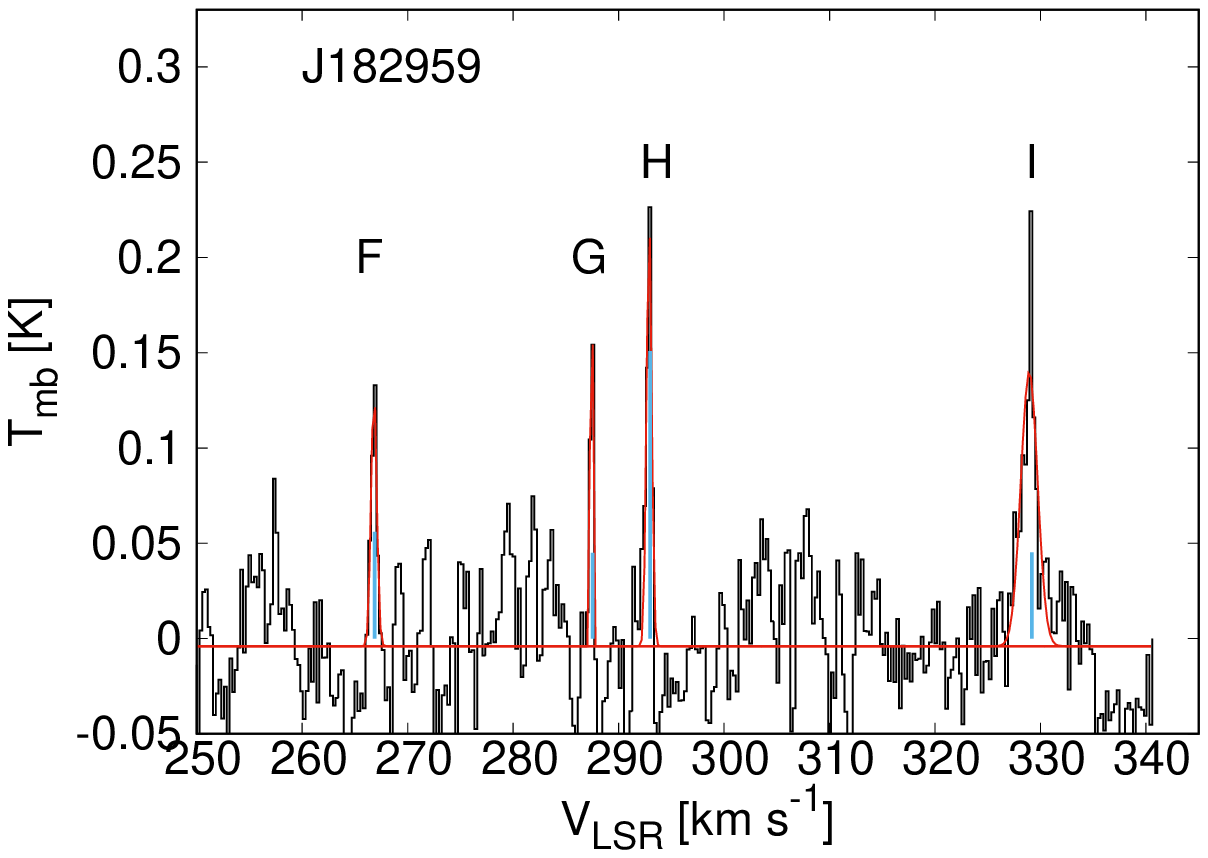}      \\       
     \caption{The observed CN (2-1) spectra (black) with the Gaussian fits (red). Top panel shows the hyperfine structure expected under optically thin LTE conditions with the main hyperfine lines labelled. Blue lines in the observed spectra mark the hyperfine components shown in the top panel, and have been scaled to match the peak intensity in the `D' component of the observed spectrum. Black dashed line marks the cloud systemic velocity of $\sim$8 km/s and $\sim$4.4 km/s in Serpens and Ophiuchus, respectively. }
     \label{cn-figs}  
  \end{figure*}       

\setcounter{figure}{1}    
 \begin{figure*}     
   \centering           
      \includegraphics[width=2.8in]{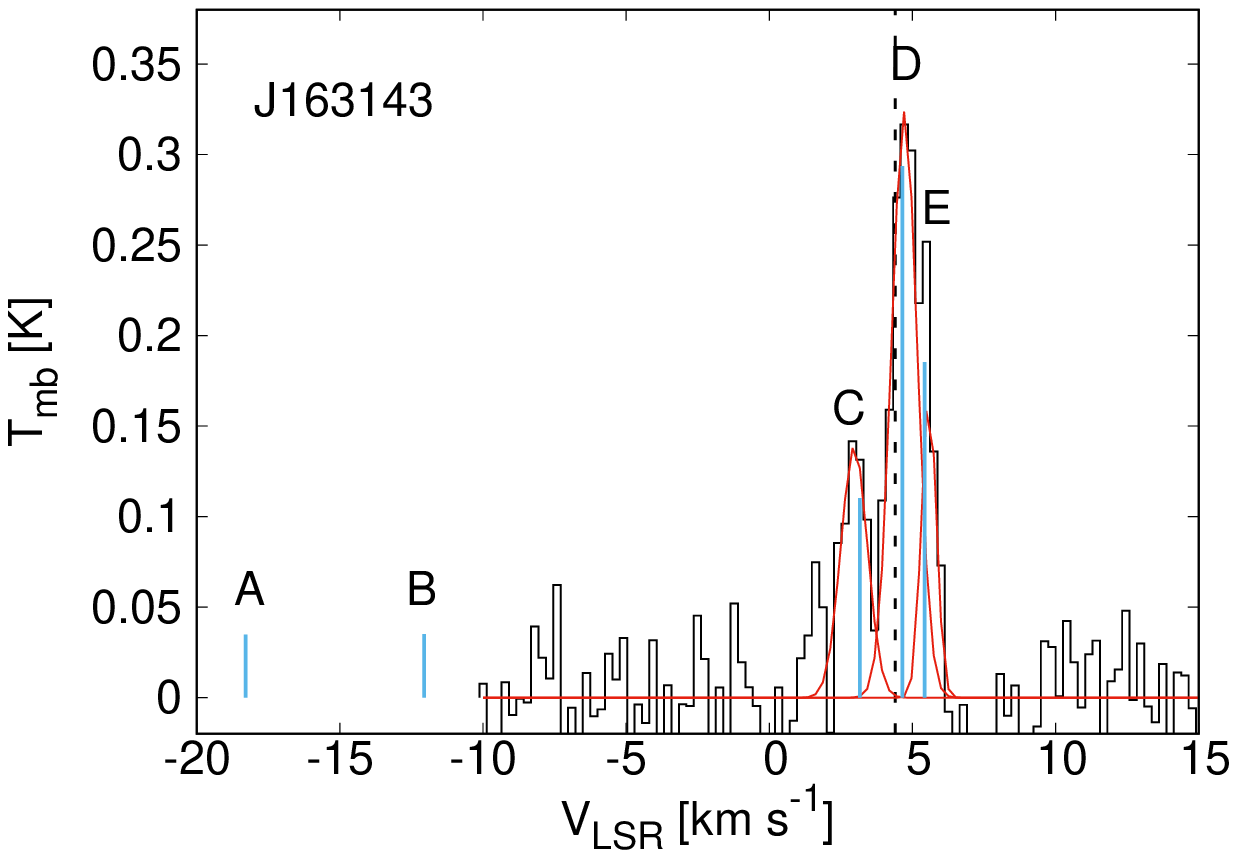}        \\     
     \includegraphics[width=2.8in]{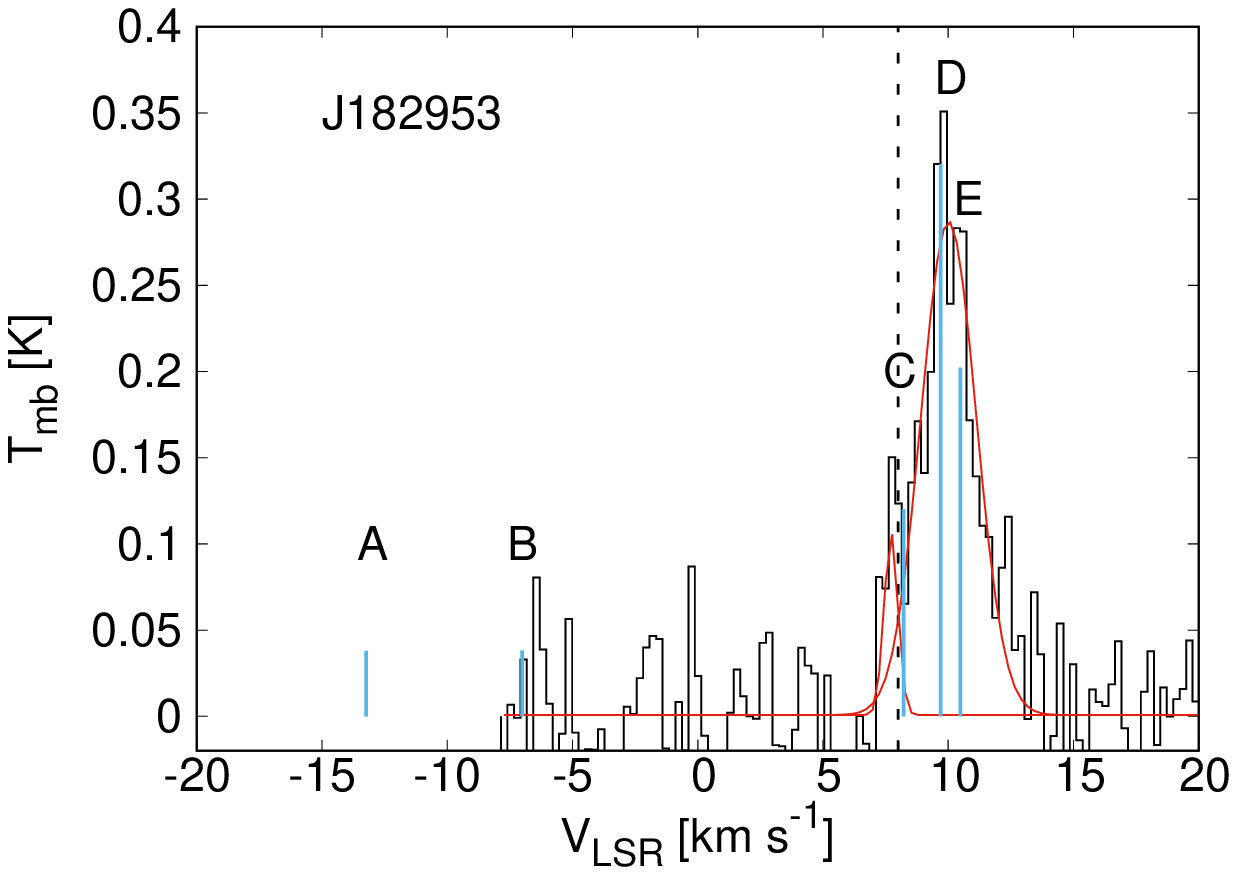}      
     \includegraphics[width=2.8in]{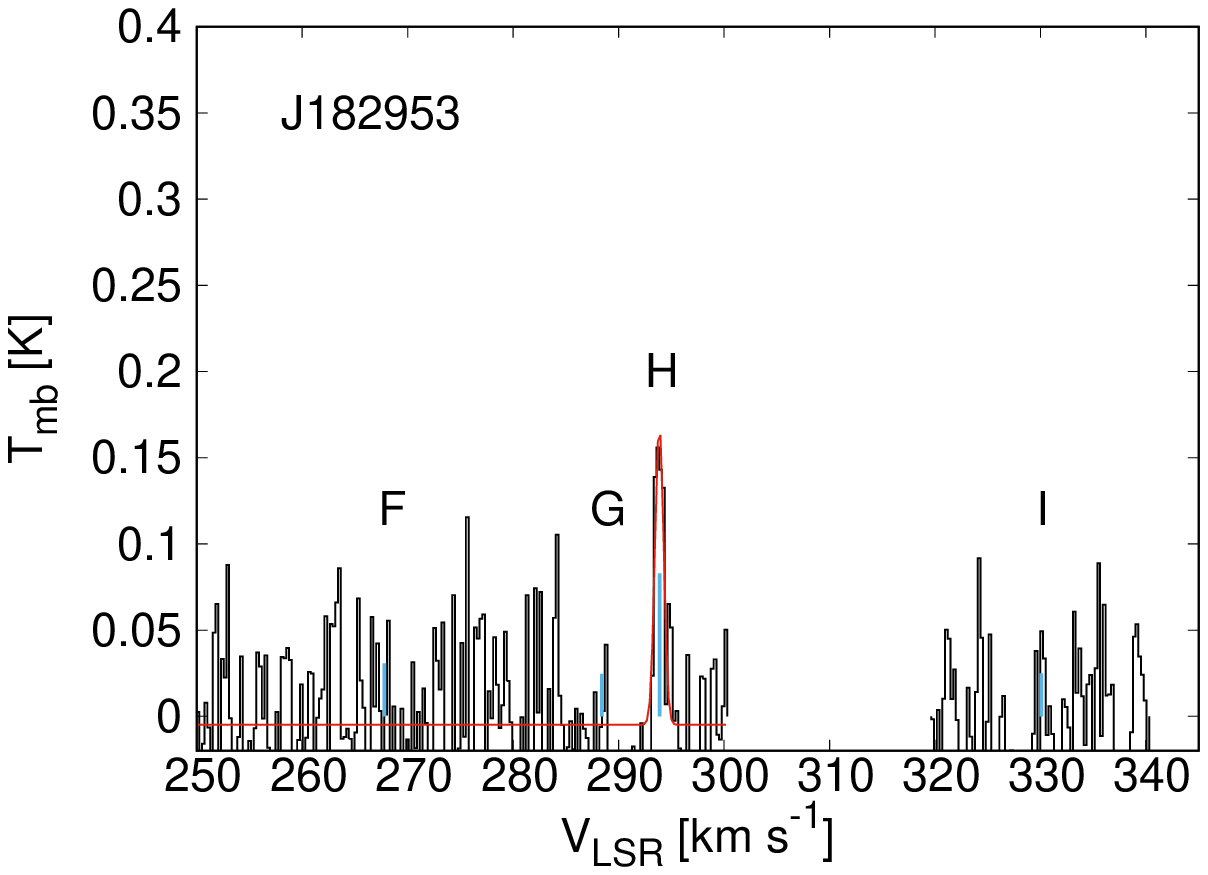}       \\       
     \includegraphics[width=2.8in]{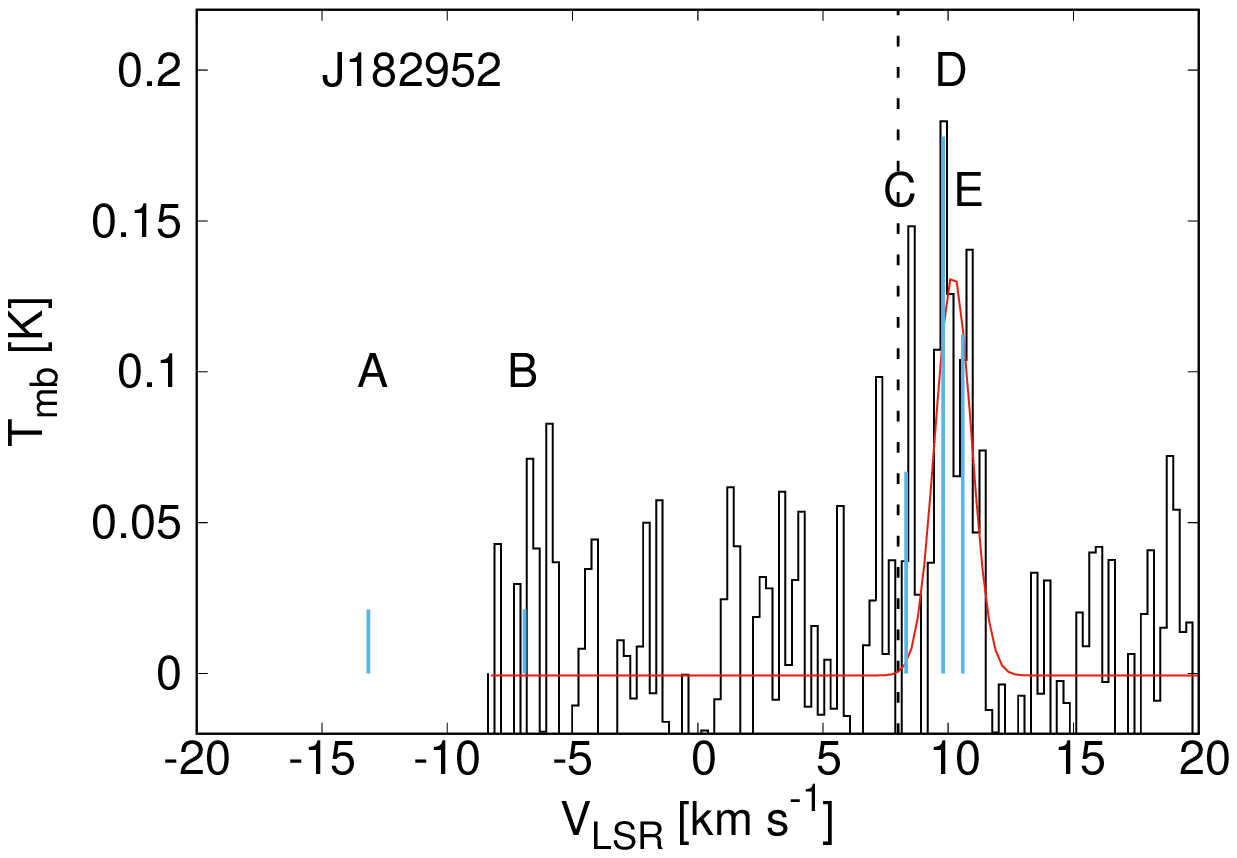}               
     \caption{Continued. }
  \end{figure*}

\subsubsection{HNC}

\begin{figure*}
  \centering              
     \includegraphics[width=2.2in]{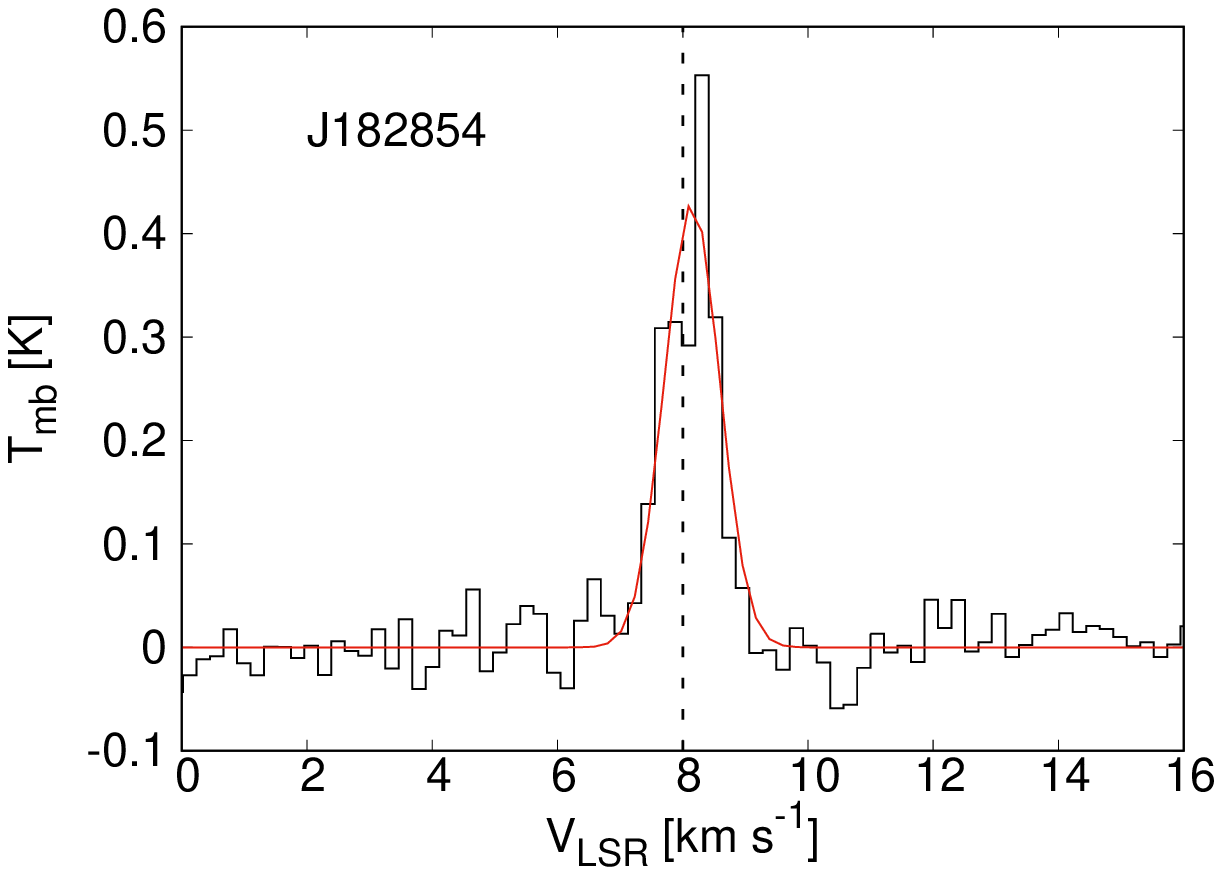}
     \includegraphics[width=2.2in]{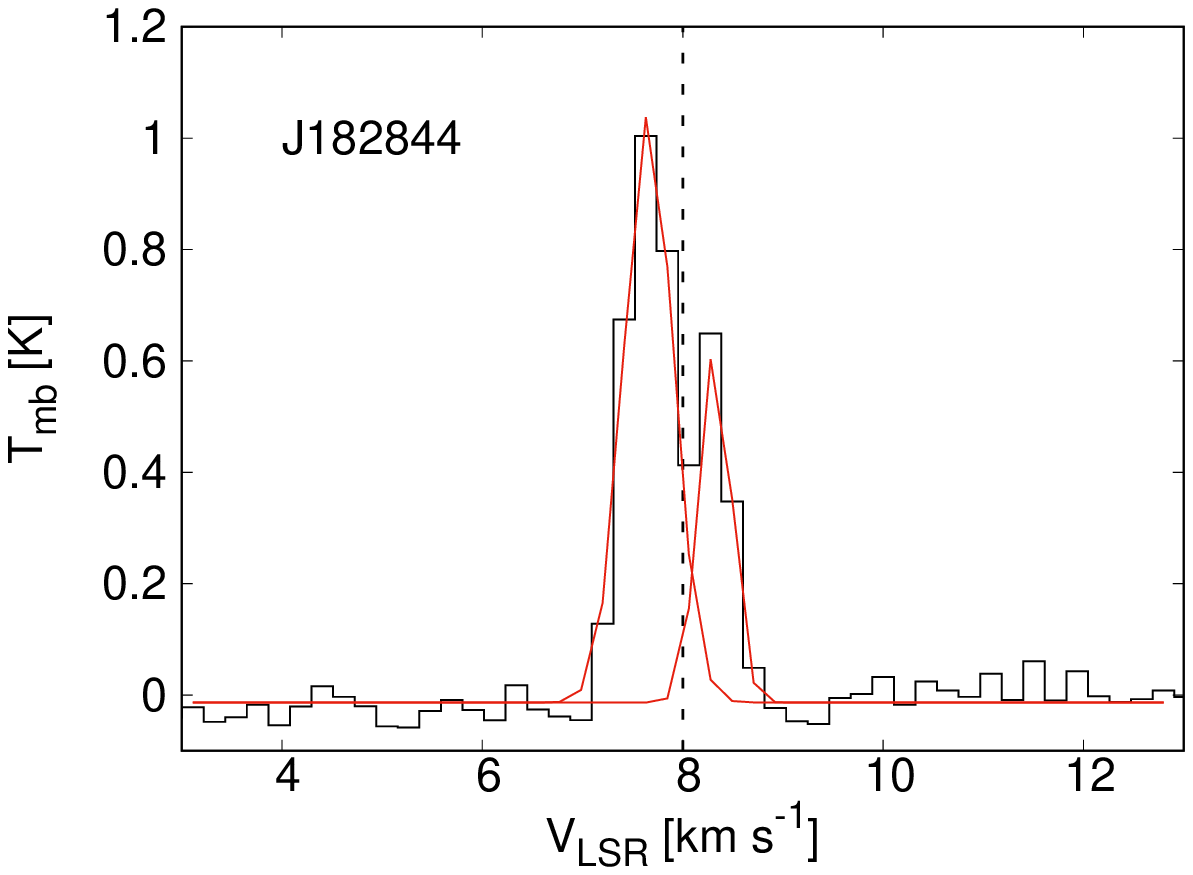}
     \includegraphics[width=2.2in]{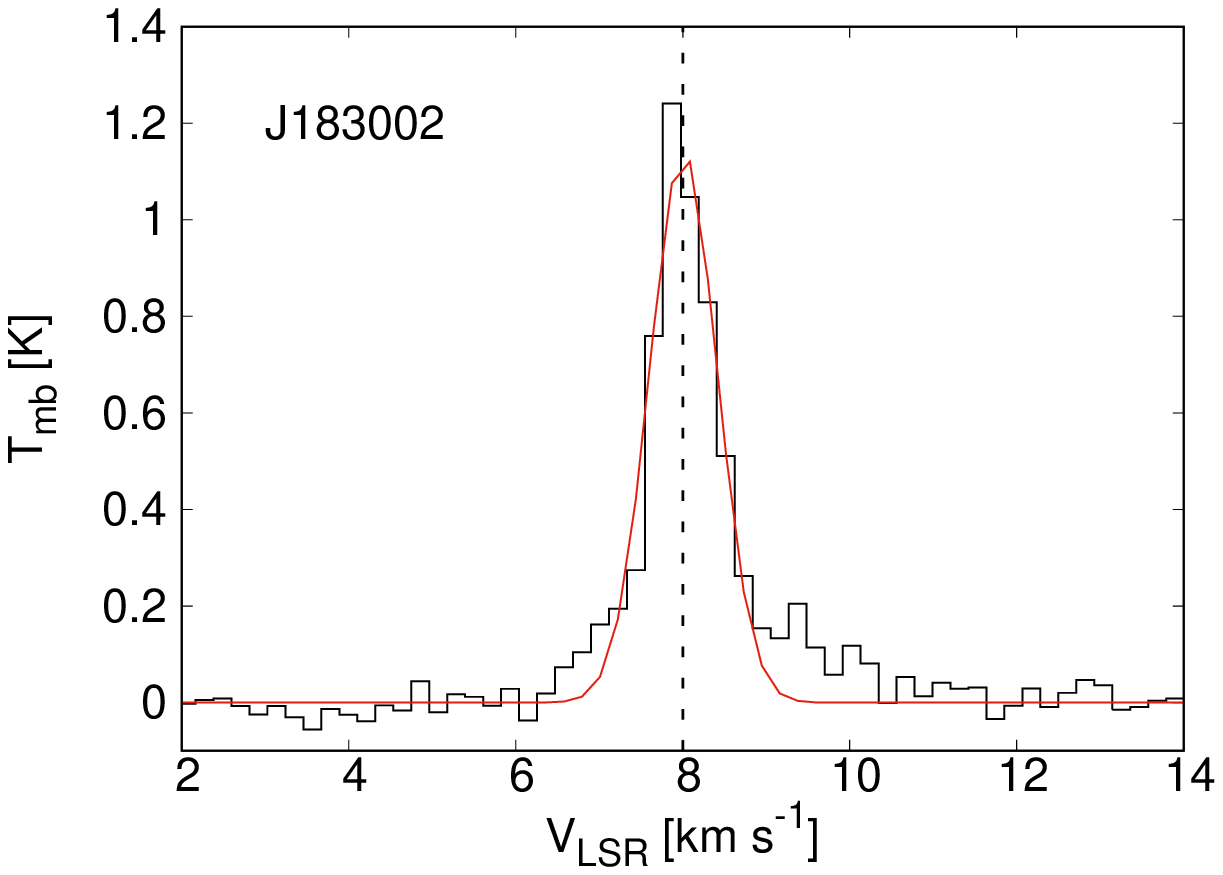}     
     \includegraphics[width=2.2in]{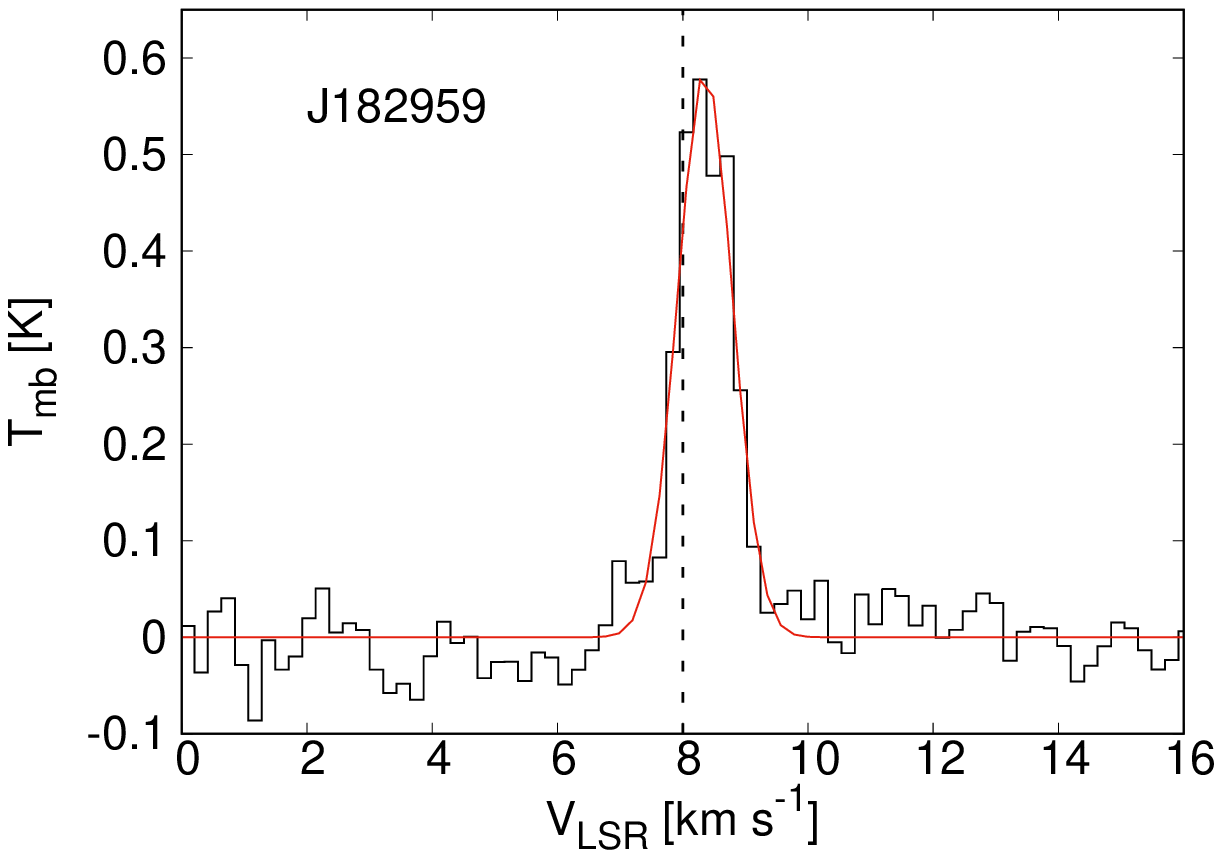}    
     \includegraphics[width=2.2in]{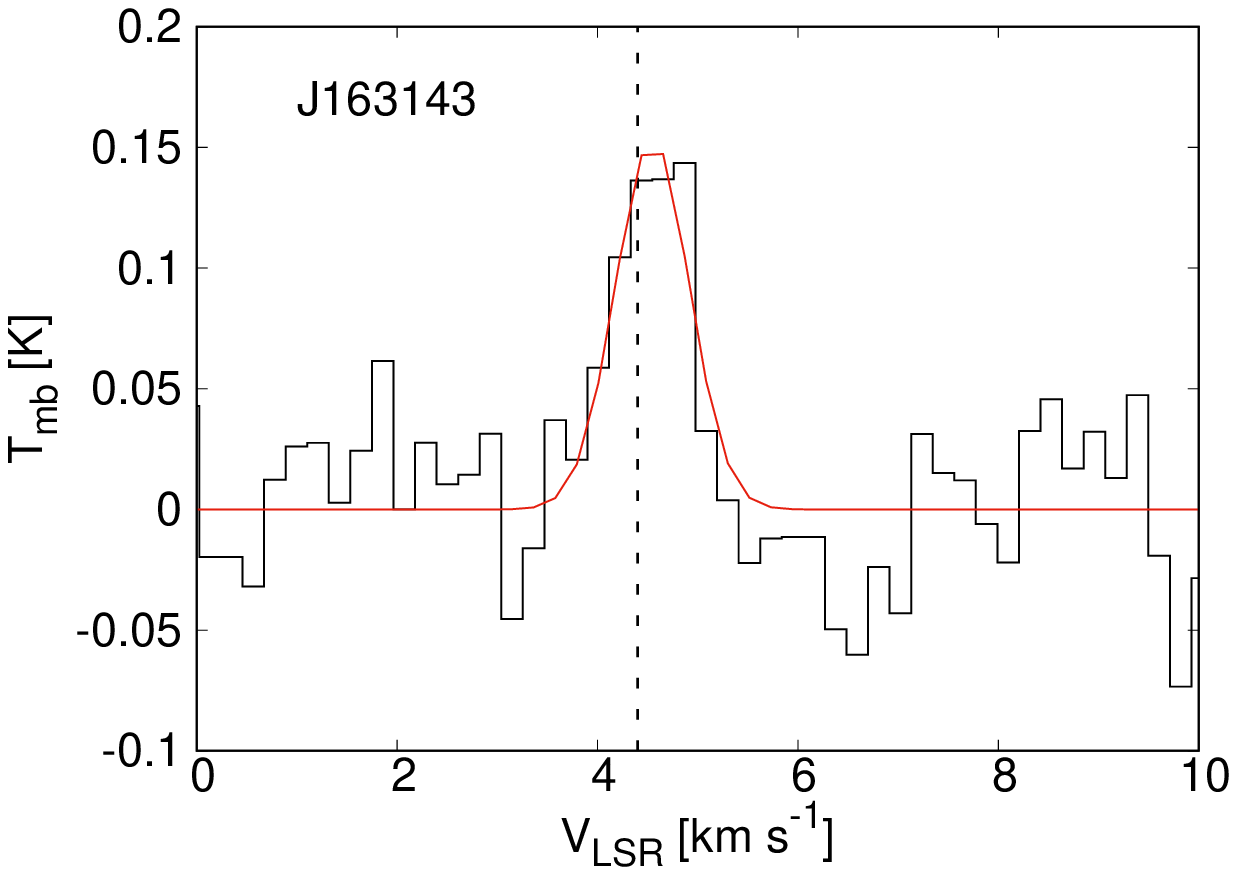}           
     \includegraphics[width=2.2in]{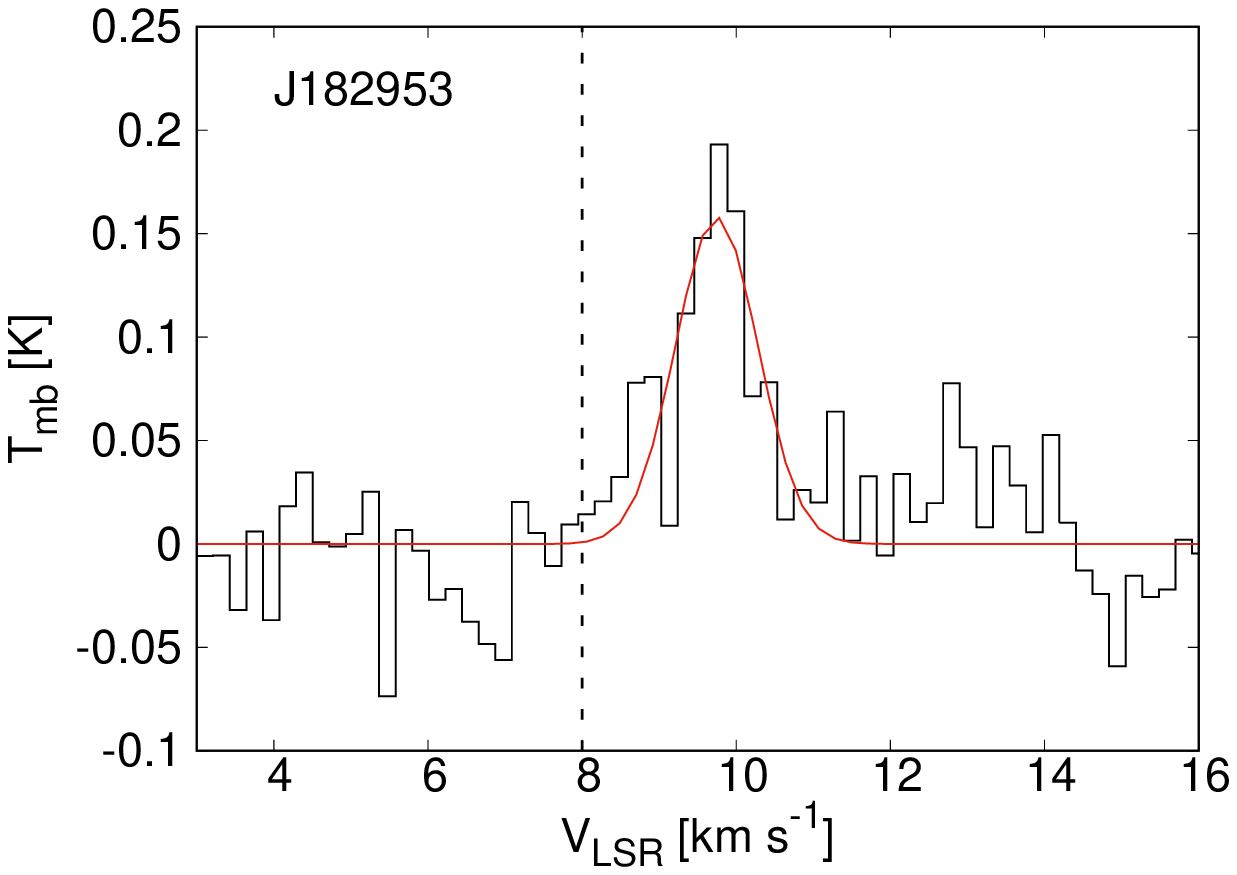}
     \caption{The observed HNC (3-2) spectra (black) with the Gaussian fits (red) for the proto-BDs. Black dashed line marks the cloud systemic velocity of $\sim$8 km/s and $\sim$4.4 km/s in Serpens and Ophiuchus, respectively.  }
     \label{hnc-figs}     
  \end{figure*}

There is emission detected in the HNC (3-2) line for the proto-BDs J182854, J182844, J183002, J182959, with a weak detection in J182953 and J163143 (Fig.~\ref{hnc-figs}). The HNC spectrum for J182844 shows a similar signature of self-absorption as seen for the main `A' component plus the `D' feature in its HCN spectrum (Fig.~\ref{hcn-figs}). The blue-dominated asymmetry may be indicative of an infalling envelope. The double-peaked profile with a central dip also suggests two different velocity components. On the other hand, our line overlap analysis of the optically thin HN$^{13}$C and HNC lines for this object indicates that the stronger blue-shifted peak represents the actual emission from the proto-BD core (Sect.~\ref{overlap}). J183002 shows extended wings in the HNC profile, which is indicative of a contribution from an outflow. The rest of the objects show a narrow, Gaussian-like HNC profile, indicative of emission from the inner, dense regions. An interesting result is that for the two objects J182844 and J182854 with both HCN and HNC line detection, the emission in HNC is stronger than HCN for both cases (Sect.~\ref{discussion}). 


\subsubsection{Optically thin isotopologues}

Observations of the much less abundant and thus optically thin isotopologues of HCN and CN do not exhibit anomalous intensity ratios (e.g., Tennekes et al. 2006; Hily-Blant et al. 2010) and can provide a better assessment of the abundance of these molecules in the proto-BDs. Due to time limitations, we could only obtain observations in the HN$^{13}$C (3-2), H$^{13}$CN (3-2), $^{13}$CN (2-1), and HC$^{15}$N (3-2) lines for the proto-BD J182844. There is weak emission detected only in the HN$^{13}$C (3-2) line, which shows a narrow, Gaussian shaped profile (Fig.~\ref{isotope}). None of the other isotopologues are detected in this proto-BD. The typical rms in these observations was $\sim$50 mK. The non-detections thus indicate peak intensities in the optically thin isotopologues of HCN and CN of $<$50 mK. The emission in HN$^{13}$C is also at best a $\sim$2-$\sigma$ detection, and it is likely that these lines have been beam-diluted in the single-dish observations due to their extremely weak intensities. The line parameters for HN$^{13}$C and the upper limits for the undetected lines are listed in Table~\ref{isotope-thin}. A comparison of the HN$^{13}$C, HNC, and HCN line profiles is presented in Sect.~\ref{overlap}.

 \begin{figure*}
  \centering  
     \includegraphics[width=2in]{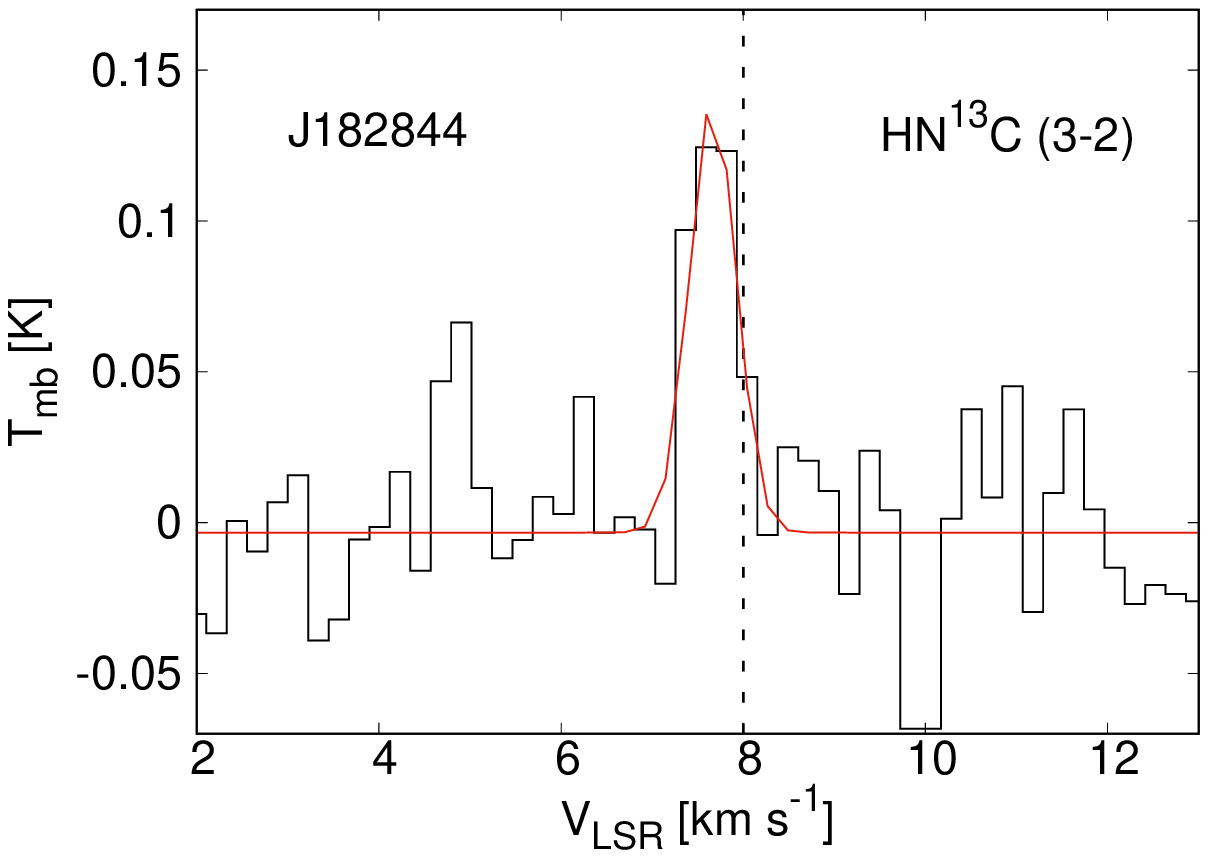} \\
     \includegraphics[width=2in]{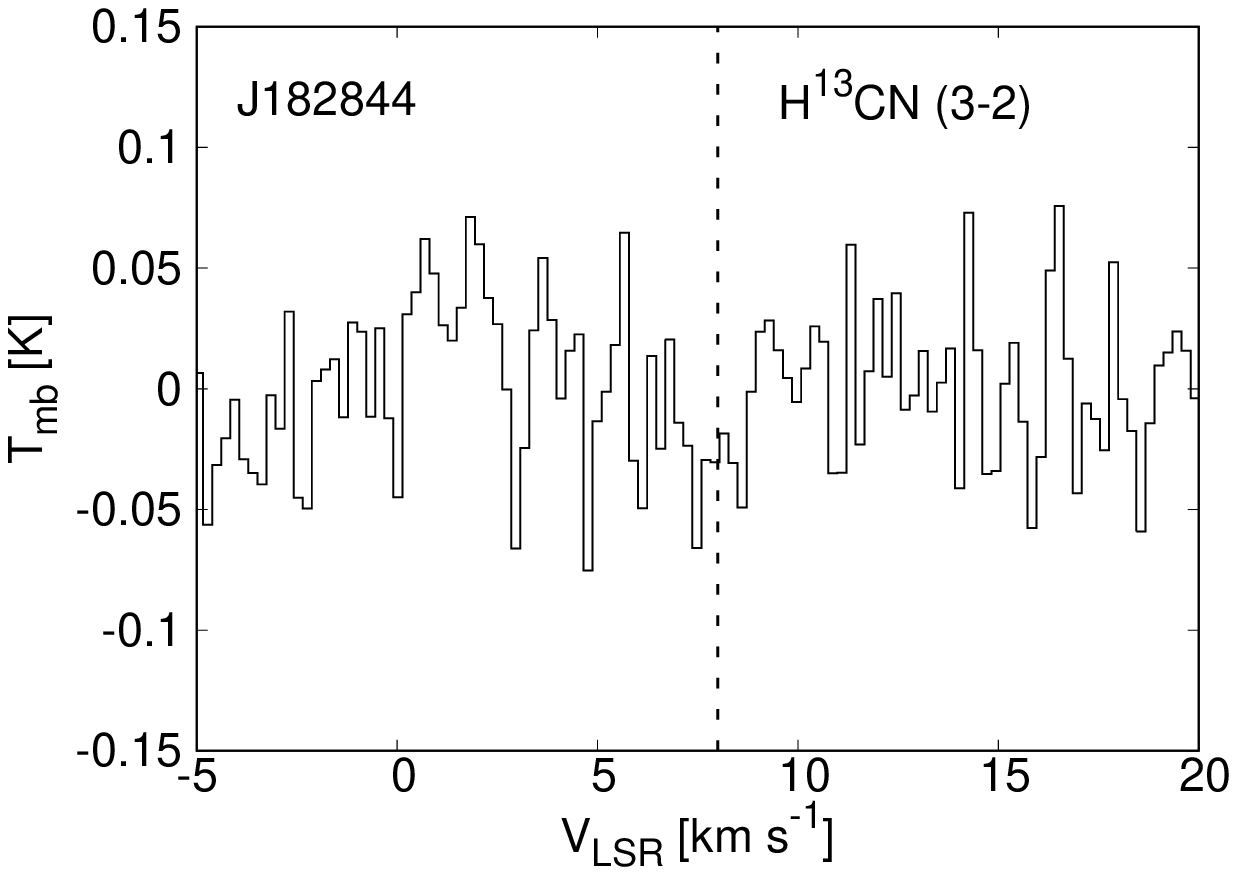}
     \includegraphics[width=2in]{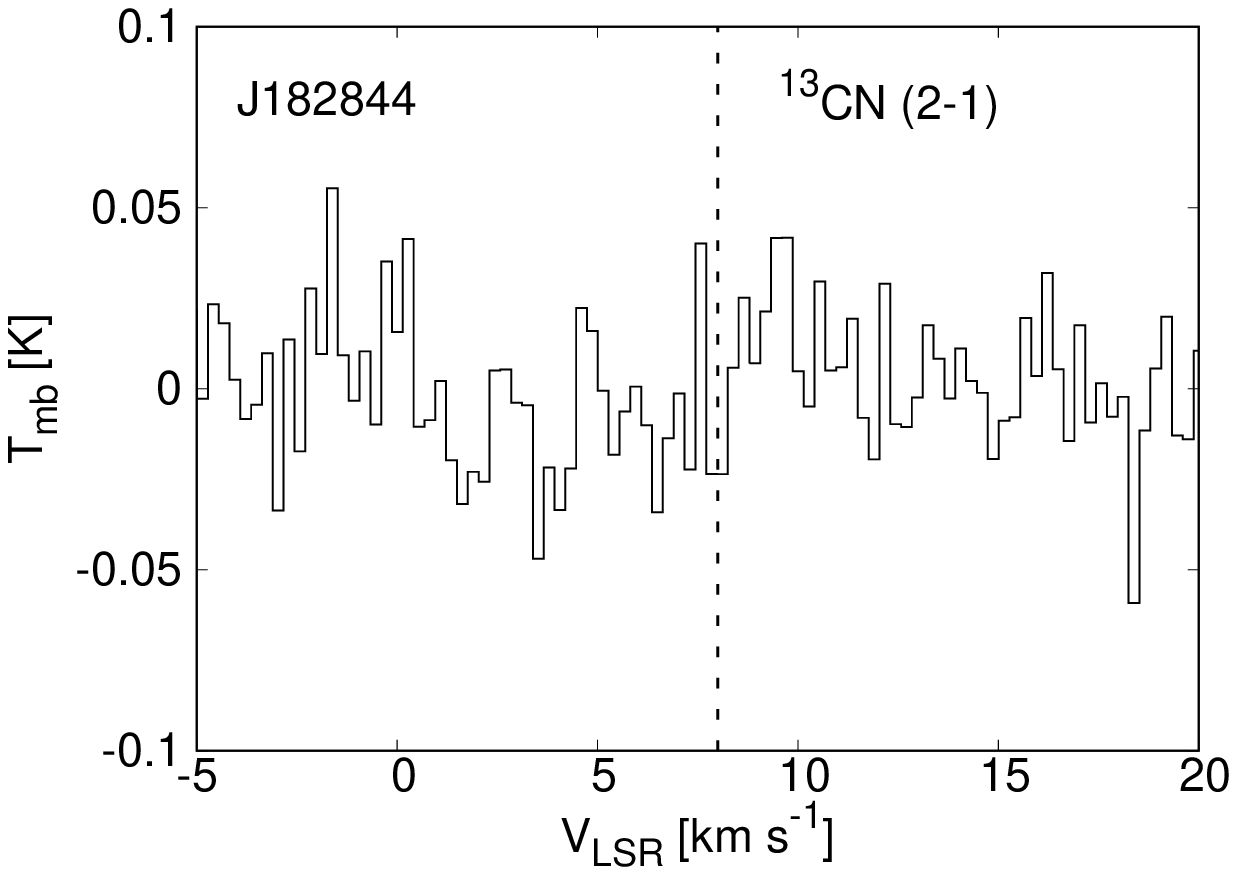}
     \includegraphics[width=2in]{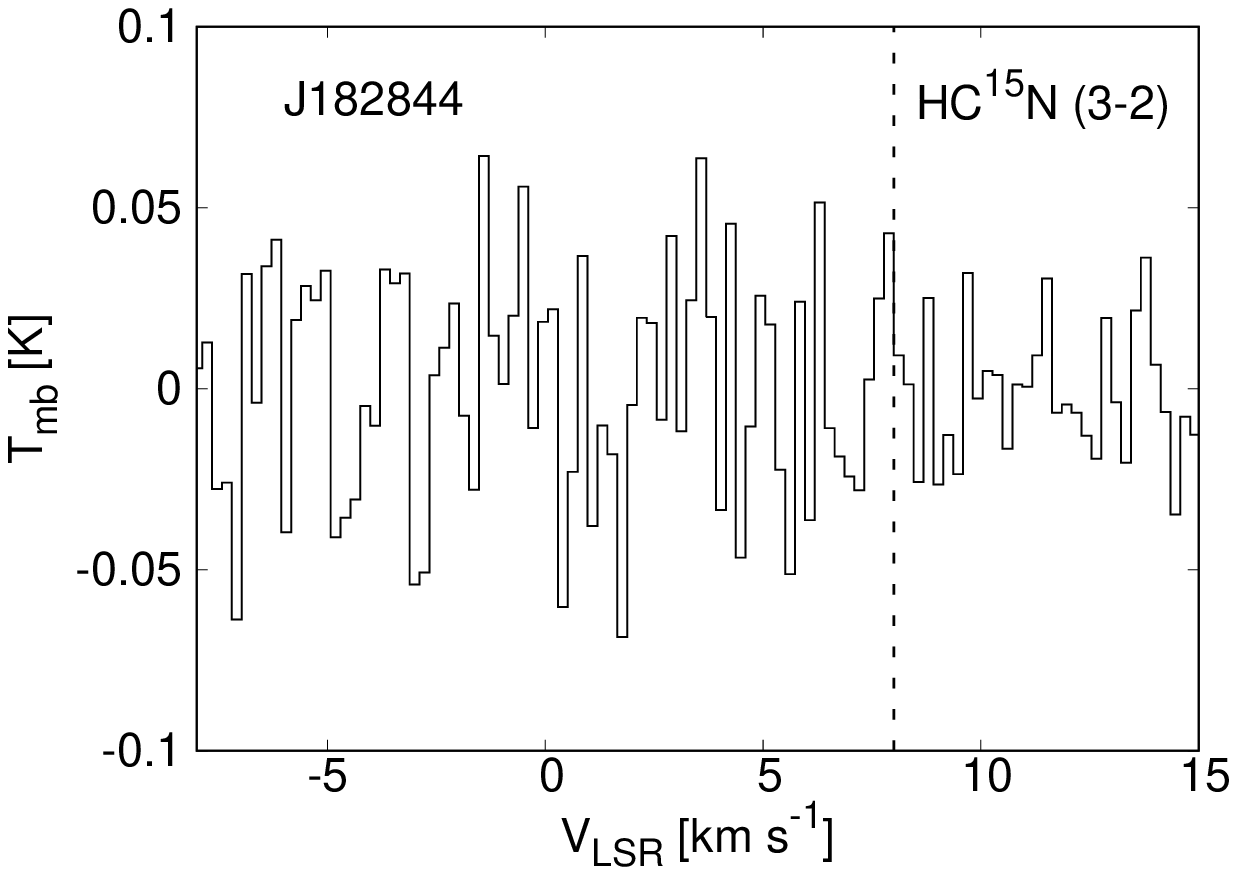}
     \caption{Observations in the HN$^{13}$C (3-2), H$^{13}$CN (3-2), $^{13}$CN (2-1), and HC$^{15}$N (3-2) lines for the proto-BD J182844. Top panel shows the observed HN$^{13}$C (3-2) spectrum (black) with the Gaussian fit (red). Black dashed line marks the cloud systemic velocity of $\sim$8 km/s in Serpens. }
  \label{isotope}       
\end{figure*}

\subsection{Column Densities and Abundances}
\label{Ncol}

\begin{table*}
\centering
\caption{HN$^{13}$C, H$^{13}$CN, HC$^{15}$N, and $^{13}$CN observations for J182844}
\label{isotope-thin}
\begin{threeparttable}
\begin{tabular}{lllllllll} 
\hline
Species & V$_{lsr}$        & T$_{mb}$ & $\int{T_{mb} dv}$ &  $\Delta$v     & $N$                            & $N(X)$/$N (H_{2})$  & $T_{ex}$ & $\tau$  \\
	     & (km s$^{-1}$) & (K) 	          & (K km s$^{-1}$)   & (km s$^{-1}$) & x10$^{12}$ (cm$^{-2}$) &   x10$^{-10}$  & (K) &  \\
\hline

HN$^{13}$C (3-2) & 7.6 & 0.1 &  0.07  & 0.5 & 2.4 & 0.6 & 4.4 & 0.2  \\

H$^{13}$CN (3-2) & 8.0 & $<$0.05 &  $<$0.05  & 1.0 & $<$1.3 & $<$0.4 & -- & --  \\

HC$^{15}$N (3-2) & 8.0 & $<$0.05 &  $<$0.05  & 1.0 & $<$1.3 & $<$0.4 & -- & --  \\

$^{13}$CN (2-1) & 8.0 & $<$0.05 &  $<$0.05  & 1.0 & $<$1.2 & $<$0.3 & -- & --  \\ 

\hline \hline
\end{tabular}
\begin{tablenotes}
 \item We estimate 1-$\sigma$ uncertainties of $\sim$15\%-20\% for the integrated intensity, $\sim$0.1 km s$^{-1}$ in V$_{lsr}$, $\sim$10\% on T$_{mb}$ and $\Delta$v, $\sim$20\% on the column densities and $\sim$40\% on the abundances. The molecular abundances are relative to H$_{2}$, [N(X)/N(H$_{2}$)].  
\end{tablenotes}
\end{threeparttable}
\end{table*}

\begin{table*}
\centering
\caption{HCN, HNC, and CN Column Densities and Abundances}
\label{column2}
\begin{threeparttable}
\begin{tabular}{llllllllllllllllllll} 
\hline
Object & $n_{H_{2}}$ & $N (H_{2})$  & $N$ (HCN) & [HCN] &  $N$ (HNC) & [HNC] & $N$ (CN) & [CN]   \\

	   & x10$^{6}$ (cm$^{-3}$) & x10$^{22}$ (cm$^{-2}$) & x10$^{12}$ (cm$^{-2}$)  &   x10$^{-10}$ &  x10$^{12}$ (cm$^{-2}$) &   x10$^{-10}$ &  x10$^{14}$ (cm$^{-2}$) &   x10$^{-10}$   \\

\hline

J182854 & 1.0$\pm$0.4	& 3.7$\pm$0.6 		& 1.5   & 0.2   			& 6.5	   	& 1.7   	& $<$0.05      & $<$1.3      \\

J182844 & 0.7$\pm$0.3	& 3.6$\pm$0.5 		& 3.4   & 0.9		& 6.9 	& 1.9 	& 1.4 & 38.8 	 \\

J183002 & 2.1$\pm$0.2	& 7.6$\pm$0.5 		&  --  &  -- 		& 8.5  	& 1.2  	& 0.16   & 	2.1	  \\

J182959 & 0.8$\pm$0.5	& 4.5$\pm$0.5		& $<$2.0     & $<$0.4      		& 12.0    	& 2.6  	& 0.4  & 8.8 		 \\

J163143 & 0.3$\pm$0.2	& 1.8$\pm$0.2  		& $<$2.0	   & $<$1.1 		& 6.0      	&  3.3	& 1.2  & 66.6 	  \\

J182953 &$<$0.5	 & $<$2.0  	& $<$2.0	   & $<$1.0 		& 8.0	   	& 4.0		&  0.4   & 20.0	    \\

J163136 &0.6$\pm$0.1	& 2.9$\pm$0.9  		& $<$2.0	   & $<$0.7		& $<$2.0	   & $<$0.7 & $<$0.05    & $<$1.6 \\

J182940 & 0.05$\pm$0.04 	& 0.4$\pm$0.07 &   $<$1.6    & $<$4.0      & $<$1.6     	& $<$4.0     	&  $<$0.05     & $<$12.5 	   \\

J182927 &$<$0.1	& $<$0.6  	&  $<$2.0   & $<$3.3     		& $<$2.0   & $<$3.3  &  $<$0.05     & $<$8.3	   \\ 

J182952 &$<$0.9	 & $<$3.0  	& $<$2.0	   & $<$0.6 		& $<$2.0	   & $<$0.6  &  0.04    & 1.3	   \\

\hline
\end{tabular}
\begin{tablenotes}
  \item We estimate 1-$\sigma$ uncertainty of $\sim$20\% on the column densities and $\sim$40\% on the abundances. The molecular abundances are relative to H$_{2}$, [N(X)/N(H$_{2}$)].  
\end{tablenotes}
\end{threeparttable}
\end{table*}

We have used the non-LTE radiative transfer code RADEX (van der Tak et al. 2007) to estimate the column densities for all species detected towards the proto-BDs. RADEX considers a static, uniform density sphere, with the CMB as the only background radiation field, and with H$_{2}$ as the dominant collision partner for all calculations. There could be a contribution from an internal heating source that can produce an excess continuum emission, but we do not have any measurement of it. The kinetic temperature, $T_{kin}$, of the gas, the H$_{2}$ number density, $n_{H_{2}}$, the turbulent line width, and the column density for a given species are used as input. The kinetic temperature is expected to be relatively low ($\sim$10 K) throughout the outer and inner envelope layers in the proto-BDs (Machida et al. 2009). A low kinetic temperature of $\sim$10 K can also be deduced from the NH$_{3}$ (1,1) and (2,2) line maps in the Serpens region close to the location of our objects (Freisen et al. 2016). However, we note that there is no source detection in the NH$_{3}$ (1,1) and (2,2) line maps for any of our targets from these previous surveys and much deeper NH$_{3}$ observations are required. The H$_{2}$ number density was calculated from the total mass M$_{total}^{d+g}$ (Sect.~\ref{sample}) and size of the source. The source size was approximated by the outer radius of the envelope, $R_{env}$, estimated from SED modelling. We find $R_{env}$ in the range of $\sim$1000-1800 AU (Riaz et al., {\it in prep}). The H$_{2}$ column density, $N (H_{2})$, was derived as $n_{H_{2}}$ * 2 * $R_{env}$. Tables~\ref{column2} lists the values for $n_{H_{2}}$ and $N (H_{2})$ derived for the sources. We note that the number density of the sources have been derived from single-dish observations integrated over a beam size of $\sim$14.5$\arcsec$. Due to this, the inner, dense regions ($\leq$500 AU) in the proto-BDs where $n_{H_{2}}$ is expected to reach values of $>$10$^{6}$ cm$^{-3}$ (e.g., Machida 2014) are beam-diluted in the dust continuum observations, but can be probed with these dense gas tracers.

RADEX uses the collision rates for CN from Lique et al. (2010) and for HCN from Dumouchel et al. (2010). This is the CN and HCN fine-structure with He scaled to H$_{2}$. Since RADEX does not treat line overlaps, we are restricted to use fine-structure line splitting. RADEX provides the results for the `A' component in HCN (3-2), and for the composite `C'+`D'+`E' components and the `F'+`G'+`H' components in CN (2-1). Therefore, we have calculated the HCN column density using the `A' hyperfine component, which is a composite of four hyperfine components that make up 92.6\% of the total intensity in the HCN (3-2) transition. The integrated intensity for this composite spectral feature was divided by the sum of the relative intensities of the four hyperfine components (Table~\ref{hcn-hfs}). For CN, we have used the composite `C'+`D'+`E' components to estimate the column density. These three components are detected in all of the objects with CN emission. The sum of the integrated intensities for these three components was divided by the sum of their relative intensities (Table~\ref{cn-hfs}). We used the sum of the line width for the `C', `D', and `E' components as input in RADEX. For HNC, we have used the integrated intensity (Table~\ref{hnc}) to estimate the column densities.

For a fixed $T_{kin}$, $n_{H_{2}}$, and line width in RADEX, the input column density is varied to match the output peak intensity from RADEX with the observed peak line intensity (T$_{mb}$). The output from RADEX is the radiation temperature T$_{R}$, which was converted to T$_{mb}$ by making correction for the beam filling factor. This is the ratio of the source size to the beam size and is estimated to be $\sim$0.6-0.8. RADEX assumes a Gaussian profile that is convolved by the line width. Thus, in effect, the observed integrated intensity is matched with the model value. The column densities thus derived are listed in Tables~\ref{isotope-thin};~\ref{column2}. Also listed are the molecular abundances relative to H$_{2}$, [N(X)/N(H$_{2}$)]. We estimate an uncertainty of $\sim$20\% on the column densities. The uncertainty was estimated by varying the input column density to match the 1-$\sigma$ range in the observed integrated line intensity. The uncertainty on the derived molecular abundances is estimated to be $\sim$40\% and is propagated from the 1-$\sigma$ error on the H$_{2}$ density and the column density. For the non-detections, we have derived the upper limits on the column density and abundance using the 3-$\sigma$ upper limit on the integrated intensity, assuming a line width of 1 km s$^{-1}$ and the same kinetic temperature. There is a factor of $\sim$2 increase in the column densities if the H$_{2}$ density is increased by the uncertainty on the H$_{2}$ measurements listed in Table~\ref{column2}. A similar drop by a factor of $\sim$2 in the column density occurs if the kinetic temperature is assumed to be 8 K instead of 10 K. The output from RADEX also includes the excitation temperature $T_{ex}$, the optical depth $\tau$. These are listed in Table~\ref{tex-tau} and discussed in Sect.~\ref{tau-ncrit}. The opacities are listed for the same hyperfine components that were used to estimate the column densities, i.e. the `A' component in HCN and `C'+`D'+`E' components in CN. The RADEX estimates on line opacities are $\leq$1.5. The $T_{ex}$ for CN, HCN, and HNC is in the range of $\sim$4-5 K, which is lower than the assumed $T_{kin}$$\sim$10 K and indicates sub-thermal non-LTE conditions (Sect.~\ref{tau-ncrit}). The effects of self-absorption signatures and high opacity on the derived abundances are discussed in Sect.~\ref{overlap}.

\section{Discussion}
\label{discussion}

\subsection{Strong CN emission}

CN is more typically detected in our proto-BD sample compared to HCN, with about 70\% objects showing CN emission while HCN is only detected in 2 objects. We find the CN abundance to be at least an order of magnitude higher than HCN (Table~\ref{column2}), with a CN/HCN ratio of $>$20 for all objects with CN detection. CN and HCN can both form in reactions with the intermediaries CH and CH$_{2}$ that are produced via radiative association reactions with C$^{+}$ ions (e.g., Sternberg \& Dalgarno 1995; Hily-Blant et al. 2010; Boger \& Sternberg 2005). The CN and HCN abundances increase as the C$^{+}$ abundance increases in layers closer to the UV source where carbon is photo-ionized. The enhanced CN/HCN ratios of $>$10 is a known signature of the UV radiation field originating from the stellar accretion zone (e.g., Bergin et al. 2003; Thi et al. 2004). Infrared spectroscopy of proto-BDs shows that the accretion-associated Br-$\gamma$ line is commonly detected in these objects despite their faintness in the near-infrared, with a line luminosity that is $\sim$10 times higher than any other shock emission line, indicating that they are undergoing active accretion (e.g., Whelan et al. 2018). We find accretion rates for these proto-BDs of the order of 10$^{-8}$ M$_{\sun}$ yr$^{-1}$, which are comparable to low-mass protostars. Considering the modest bolometric luminosity of the proto-BDs, the accretion luminosity (L$_{acc}$) is very high, with an L$_{acc}$ / L$_{bol}$ of $\sim$50\%-70\%. This is consistent with these objects being very young and still accreting large amounts of circumstellar matter. In such a case, HCN will be dissociated by Lyman-$\alpha$ photons (HCN + $\nu$ $\rightarrow$ CN + H), thus producing more CN and a high CN/HCN ratio. It is difficult to destroy CN via photo-dissociation as very high energetic photons are required (e.g., Bergin et al. 2003). Two extreme cases are J163136 and J182953 that show strong CN emission but there is no HCN detection (Table~\ref{column2}), indicating that most of the HCN has been photo-dissociated. Note that we do not find a high HCN/HNC ratio for any objects, which also suggests that the large CN/HCN is likely caused by the stellar UV field.

We have argued that the large CN/HCN ratios in proto-BDs are due to their strong accretion activity and thus controlled by the stellar UV flux. It may be the case that our single-dish observations are probing the ``hot cores'' in the proto-BDs. While the angular size of the hot cores in these objects is expected to be less than 1$\arcsec$ (e.g., Machida 2014), if the abundance enhancements are sufficiently large (a factor of 10-100), they can be detected even in unresolved data (e.g., van Dishoeck 2009). It is also known from previous works on T Tauri stars that strong accretion activity produces UV fluxes that are 100-1000 times higher than the ISRF (e.g., Bergin et al. 2004). Also important is the X-ray radiation that can penetrate into the inner, dense regions of the core and control the ion chemistry and molecular dissociation in the deeper layers. Dust settling and grain coagulation can also enhance the CN/HCN ratio (e.g., Hily-Blant et al. 2010). These factors could also contribute to the observed beam-averaged abundances.



\subsection{High HNC/HCN ratio}

HNC is detected in all proto-BDs except J163136. We find a HNC/HCN abundance ratio of $\geq$1 for all objects in the sample, including the lower limits (Table~\ref{column2}). The high HNC/HCN ratio between $\sim$1--8 for the proto-BDs is likely caused by a combination of low temperature and high density. The enhanced abundance in HNC compared to HCN can be explained if we consider the main formation pathway of HNC via the neutral-neutral reaction NH$_{2}$ + C $\rightarrow$ HNC + H, where NH$_{2}$ is formed via cosmic-ray driven He-impact ionization of N$_{2}$ followed by successive hydrogenation reactions with H$_{2}$. An important intermediate reaction in this series is N$^{+}$ + H$_{2}$ $\rightarrow$ NH$^{+}$ + H, which is an endothermic reaction with an energy barrier of $\sim$85 K. However, the reaction of N$^{+}$ with ortho-H$_{2}$ can occur at low temperature, as argued in Hily-Blant et al. (2010). Most of the elemental nitrogen tends to be in the form of gaseous N$_{2}$ or N-bearing solid compounds (NH$_{3}$ or N$_{2}$ ices) as suggested by steady-state chemical models (e.g., Hily-Blant et al. 2010; Maret et al. 2006). A high N$_{2}$ density may be the key to forming more HNC relative to HCN, and it may be the case that HNC in these very low-luminosity, dense proto-BD cores forms in reactions with N$_{2}$ ice at very low temperature. On the other hand, if most N$_{2}$ is in the ice form, some form of desorption may be occurring, which could be related to the high accretion energy towards the inner, denser regions as also explained for the strong CN emission. The high HNC/HCN ratio could be due to an efficient non-thermal desorption mechanism of HNC. 


As noted in chemical modelling studies, there is a rapid increase in the HNC abundance by 2-3 orders of magnitude as $T_{kin}$ drops from just $\sim$15 K to $\leq$10 K (e.g., Graninger et al. 2014; Fuente et al. 1993). A recent study of the low temperature chemistry of HNC and HCN isomers using updated collisional coefficients for excitation has shown that the HNC/HCN intensity ratios increase with decreasing kinetic temperature (Hern\'{a}ndez et al. 2017); the largest intensity ratios are seen at T$_{kin} \leq$ 10 K in the optically thin regime ($N$ = 10$^{12}$ cm$^{-2}$), and for the high-$J$ lines of (2-1) as compared to the (1-0) transition. Even assuming the same column density for both isomers, HNC is more easily excited than HCN due to larger collisional rate coefficients, and will have a line intensity stronger than HCN (Hern\'{a}ndez et al. 2017). Our results appear consistent with chemical modelling studies.





For the proto-BD J182854 where the HNC/HCN ratio is the largest ($\sim$7) but CN/HCN is the lowest ($<$6) in the sample, it may be more deeply embedded in a cold environment of perhaps $<$10 K that has produced the anomalously high HNC/HCN ratio. Note that HN$^{13}$C is detected in J182844 but H$^{13}$CN is undetected, which also implies a HNC/HCN abundance ratio of $>$1.8 (Fig.~\ref{isotope}; Table~\ref{isotope-thin}). The interesting test cases for HNC being formed at very cold temperatures independent of the formation/destruction of HCN are the proto-BDs J182959, J182953, and J163143 that show CN emission due to a high stellar UV dose and HNC emission due to low temperature, but there is no HCN detection. If even a moderate UV flux was driving the destruction of HCN and HNC then both would be destroyed equally. Some HNC thus seems to have been retained due to its enhanced production at low temperature. The detection of HNC in all but one of the proto-BDs indicates that there is a large reservoir of cold chemistry in these early stage brown dwarfs.


 \begin{figure*}
  \centering              
     \includegraphics[width=3in]{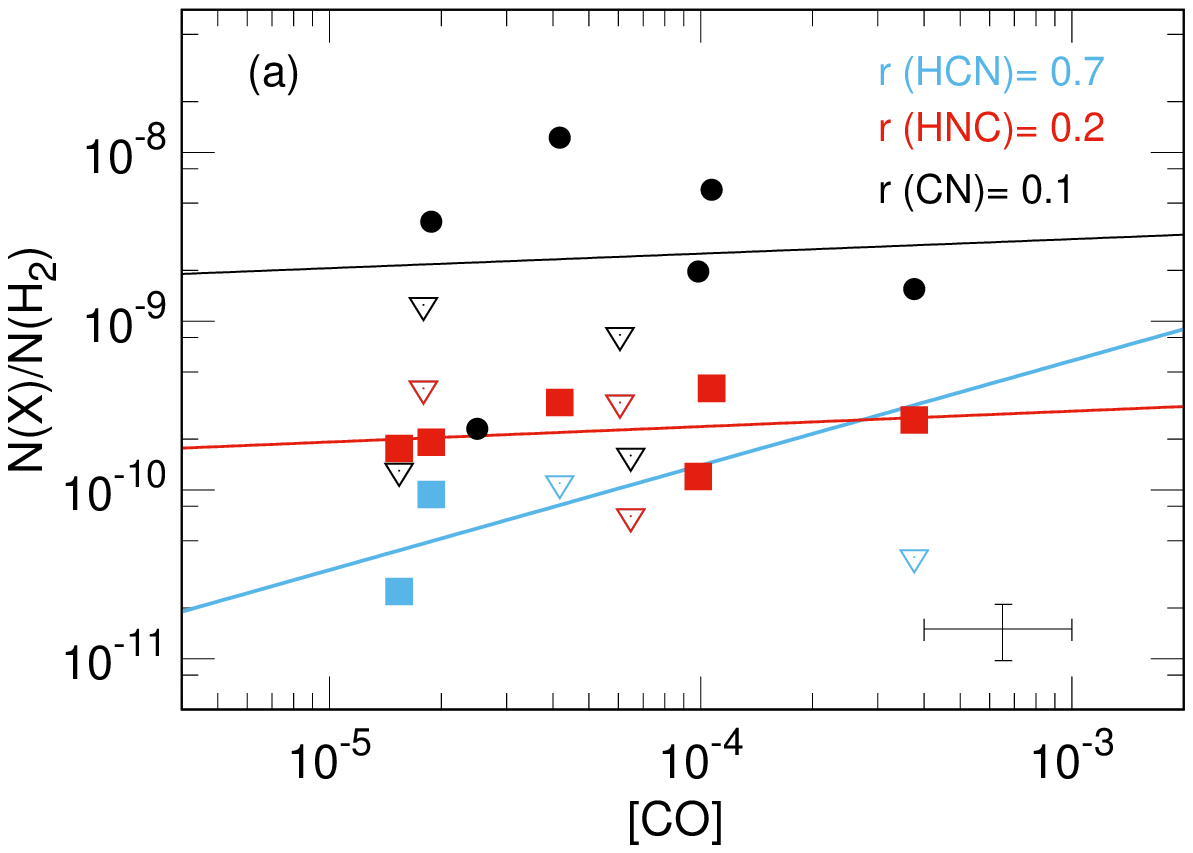}              
     \includegraphics[width=3in]{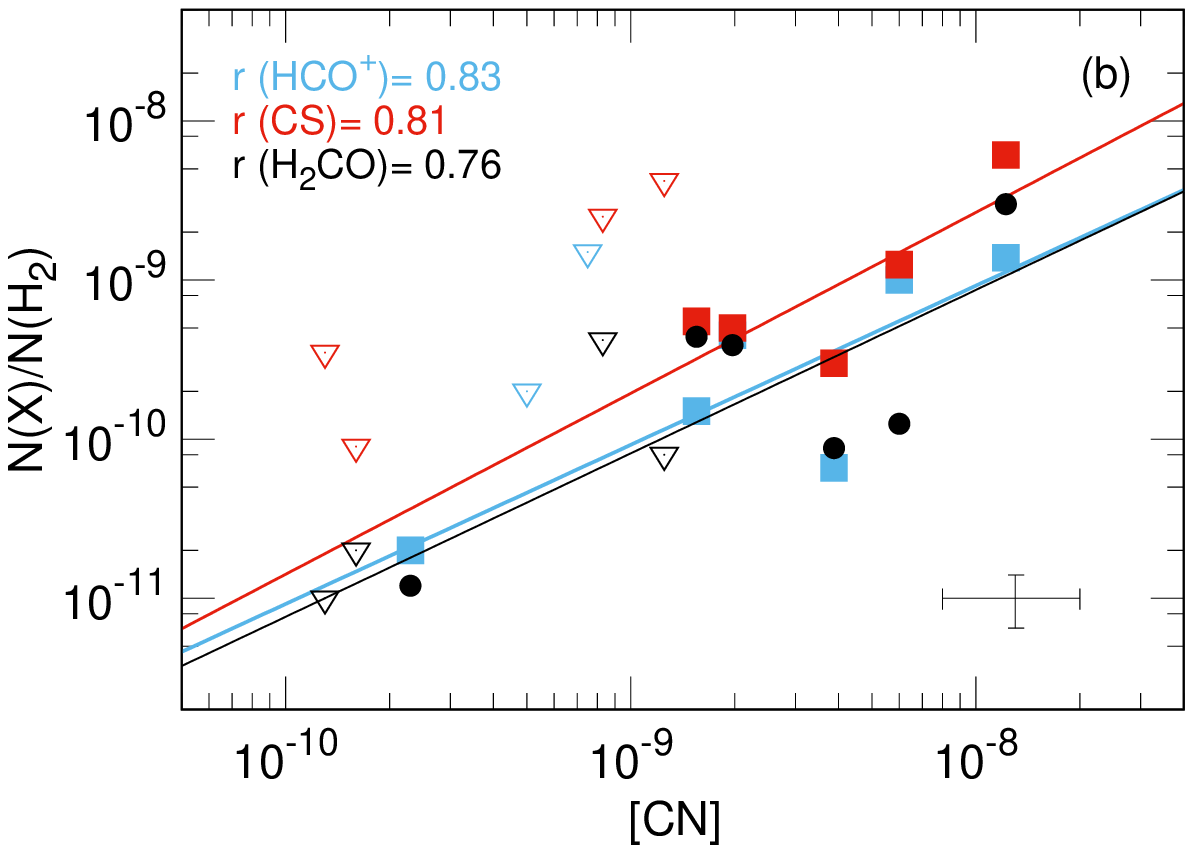}  
          \includegraphics[width=3in]{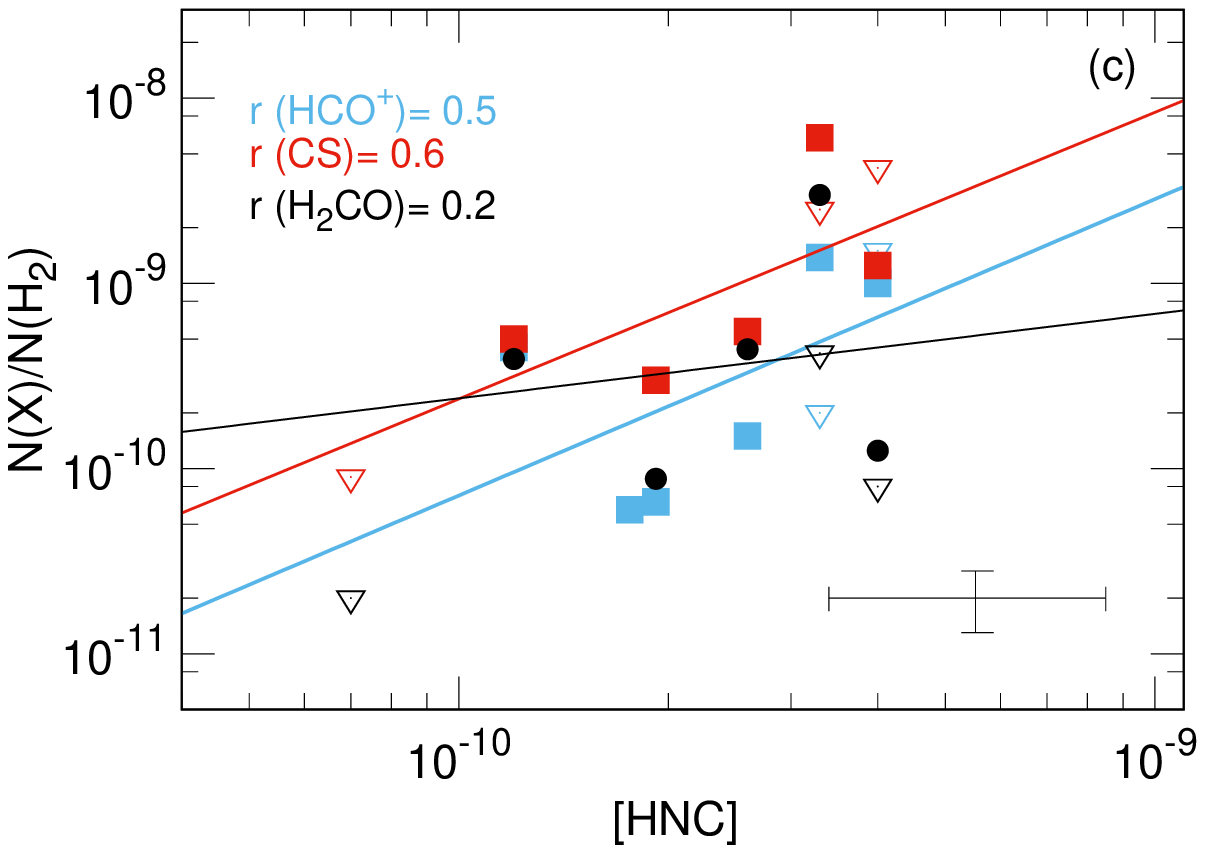}  
     \includegraphics[width=3in]{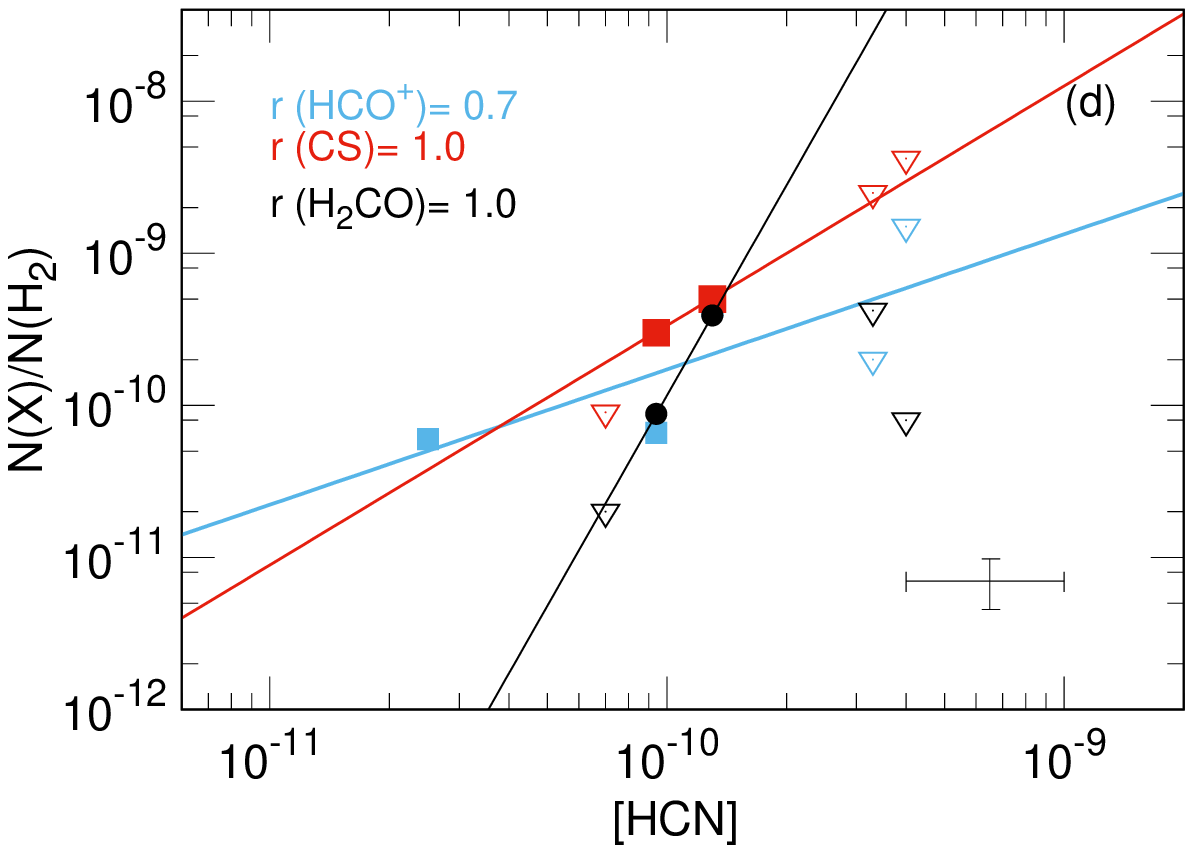}               
     \caption{Correlations between the CN, HCN, HNC abundances (with respect to H$_{2}$) with CO and other high-density tracers for the proto-BDs. The correlation coefficients are noted in each plot. Diamonds denote the non-detections, which have not been included in the correlations. Typical uncertainties in the abundances are estimated to be $\sim$40\% and are plotted in the bottom, right corner of each plot.   }
  \label{cn-hcn-hnc}       
\end{figure*}

\subsection{Trends in molecular abundances}

Figure~\ref{cn-hcn-hnc} explores the correlations among the CN, HCN, and HNC abundances with CO and other high-density tracers of HCO$^{+}$, CS, and H$_{2}$CO. The CO, HCO$^{+}$, CS, and H$_{2}$CO data is from Riaz et al. ({\it in prep}). The error bars on the molecular abundances are estimated to be $\sim$40\%-45\% and have been taken into consideration in the measurement of the correlation coefficients. We have not included the lower/upper limits for the non-detections in the correlations. The typical uncertainty on the correlation coefficients is $\sim$20\%. Both CN and HNC show a flat distribution with CO (Fig.~\ref{cn-hcn-hnc}a), and the abundances are nearly constant with declining CO abundance. The lack of correlation between CN and CO for proto-BDs is consistent with previous results on pre-stellar cores where CN appears to avoid depletion in regions where CO is depleted, suggesting that CN can survive in the gas phase at densities of the order of 10$^{6}$ cm$^{-3}$ (e.g., Hily-Blant et al. 2008; 2010; Padovani et al. 2011). Here we have shown that a similar non-correlation exists between HNC and CO for the proto-BDs. If HNC is being formed in proto-BDs through ice surface reactions then its abundance will remain large even in the inner, dense regions where CO will be depleted. The nearly constant HNC abundance in proto-BDs for [CO] between 10$^{-3}$-- 10$^{-5}$ also suggests a high initial abundance of gaseous nitrogen based on the chemical models presented in Hily-Blant et al. (2010). The correlation between HCN and CO is comparatively stronger (coefficient $\sim$0.7) (Fig.~\ref{cn-hcn-hnc}a). However, we only have two data points for HCN compared to six for CN and HNC.

Figure~\ref{cn-hcn-hnc}b shows that CN is well-correlated with HCO$^{+}$, CS, and H$_{2}$CO (coefficient $\sim$ 0.7-0.8). HNC shows a comparatively weaker correlation with HCO$^{+}$ and CS (coefficient $\sim$ 0.5-0.6), whereas a nearly flat distribution is seen with H$_{2}$CO (Fig.~\ref{cn-hcn-hnc}c). In comparison, the two data points in HCN are tightly correlated with H$_{2}$CO and CS (coefficient = 1.0) but shows a weaker correlation with HCO$^{+}$ (coefficient $\sim$0.7) (Fig.~\ref{cn-hcn-hnc}d). Based on physical+chemical modelling, we find that the HCN, CS and H$_{2}$CO emission arises from the inner dense regions ($<$400 AU) in these proto-BDs, whereas the peak in the HCO$^{+}$ and HNC emission is seen at intermediate radii ($\sim$400-800 AU) (Riaz et al. in prep). The similarities in the density and velocity structure of the physical components probed by these tracers can explain the correlations in their abundances.



\subsection{Trends in comparison with protostars and evolutionary stage}

\begin{figure*}
  \centering                              
     \includegraphics[width=2.8in]{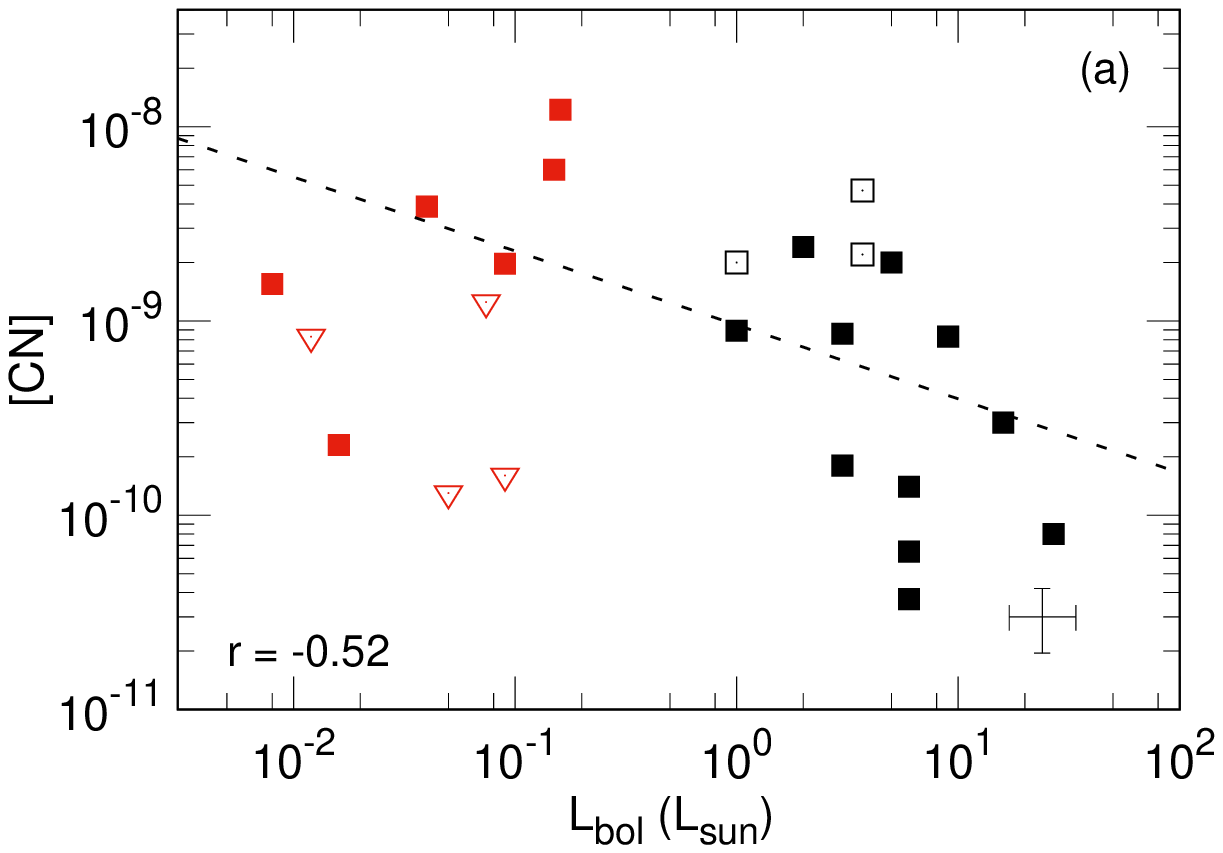}
     \includegraphics[width=2.8in]{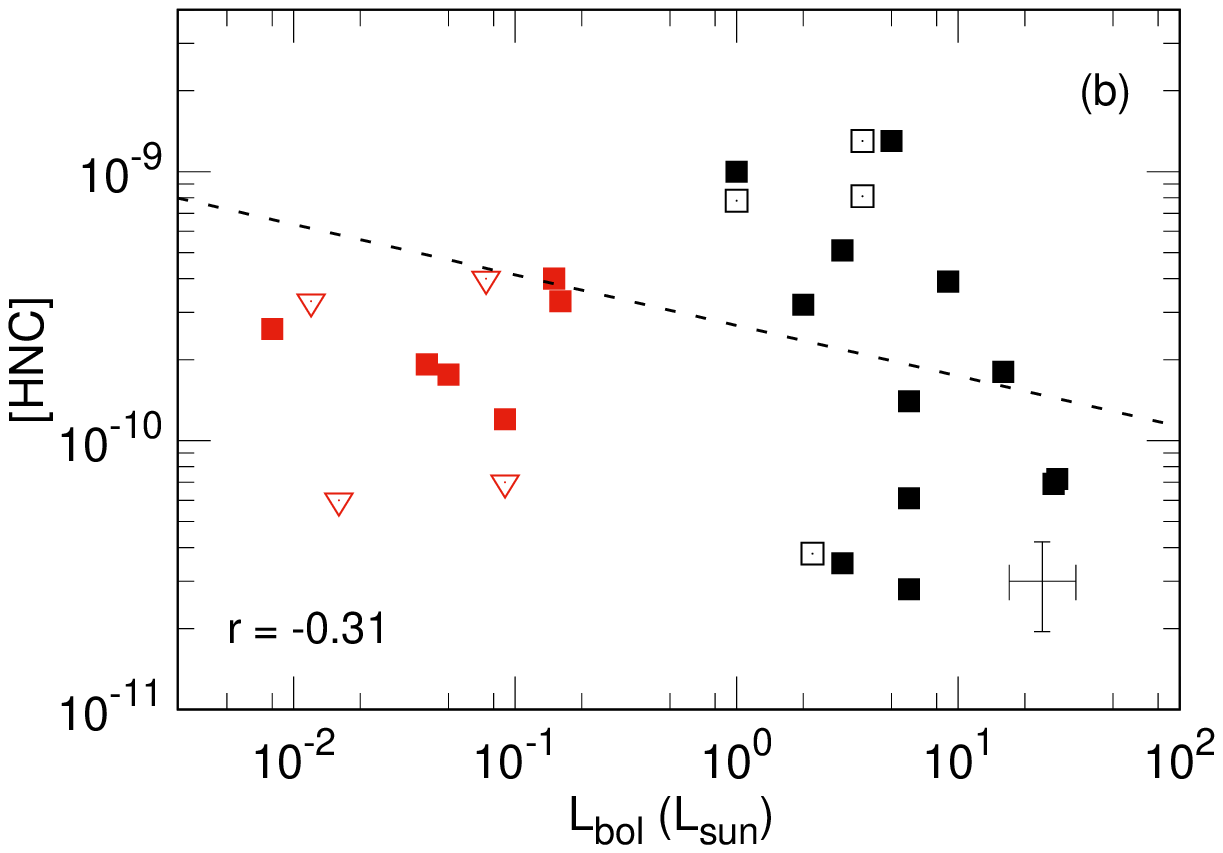}
     \includegraphics[width=2.8in]{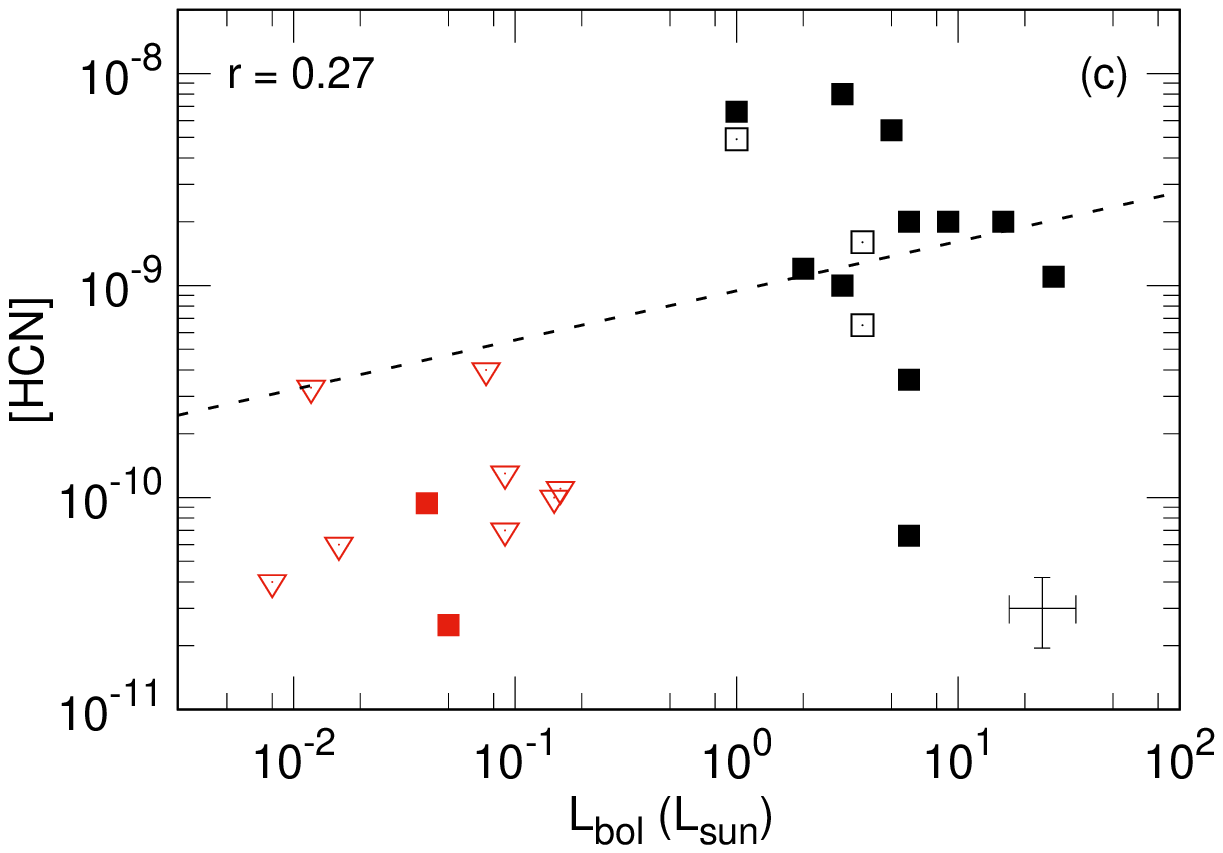}            
     \includegraphics[width=2.8in]{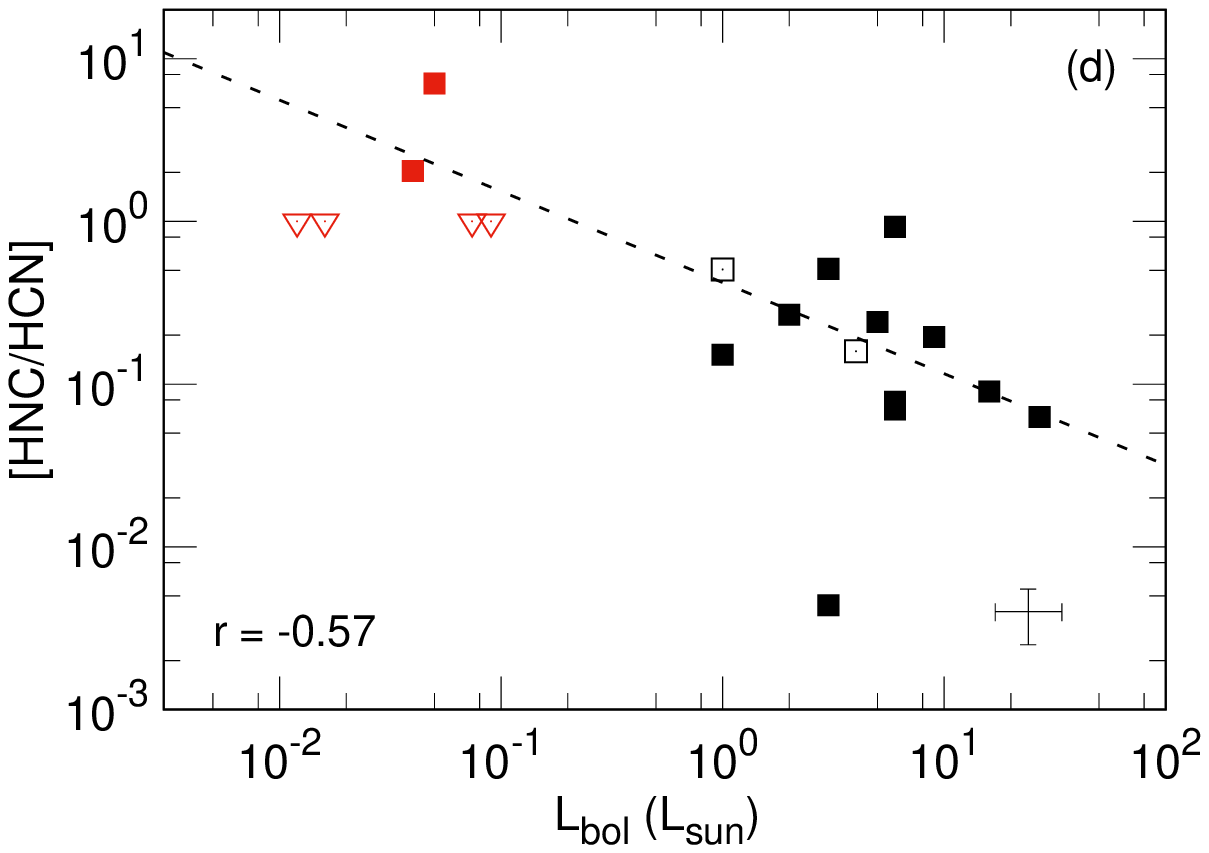}  
     \includegraphics[width=2.8in]{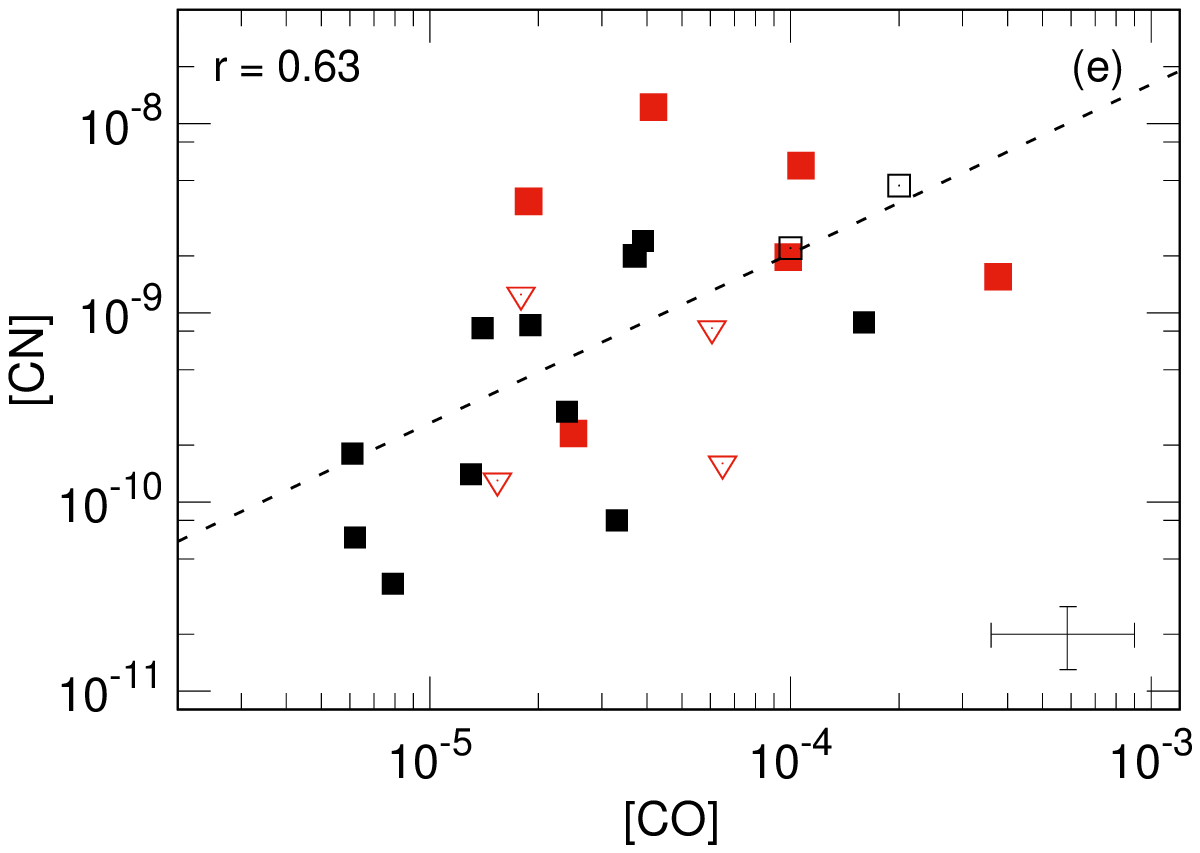}     
     \includegraphics[width=2.8in]{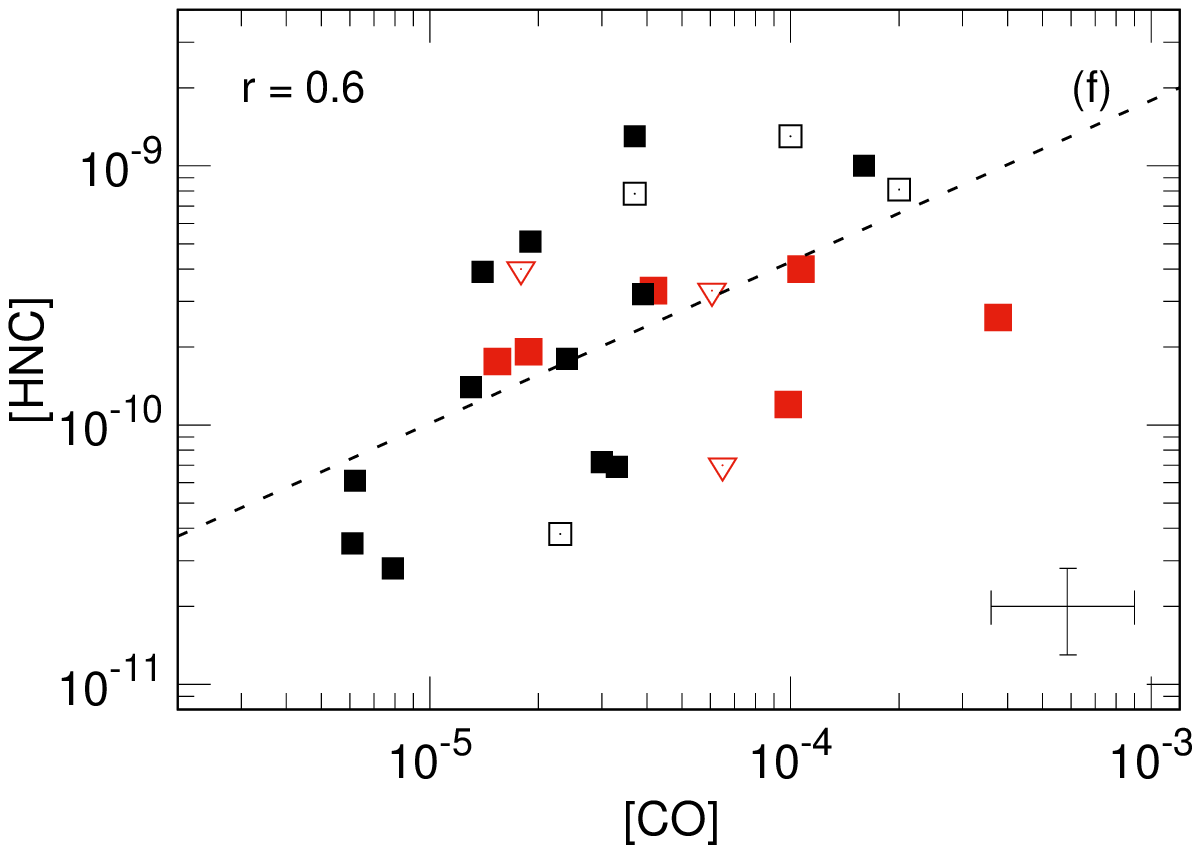}     
     \includegraphics[width=2.8in]{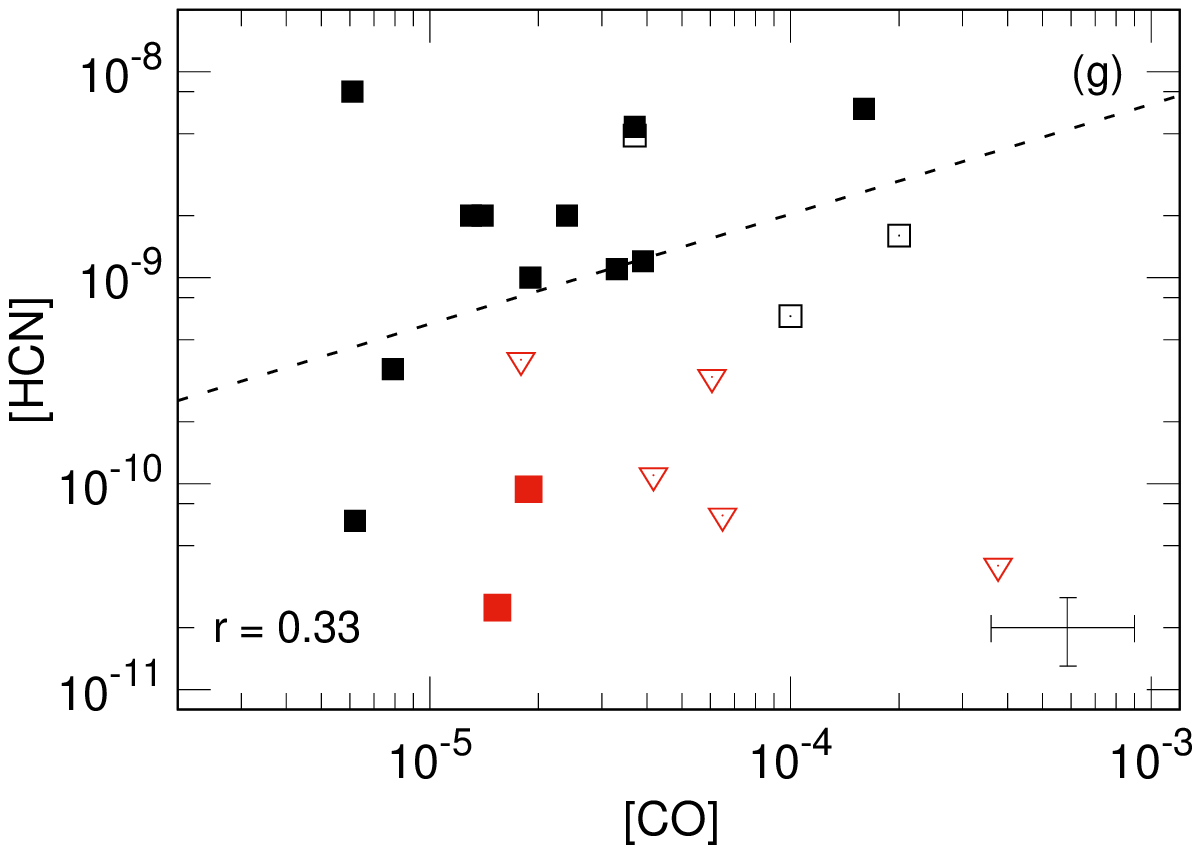}     
     \caption{The molecular abundances for the proto-BDs (filled red points) compared with pre-stellar cores and Class 0 protostars (filled black points), and Class I protostars (open black points). Red diamonds denote the non-detections, which have not been included in the correlations. Typical uncertainties are estimated to be $\sim$40\% on the abundances and $\sim$30\% on L$_{bol}$, and are plotted in the bottom, right corner of each plot.  }
  \label{lbol}       
\end{figure*}

Figure~\ref{lbol} compares the molecular abundances for the proto-BDs with pre-stellar cores and Class 0/I protostars, the data for which has been compiled from the works of Joergensen et al. (2002; 2004). The abundance measurements in these works have also been derived from the application of a 1D model (RATRAN) to single-pointing, single-dish observations. We find a trend of higher CN abundances in the proto-BDs compared to protostars, such that the mean [CN] value for the proto-BDs is (7.3$\pm$0.6)x10$^{-9}$ compared to (1.2$\pm$0.9)x10$^{-9}$ for the protostars (Fig.~\ref{lbol}a). The CN abundances for the proto-BDs also show a spread of more than an order of magnitude. In comparison, HNC for the proto-BDs shows a nearly constant abundance between $\sim$0.01--0.1 L$_{\sun}$ (Fig.~\ref{lbol}b). This is unlike the large spread of $\sim$2 orders of magnitude in [HNC] seen for the protostars spanning L$_{bol}$$\sim$1--30 L$_{\sun}$. Due to the different trends, the mean [HNC] for the proto-BDs is (2.5$\pm$1.0)x10$^{-10}$, similar to (3.4$\pm$1.4)x10$^{-10}$ for the protostars. In contrast, there is a decline in the HCN abundance with the bolometric luminosity (Fig.~\ref{lbol}c), with a mean [HCN] value of (9.2$\pm$3.7)x10$^{-11}$ for the proto-BDs, which is an order of magnitude lower than the mean value (3.3$\pm$1.3)x10$^{-10}$ for the protostars. Other than one single object, the full protostellar sample has HCN abundance higher than the proto-BDs. This can be related to the stronger CN emission and hence most of the HCN being photo-dissociated in the proto-BDs. The most notable trend is seen for the HNC/HCN ratio that shows a clear rise for the lowest luminosity objects (Fig.~\ref{lbol}d), suggesting that this ratio is higher under low-temperature environments. While the sample size is small, all of the proto-BDs show a higher HNC/HCN ratio than the full protostellar sample.

Overall, these trends suggest that HNC is detected in sub-stellar embedded objects even at L$_{bol}$$\sim$0.01 L$_{\sun}$, with a similar (not lower) abundance as measured in a 0.1 L$_{\sun}$ object. The nearly constant HNC abundance for the proto-BDs suggests that the same formation mechanism can be expected for the lowest luminosity objects, unlike in $>$1 L$_{\sun}$ protostars where competing formation/destruction mechanisms could result in an abundance anywhere between 1x10$^{-11}$ to 1x10$^{-9}$ (e.g., Joergensen et al. 2004). Figures~\ref{lbol}e--g compare the CN, HCN, and HNC abundance with the CO abundance for both protostars and proto-BDs. While in Fig.~\ref{cn-hcn-hnc}, it was noted that CN and HNC for the proto-BDs show a flat distribution with CO, the combined trend indicates that both CN and HNC abundances decline by $\sim$2 orders of magnitude with a similar decline in the CO abundance, which suggests that these species deplete on a similar timescale. On the other hand, the combined trend for HCN shows no particular dependence on CO depletion. The combined trend also indicates that CN, HCN, and HNC are depleted by only a factor of $\sim$2--3 less than CO. Previous observations of pre-stellar cores have noted a differential freeze-out of these species and CO indicative of a low initial abundance of gaseous nitrogen (e.g., Padovani et al. 2009). We also find a lack of proto-BDs with [CO]$\leq$10$^{-5}$, which could be due to UV induced photo-desorption that can increase the CO abundance in the freeze-out region towards the central core.

One notable difference is the lack of detection in HC$_{3}$N in proto-BDs, unlike protostars where HC$_{3}$N is commonly detected (Joergensen et al. 2004). There are several HC$_{3}$N lines observed in bonus within the frequency range covered by our observations (Sect.~\ref{observations}). No emission is detected in any of these HC$_{3}$N lines for any object in our sample. The HC$_{3}$N abundance is known to peak either at an early stage of evolution when a large amount of atomic carbon is in the gas phase, or at a later stage in evolution when depletion of atomic oxygen from the gas phase is significant and can aid in the formation of HC$_{3}$N (e.g., Ruffle et al. 1997). It may be that processes such as UV-induced photo-desoprtion and/or grain growth have hampered depletion of CO resulting in a non-detection of HC$_{3}$N in the proto-BDs. Lower-$J$ transitions of HC$_{3}$N with $E_{j}$/k $<$ 30 K may be more suited for the proto-BD targets.


The spread in the CN abundances for the proto-BDs could be related to the range in L$_{acc}$/L$_{bol}$ for these objects, and suggests that these may be at different chemical evolutionary stages. A possible evolutionary trend could be a high HNC/HCN ratio for the earliest stage objects. Thus, J182854 and J182844 could be in the pre-substellar phase. As they evolve, the accretion energy dissipated increases and so the [CN/HCN] ratio increases. This will also raise the dust temperature in the inner, dense region close to the accretion zone, and if indeed HNC is forming via ice grain reactions, then the HNC/HCN ratio will decrease with time. Also, with increasing accretion energy, more of HCN will be photo-dissociated which will further decrease the HNC/HCN ratio. So the more evolved sources would show a smaller HNC/HCN ratio. The individual line strengths or shapes may not change significantly, but the abundance ratios could show an effect of the evolutionary stage. 

Among the more evolved Stage I-T/Stage II objects in the sample, J182952 shows CN emission but there is no HCN or HNC line detection, while none of these tracers are detected in J182940 and J182927. The disk sizes in the more evolved Class II brown dwarfs are estimated to be $<$100 AU (Riaz et al. 2012), which would be beam-diluted in our single-dish spectra. These relatively evolved sub-stellar objects provide an interesting comparison to the more embedded proto-BDs, in particular, the HNC line is undetected in the evolved objects but is detected in all of the Stage 0/I proto-BDs. This species can thus be used as an efficient tracer to search and identify early stage very low luminosity and sub-stellar mass objects. Future interferometric observations and line radiative transfer modelling of a larger sample can provide more robust measurements on the molecular abundances for these species in proto-BDs.

\section{Conclusions}

We present results from the first investigation of nitrogen chemistry in Class 0/I proto-brown dwarf (proto-BD) cores. We have used the IRAM 30 m telescope to observe the CN (2-1), HCN (3-2), and HNC (3-2) lines in 7 proto-BDs for which physical properties have previously been derived from radiative transfer modelling of dust continuum observations. The proto-BDs show a large CN/HCN abundance ratio of $>$20, and a HNC/HCN abundance ratio close to or larger than unity. The enhanced CN/HCN ratios can be explained by strong accretion activity, while the larger than unity HNC/HCN ratio is likely caused by a combination of low temperature and high density. The abundance ratios indicate CN and HCN emission to have an origin close to the accretion zone in the proto-BDs, while HNC likely has an origin in the cooler, outer envelope layers. We have investigated the correlations in the molecular abundances of these species for the proto-BDs with Class 0/I protostars, covering a range in the bolometric luminosities from $\sim$40 L$_{\sun}$ down to $\sim$0.01 L$_{\sun}$. We find tentative trends of CN (HCN) abundances being about an order of magnitude higher (lower) in the proto-BDs compared to the protostars. HNC for the proto-BDs shows a nearly constant abundance between $\sim$0.01--0.1 L$_{\sun}$, unlike the large spread of $\sim$2 orders of magnitude seen for the protostars. Also notable is a rise in the HNC/HCN abundance ratio for the lowest luminosity objects, suggesting that this ratio is higher under low-temperature environments. None of the Class Flat/Class II brown dwarfs in our sample show emission in these high-density tracers, consistent with their evolved stage. The detection of HNC in all of the proto-BDs indicates that this species can be used as an additional tracer to identify early stage brown dwarfs. 



\section*{Acknowledgements}

B.R. acknowledges funding from the Marie Sklodowska-Curie Individual Fellowship (Grant Agreement No. 659383). We thank T. Hirota for a detailed review and valuable comments. This work is based on observations carried out under project number 139-15 and 120-17 with the IRAM 30m telescope. IRAM is supported by INSU/CNRS (France), MPG (Germany) and IGN (Spain). B.R. would like to thank the IRAM staff for help provided during the observations.









\appendix

\section{CLASS Hyperfine Fit}
\label{class}

\begin{figure}
  \centering              
     \includegraphics[width=3in]{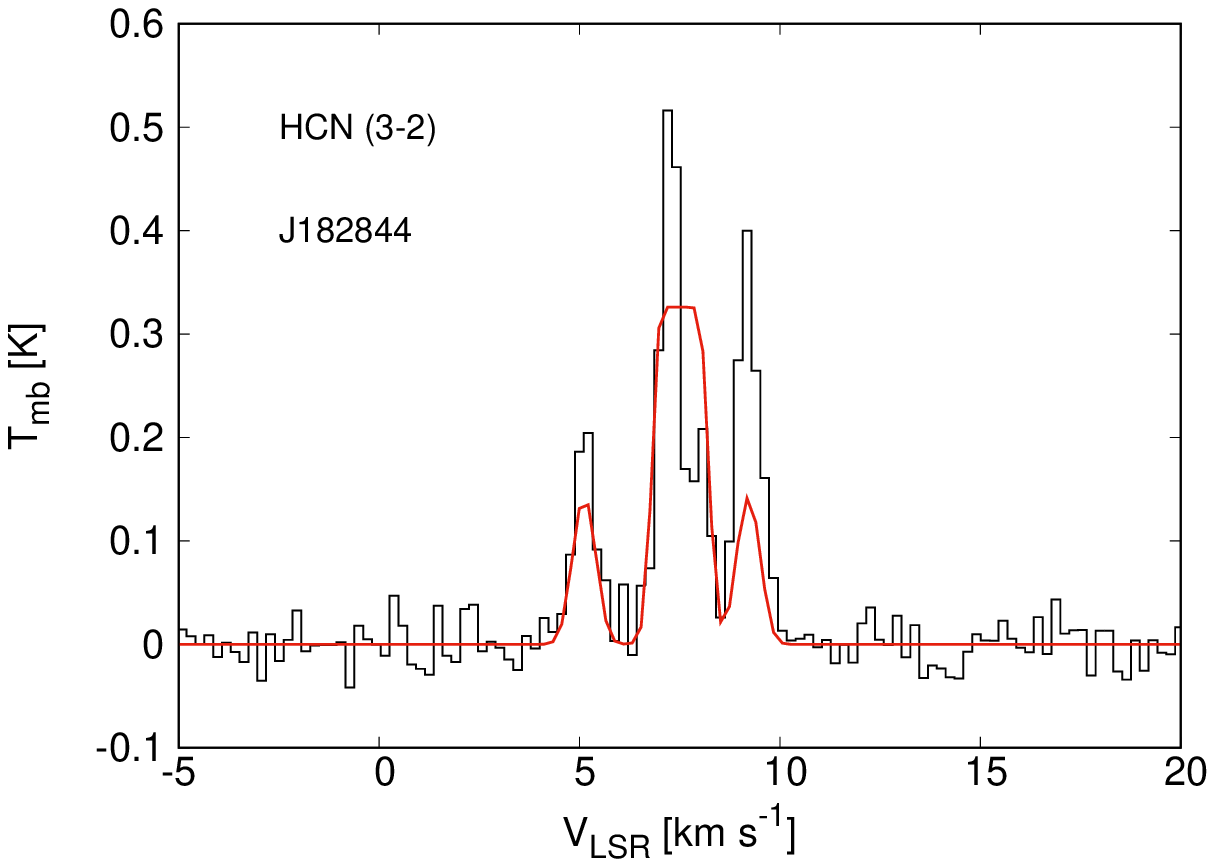}
     \includegraphics[width=3in]{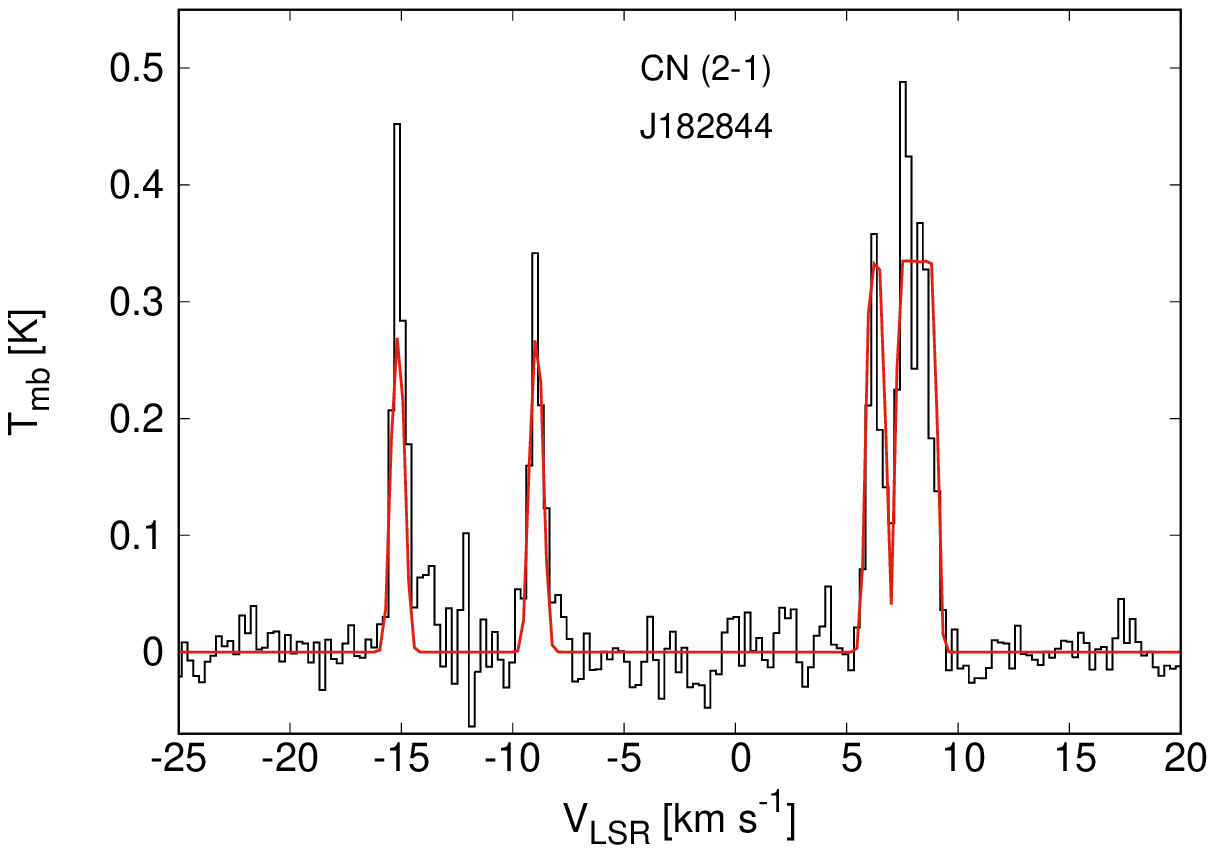}                           
     \caption{The fits to HCN (3-2) (top) and CN (2-1) (bottom) spectra for J182844 using the CLASS hyperfine fitting routine. }
  \label{class-hfs}       
\end{figure}

We have opted to fit multiple Gaussians to the HCN and CN hyperfine lines to measure the line parameters. Another option is to use the hyperfine structure fitting routine in the GILDAS/CLASS package (Hily-Blant et al. 2005). This routine assumes that all hyperfine lines have the same width and the components do not overlap, and then fits Gaussian profiles for the opacity as a function of frequency. The results from the CLASS routine are shown in Fig.~\ref{class-hfs} for the HCN and CN spectra of J182844. The fits are flat-topped and the peak intensities are under estimated. The CLASS hyperfine routine thus fails to provide a good fit to the full hyperfine structure as it cannot correctly interpret the observed anomalies in the relative strength and width of the individual components.

\section{Excitation temperature, optical depth, and critical density}
\label{tau-ncrit}

\begin{figure}
  \centering              
     \includegraphics[width=3in]{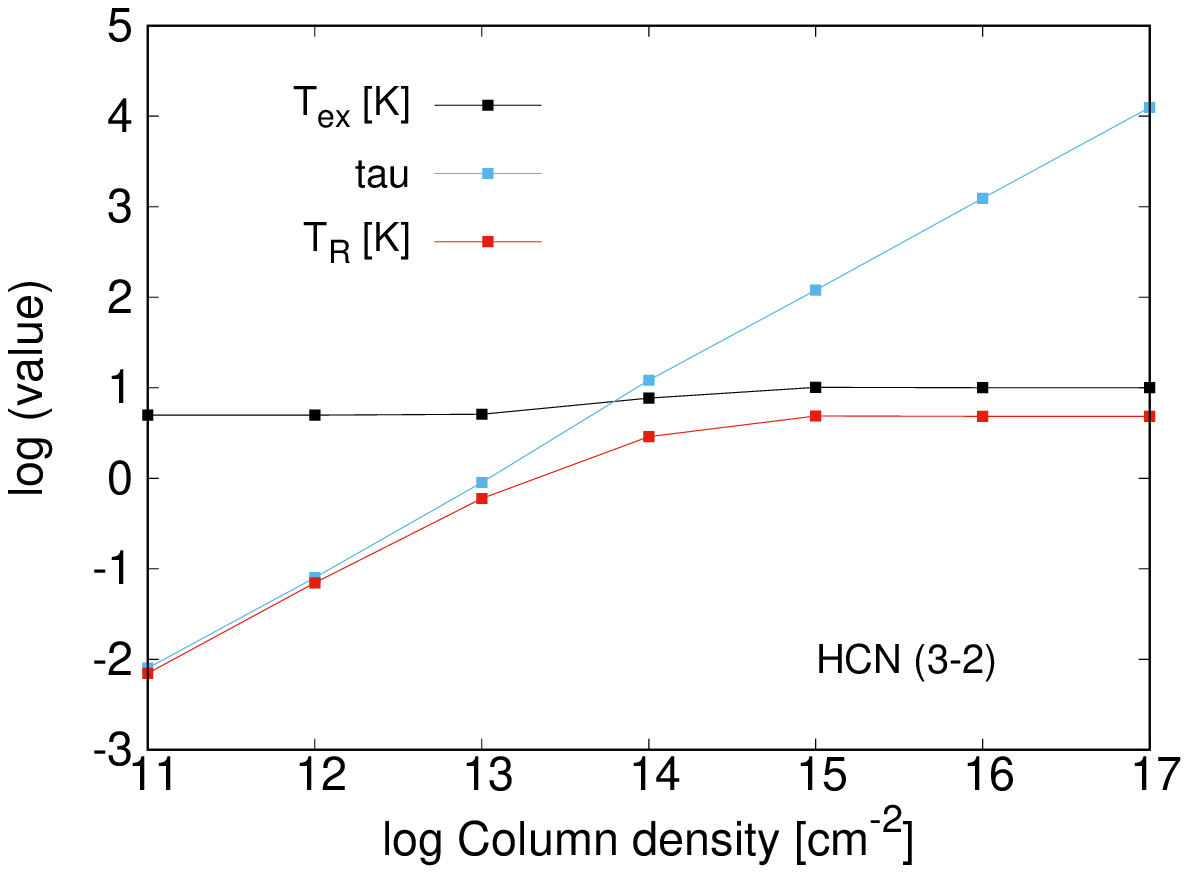}   
     \caption{An example of the output from RADEX on $T_{R}$, $T_{ex}$, and $\tau$ plotted against increasing column density for the `A' component of HCN (3-2). }
  \label{radex}       
\end{figure}

Table~\ref{tex-tau} lists the output on $T_{ex}$ and $\tau$ from RADEX for the best estimates on the column densities. The uncertainties are estimated to be $\sim$20\% on $T_{ex}$ and $\tau$. The $T_{ex}$ for CN, HCN, and HNC is in the range of $\sim$4-5 K, which is lower than the assumed $T_{kin}$$\sim$10 K. This indicates (slightly) sub-thermal non-LTE conditions. The RADEX estimates on line opacities are $\leq$1.5, indicating that the lines are marginally optically thick. Fig.~\ref{radex} shows an example from RADEX on the dependence of the line intensity in the `A' component of HCN (3-2) for J182844 on the column density with the optical depth. For small $\tau <$ 1, the line intensity increases linearly with the column density, and then flattens as $\tau$ increases to larger values. The flattening in intensity is gradual as the main hyperfine component `A' in HCN (3-2) becomes optically thick while the satellite components are still optically thin. The main effect of saturated line emission due to a large optical depth is that the column density could be under-estimated. For our sources, the HCN column densities estimated by RADEX are in the range of 10$^{12}$--10$^{13}$ cm$^{-2}$ where $\tau \leq$1, thus we do not expect a high inaccuracy due to deviation from a linear approximation.

Table~\ref{line-obs} lists the critical density under optically thin LTE conditions, $n_{crit}^{thin}$, defined as the ratio of the Einstein $A$ coefficients and the sum over collisional rates (e.g., Shirley 2015). In principle, the critical density describes the competition between radiative and collisional excitation. The value for $n_{crit}^{thin}$ can indicate the density of the emitting gas. A comparison of $n_{crit}^{thin}$ with the number density of the sources (Table~\ref{column2}) indicates that none of the dense gas tracers can be detected. A large optical depth can also drive the level population into LTE, even if the ambient density is lower than the critical density. For a large optical depth ($\tau \geq$ 5), the effective spontaneous transition rate is reduced by the escape fraction and can be approximated by the ratio of the Einstein coefficient $A_{jk}$ to the optical depth. This has an effect of reducing the critical density from its optically thin value (e.g., Shirley 2015). The effective critical density in the optically thick case is $\sim$1-2 orders of magnitude lower than $n_{crit}^{thin}$, which makes it possible to detect even a modest line intensity from the dense gas tracers, such as HCN, that are typically both sub-thermally populated ($T_{ex} < T_{kin}$) and optically thick. However, results from RADEX do not indicate a large opacity towards our objects (Table~\ref{tex-tau}). As noted, the number densities are expected to reach values of $>$10$^{6}$ cm$^{-3}$ towards the inner ($<$500 AU) dense regions in the proto-BD cores, as predicted by the core collapse simulations for brown dwarfs (Machida 2014). These inner, dense regions can be probed by the high density chemical tracers of CN, HCN, and HNC. Assuming a distance of $\sim$260--430 pc to Serpens (e.g., Winston et al. 2018; Dzib et al. 2010), the inner regions ($<$2$\arcsec$) would be beam diluted in the single-dish SCUBA-2 continuum maps (beam size $\sim$14.5$\arcsec$) that have been used to estimate the number densities. We can thus consider the $n_{H_{2}}$ values in Table~\ref{column2} to be lower limits on the number densities. Interferometric mapping can provide a better estimate on  $n_{H_{2}}$ for these compact proto-BDs.


\section{Line overlap analysis}
\label{overlap}

The central `A' component in the HCN (3-2) spectra (Fig.~\ref{hcn-figs}) shows a self-absorbed profile that is typically seen in dense collapsing cores and is suggestive of an interplay between infall activity and a large optical depth (e.g., Evans 1999). To understand if the self-absorption seen in the HCN and HNC spectra is real, we have conducted a simple line analysis based on overlapping the HCN (3-2), HNC (3-2), and HN$^{13}$C (3-2) spectra for the proto-BD J182844 (Fig.~\ref{isotope}). All spectra have been normalized to unity. For HCN, we have modified the frequency (velocity) when extracting the spectrum to the `A' and `C' hyperfine components, and then compared them with HNC and HN$^{13}$C. The optically thin isotopologue HN$^{13}$C mainly probes the dense gas towards the proto-BD core and should be less contaminated by the surrounding material compared to the main isotope. As can be seen in Figs.~\ref{isotope}a,b, the HN$^{13}$C, HNC, and both the `A' (left panel) and `C' (right panel) HCN components show a peak at $\sim$7.6 km s$^{-1}$. This indicates that the red-shifted emission seen in the HNC and HCN `A' component with a peak at $\sim$7.6 km s$^{-1}$ represents the actual contribution from the proto-BD core, while the emission at $\sim$8.3 km s$^{-1}$ (Fig.~\ref{hnc-figs}) is likely contamination from the unrelated surrounding (foreground/background) material. 


At present, we cannot reject the possibility that the `D' peak could be emission from an outlying layer in the outer envelope regions within the proto-BD systems, or from unrelated cloud material surrounding the proto-BDs, which is in the line of sight and produces this deep self-absorption in the `A' hyperfine component. However, we do not see a similar ``self-absorption'' in the `B' and `C' components for the proto-BDs, particularly in `C' which is about the same intensity and width as `A' and the signal-to-noise ratio is similar ($\sim$10-12) in both of these lines. This is unlike what is typically seen in low-mass dense cores, such as, L1521, L1197, and L1495, where {\it all} of the individual hyperfine line profiles show a narrow self-absorption at the velocity centroid of the core, and are explained to be significantly affected by a large optical depth (e.g., Hily-Blant et al. 2010). We also do not see similar deep self-absorption in the CN spectra for these proto-BDs. Note that we can clearly distinguish between the `A' hyperfine component and the `D' feature using separate Gaussian fits (Fig.~\ref{hcn-figs}), and none of the hyperfine lines that make up the composite `A' component lie within the Gaussian fit to the `D' feature. This kind of multiple Gaussian fitting analysis can thus still provide a good measurement on the contribution from the individual hyperfine components and minimizes the contamination from the unrelated features. Future mapping observations at high resolution will be important to probe the actual contribution from the proto-BDs.



We can roughly estimate the possible systematic effects due to high opacity on the column densities and derived abundances. As noted, the main `A' hyperfine component in the HCN spectra for the proto-BDs may be affected by self-absorption, the signatures of which are also seen in the HNC line profile (Figs.~\ref{hcn-figs};~\ref{hnc-figs}). There are also opacity effects that result in the relative line intensity ratios for the HCN and CN hyperfine components higher than the optically thin limit. Such effects cannot be modelled with RADEX. If instead we consider the `A'+`D' components of HCN (3-2) for J182844, then the column density is 1.4x10$^{13}$ cm$^{-2}$ and the derived abundance is 3.8x10$^{-10}$, a factor of $\sim$4 higher than the abundance derived from using only the `A' component. Likewise, the HCN column density for J183002 derived using the sum of the integrated intensities of the `A'+`D' components is 3.7x10$^{12}$ cm$^{-2}$ and the derived abundance is 1.0x10$^{-10}$, which is $\sim$5 times higher than the abundance derived from using only the `A' component. However, the `D' feature does not coincide with any of the known HCN hyperfine components (marked by blue line sin Fig.~\ref{hcn-figs}), which makes it difficult to determine if `A'+`D' is the central, bright hyperfine component with effects of deep self-absorption, or if `D' is contamination from the (unrelated) foreground cloud material in the line of sight. The molecular data for the optically thin satellite components `B' and `C' of HCN that are unaffected by such self-absorption effects is not available in RADEX. 


For CN, we can check the column density using the fainter `F'+`G'+`H' components, the molecular data for which is available in RADEX. For the same proto-BD J182844, the CN column density derived using the `F'+`G'+`H' components is 0.75x10$^{14}$ cm$^{-2}$ and the derived abundance is 20.8x10$^{-10}$, which is a factor of $\sim$2 lower than the estimate derived from the central `C'+`D'+`E' components. Note that the SNR in the `F', `G', and `H' lines is lower than the central components. The estimate on $\tau$ for the fainter `F'+`G'+`H' components is 1.0, which is comparable to the optical depth estimated for the central `C'+`D'+`E' components. 

Considering the upper limit on the H$^{13}$CN (3-2) abundance for J182844 (Table~\ref{isotope-thin}), the [HCN/H$^{13}$CN] abundance ratio is measured to be 2.5 using the HCN `A' component. The measured [HNC/HN$^{13}$C] abundance ratio for J182844 is 5.0. Both values are much lower than the ISM isotopic ratio of 77. This suggests that the effects of self-absorption can result in under estimating the HCN and HNC abundances using RADEX. The [CN/$^{13}$CN] abundance ratio for J182844 calculated using the $^{13}$CN (2-1) upper limit (Table~\ref{isotope-thin}) is $\sim$130, higher the ISM isotopic ratio of 77 and suggesting an over estimate of the CN abundance. The measured ratios are uncertain due to the upper limits on H$^{13}$CN and $^{13}$CN and the weak $\sim$2-$\sigma$ detection in HN$^{13}$C. A proper comparison between the brightest and the faintest components and the isotopologues will be conducted once high-sensitivity and high-resolution observations in these lines become available for the full sample.






 \begin{figure*}
  \centering  
  \includegraphics[width=3in]{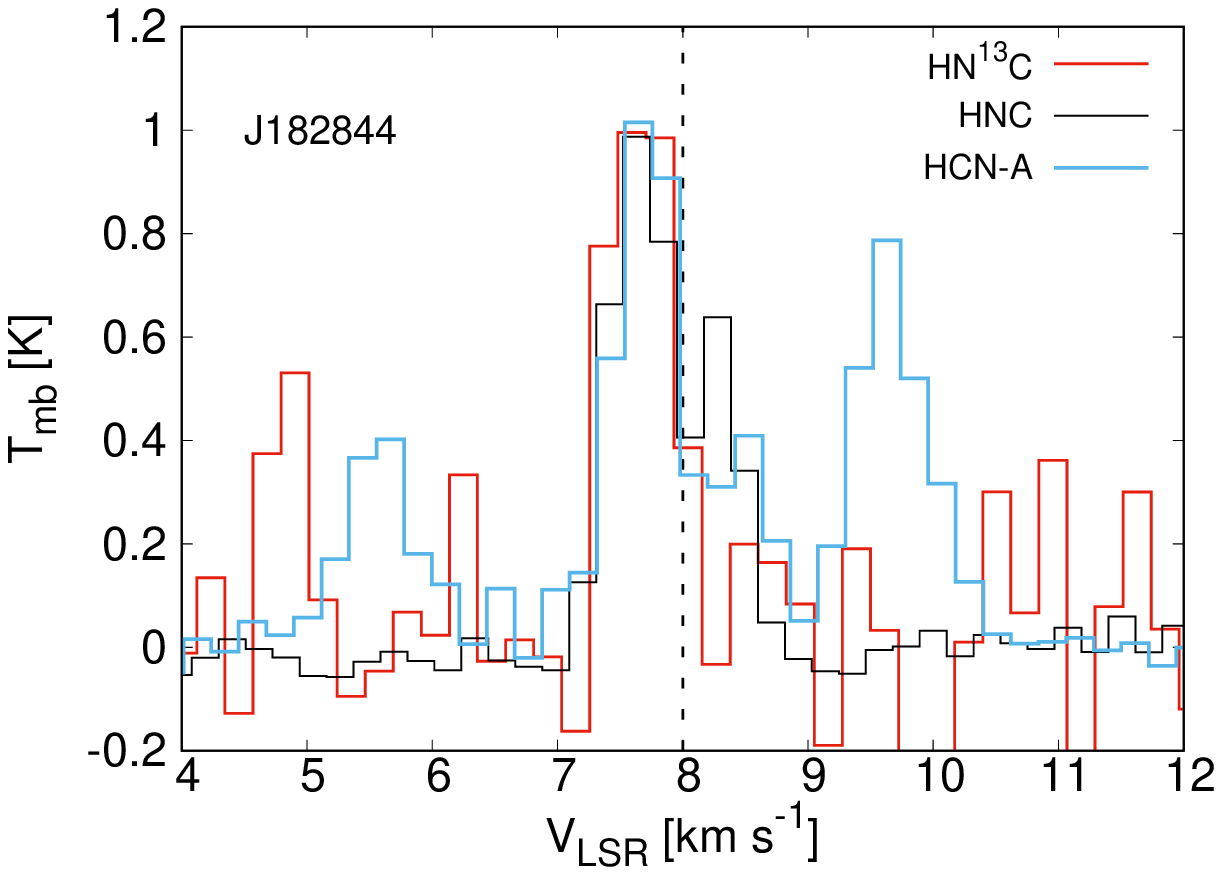}
  \includegraphics[width=3in]{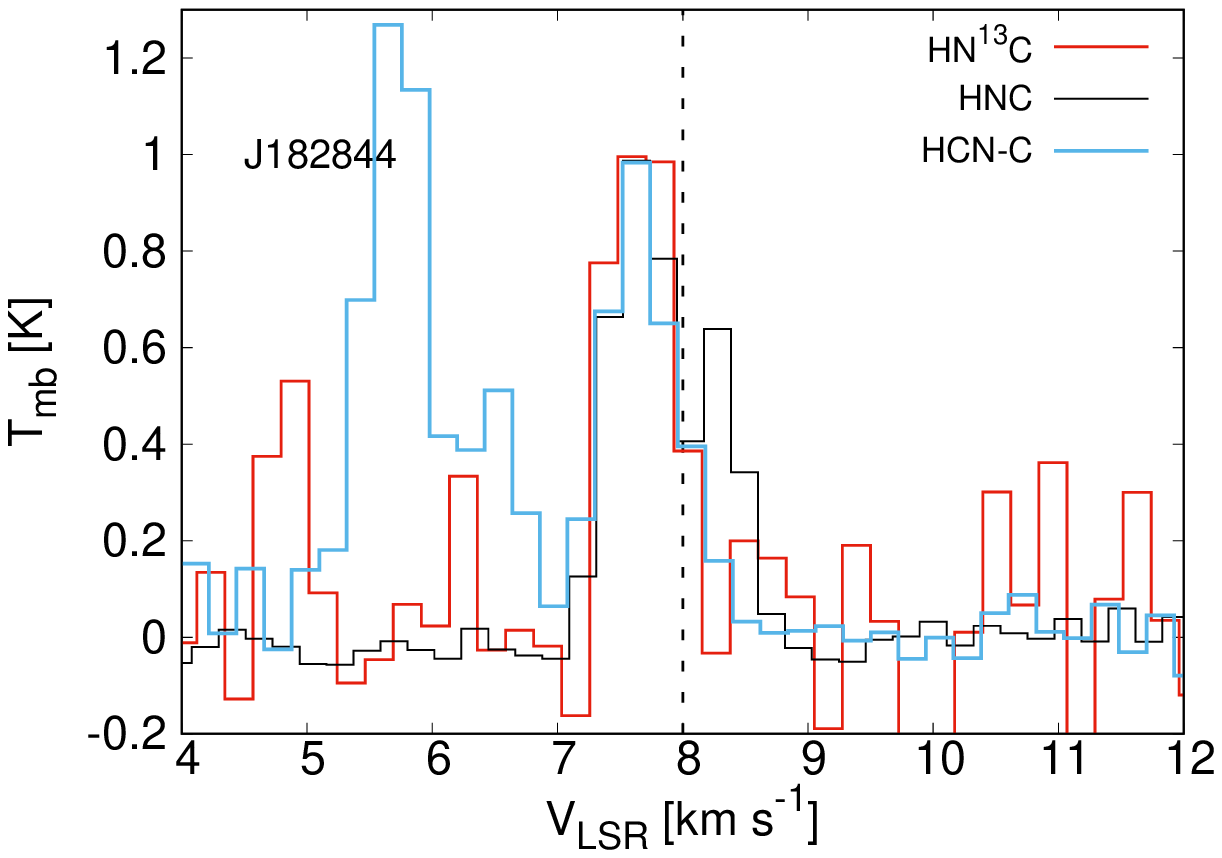}
     \caption{A comparison of the HN$^{13}$C (red), HNC (black), and HCN (blue) spectra for the proto-BD J182844. Left panel shows the HCN `A' component, right panel shows the satellite `C' component. All spectra have been normalized to unity. Black dashed line marks the cloud systemic velocity of $\sim$8 km/s in Serpens. }
  \label{isotope}       
\end{figure*}

\section{Tables}
\label{data}

The line parameters derived from the HCN (3-2), CN (2-1), and HNC (3-2) spectra are listed in Tables~\ref{hcn},~\ref{cn}, and~\ref{hnc}, respectively. The rest frequencies and relative intensities of the HCN (3-2) and CN (2-1) hyperfine components are listed in Tables~\ref{hcn-hfs} and ~\ref{cn-hfs}, respectively. Tables~\ref{hcn-ratios} and~\ref{cn-ratios} list the HCN and CN integrated intensity ratios, respectively, for the targets. Table~\ref{tex-tau} lists the values for T$_{ex}$ and $\tau$ as estimated by RADEX.

\begin{table*}
\centering
\caption{HCN (3-2) Line Parameters}
\label{hcn}
\begin{threeparttable}
\begin{tabular}{llllllllll} 
\hline
Object   &  Hyperfine Component & V$_{lsr}$ (km s$^{-1}$) & T$_{mb}$ (K) & $\int{T_{mb} dv}$ (K km s$^{-1}$) &  $\Delta$v (km s$^{-1}$)  & SNR \\
\hline \hline

J182854	& B 			& 4.95 & $<$0.04 	& $<$0.04 & 1.0 & 	-- 	\\
		& A			& 7.27 & 0.20 		& 0.13 	& 0.51 & 	5.0 	\\
		& C 			& 9.35 & 0.12 		& 0.11 	& 0.73  & 	3.1	\\
		& D\tnote{a} 	& 8.08 & 0.11 		& 0.09 	& 0.67  &	 --	\\		
		& E\tnote{a}	& 10.94 & 0.05 		& 0.05 	& 0.75  & 	--	\\
\hline

J182844	& B 			& 5.12 & 0.20 & 0.14 & 0.56  &	 5.0	\\
		& A 			& 7.26 & 0.53 & 0.33 & 0.49  & 	13.0	\\
		& C 			& 9.20 & 0.37 & 0.27 & 0.59  & 	10.5	\\
		& D\tnote{a}	& 8.06 & 0.21 & 0.09 & 0.37  & 	--	\\
		
\hline

J182959	& -- & 8.0  & $<$0.15  & $<$0.15  & 1.0 	& -- 	\\
J163143	& -- & 4.4  & $<$0.15  & $<$0.15  & 1.0	& -- 	 \\
J182953	& -- & 8.0  & $<$0.15  & $<$0.15  & 1.0	& -- 	 \\  
J163136	& -- & 4.4  & $<$0.15  & $<$0.15  & 1.0	& --  \\
J182940	& -- & 8.0  & $<$0.12  & $<$0.12  & 1.0 	& -- 	\\
J182927	& -- & 8.0  & $<$0.15  & $<$0.15  & 1.0 	& -- 	\\
J182952	& -- & 8.0  & $<$0.15  & $<$0.15  & 1.0 	& -- 	\\  

\hline \hline
\end{tabular}
\begin{tablenotes}
  \item We estimate uncertainties of $\sim$15\%-20\% for the integrated intensity, $\sim$0.1 km s$^{-1}$ in V$_{lsr}$, $\sim$10\% on T$_{mb}$ and $\Delta$v. 
  \item[a] This peak does not coincide with any of the hyperfine components.
\end{tablenotes}
\end{threeparttable}
\end{table*}

\begin{table*}
\centering
\caption{CN (2-1) Line Parameters}
\label{cn}
\begin{threeparttable}
\begin{tabular}{lllllllll} 
\hline
Object   & Hyperfine Component & V$_{lsr}$ (km s$^{-1}$) & T$_{mb}$ (K) & $\int{T_{mb} dv}$ (K km s$^{-1}$) &  $\Delta$v (km s$^{-1}$) & SNR \\
\hline \hline

J182854	& C+D+E &  8.0 & $<$0.15 & $<$0.15 & 1.0 & -- \\
\hline
J182844	& A &  -15.12 & 0.41 & 0.30 & 0.58 & 9.0	 \\
		& B &  -8.94 & 0.30 & 0.23 & 0.61 & 	7.3	 \\
		& C &  6.25 & 0.31 & 0.26 & 0.66  & 	7.3	\\
		& D & 7.56 & 0.46 & 0.28 & 0.50  & 	9.6	\\
		& E & 8.43 & 0.34 & 0.33 & 0.78  & 	7.3	\\
		& F &  266.02 & 0.27 & 0.20 & 0.58  & 5.0		\\
		& G &  286.66 & 0.27 & 0.18 & 0.55 	 & 5.0	\\
		& H &  292.13 & 0.36 & 0.32 & 0.69 	 & 7.6	\\
		& I &  328.35 & 0.20 & 0.22 & 0.59 	 & 6.0	\\
\hline
J183002	& C &  6.47 & 0.24 & 0.20 & 0.70 & 	19.3 \\
		& D+E &  8.09 & 0.50 & 0.88 & 1.53  & 10.0	\\			
		& F & 266.20 & 0.13 & 0.10 & 0.63  & 3.2	\\
		& H & 292.36 & 0.26 &  0.27 & 0.82 	 & 9.3	\\
		& I  &  328.45 & 0.17 & 0.25 & 1.44 	 & 3.2	\\
\hline
J182959	& A &  -14.45 & 0.23 & 0.10 & 0.36 	 & 3.7	\\
		& B &  -8.16 & 0.24 & 0.11 & 0.35 	 & 3.7	\\
		& C &  7.05 & 0.24 & 0.20 & 0.68 	 & 3.7	\\
		& D &  8.47 & 0.25 & 0.31 & 0.99 	 & 3.7	\\
		& E &  9.35 & 0.18 & 0.09 & 0.43 	 & 3.7	\\
		& F &  266.79 & 0.13 & 0.08 & 0.54 	 & 1.7	\\
		& G & 287.42 & 0.15 & 0.09 & 0.55	 & 2.2	 \\
		& H &  292.89 & 0.22 & 0.13 & 0.48 	 & 3.5	\\
		& I &  328.93 & 0.14 & 0.28 & 1.57 	 & 3.5	\\
\hline
J163143	& C & 2.96 & 0.14 & 0.16 & 0.92	 & 2.0	 \\
		& D & 4.72 & 0.32 & 0.36 & 0.89 	 & 5.0	\\
		& E & 5.57 & 0.16 & 0.11 & 0.51 	 & 3.7	\\		
\hline
J182953	& C 	   & 7.71 	  & 0.10 & 0.07 & 0.55 	 & 2.2	\\
		& D+E & 10.0 	  & 0.28 & 0.8 & 2.20 	 & 4.4	\\		
		& H 	   & 293.86 & 0.16 & 0.13 & 0.80 	 & 1.8	\\
\hline		
J163136	& C+D+E & 4.4 & $<$0.15 & $<$0.15 & 1.0  & --\\ 
\hline
J182940	& C+D+E &  8.0 & $<$0.15 & $<$0.15 & 1.0 & -- \\ 
\hline
J182927	& C+D+E & 8.0 & $<$0.15 & $<$0.15 & 1.0 & -- \\
\hline
J182952	& C+D+E & 10.28 & 0.13 & 0.18 & 1.44  & 2.0	\\
\hline \hline
\end{tabular}
\begin{tablenotes}
 \item We estimate uncertainties of $\sim$15\%-20\% for the integrated intensity, $\sim$0.1 km s$^{-1}$ in V$_{lsr}$, $\sim$10\% on T$_{mb}$ and $\Delta$v. 
\end{tablenotes}
\end{threeparttable}
\end{table*}

\begin{table}
\centering
\caption{HNC (3-2) Line Parameters}
\label{hnc}
\begin{threeparttable}
\begin{tabular}{lllllllll} 
\hline
Object   & Peak & V$_{lsr}$ (km s$^{-1}$) & T$_{mb}$ (K) & $\int{T_{mb} dv}$ (K km s$^{-1}$) &  $\Delta$v (km s$^{-1}$) & SNR \\
\hline

J182854 & 1 & 8.15 & 0.42 	& 0.46 	& 1.03 & 18.3	 \\
J182844 & 1 & 7.65 & 1.04 	& 0.63 	& 0.48 & 33.0	 \\
	      & 2 & 8.32 & 0.63 	& 0.25 	& 0.32 & 20.0	 \\
J183002 & 1 & 8.00 & 1.12 	& 1.13 	& 0.95 & 40.0	 \\
J182959 & 1 & 8.35 & 0.57 	& 0.63 	& 1.03 & 12.0	 \\
J163143 & 1 & 4.55	& 0.15 	& 0.14 	& 0.867 & 3.0	  \\
J182953 & 1 & 9.75  & 0.16 	& 0.21     	 & 1.26 & 4.0	  \\
J163136 & -- & 4.4	& $<$0.15 & $<$0.15 & 1.0 & --	  \\
J182940 & -- & 8.0	& $<$0.12 & $<$0.12 & 1.0 & --	  \\
J182927 & -- & 8.0	& $<$0.15 & $<$0.15 & 1.0 & --	  \\	      
J182952 & -- & 8.0    & $<$0.15 & $<$0.15 & 1.0  & --	 \\      
	     
\hline
\end{tabular}
\begin{tablenotes}
 \item We estimate uncertainties of $\sim$15\%-20\% for the integrated intensity, $\sim$0.1 km s$^{-1}$ in V$_{lsr}$, $\sim$10\% on T$_{mb}$ and $\Delta$v. 
\end{tablenotes}
\end{threeparttable}
\end{table}

\begin{table}
\centering
\caption{HCN (3-2) hyperfine components}
\label{hcn-hfs}
\begin{threeparttable}
\begin{tabular}{ccccc} 
\hline
\multicolumn{2}{l}{Hyperfine Component} ($F \rightarrow F^{\prime}$) & Rest Frequency (GHz) & V$_{lsr}$ (km s$^{-1}$)\tnote{a} & Normalized Intensity\tnote{b} 	\\
\hline
B 		&  2 $\rightarrow$ 2   	& 265.888519 		& -2.28142  & 	0.0370	\\
A		 & 2 $\rightarrow$ 3 		& 265.886976		& -0.54045  & 	0.0011	\\
A\tnote{c} 	& 4 $\rightarrow$ 3 		& 265.886497		& 0         	   & 	0.4286	\\
A 		& 3 $\rightarrow$ 2 		& 265.886431		& 0.07446   & 	0.2963	\\	
A 		& 2 $\rightarrow$ 1		& 265.886185		& 0.35203   & 	0.2000	\\
C 		& 3 $\rightarrow$ 3 		& 265.884887		& 1.81656   & 	0.0370	\\
\hline
\end{tabular}
\begin{tablenotes}
\item[a] The velocity shifts have been calculated relative to the brightest component A ($F \rightarrow F^{\prime}$=4 $\rightarrow$ 3). 
\item[b] The intensity of each component is normalized to the sum of the intensities of all components.
\item[c] The central broad A component is a composite of four hyperfine components ($F \rightarrow F^{\prime}$ = 2 $\rightarrow$ 1; 3 $\rightarrow$ 2; 4 $\rightarrow$ 3; 2 $\rightarrow$ 3). 
\end{tablenotes}
\end{threeparttable}
\end{table}

\begin{table}
\centering
\caption{Integrated intensity ratios for the HCN (3-2) hyperfine components}
\label{hcn-ratios}
\begin{threeparttable}
\begin{tabular}{lcccc} 
\hline
Object 	& 	B/A\tnote{a} 		& 	C/A\tnote{a}		&  B/C\tnote{b}	\\
\hline
J182854	&	$<$0.2  			&	0.3				&	$<$0.4	\\
J182844	&	0.37	  			&	0.54				&	0.5		\\	 
\hline
\end{tabular}
\begin{tablenotes}
\item[a] Ratio of the integrated intensity for the `B' and `C' components relative to the `A' component. The ratios under optically thin LTE conditions are B/A=C/A=0.04. 
\item[b] Ratio of the integrated intensity for the `B' and `C' components. The ratio under optically thin LTE conditions is B/C=1.0.  
\end{tablenotes}
\end{threeparttable}
\end{table}

\begin{table*}
\centering
\caption{CN (2-1) hyperfine components}
\label{cn-hfs}
\begin{threeparttable}
\begin{tabular}{ccccc} 
\hline
\multicolumn{2}{l}{Hyperfine Component} ($J, F \rightarrow J^{\prime}, F^{\prime}$) & Rest Frequency (GHz) & V$_{lsr}$ (km s$^{-1}$)\tnote{a} & Normalized Intensity\tnote{b}	\\
\hline
A & 5/2, 5/2 $\rightarrow$ 3/2, 5/2 &	226.892128	&  -22.9422 	& 0.0317013	\\
B & 5/2, 3/2 $\rightarrow$ 3/2, 3/2 & 226.887420 	& -16.7141 	&  0.0318844 	\\	
C & 5/2, 3/2 $\rightarrow$ 3/2, 1/2 & 226.875896	& -1.48099 	&  0.100249	\\
D & 5/2, 7/2 $\rightarrow$ 3/2, 5/2 &	226.874781	&   0            	& 0.266918  	\\
E & 5/2, 5/2 $\rightarrow$ 3/2, 3/2 &	226.874190	& 0.780166 	&  0.168492	\\
F & 3/2, 3/2 $\rightarrow$ 1/2, 1/2 &	226.679311	& 258.473 	&  0.0615571	\\
G & 3/2, 1/2 $\rightarrow$ 1/2, 1/2 &	226.663692	&  279.128 	&  0.0494628	\\
H & 3/2, 5/2 $\rightarrow$ 1/2, 3/2 &	226.659558	& 284.589 	&  0.16595	\\
I  & 3/2, 3/2 $\rightarrow$ 1/2, 3/2 &	226.632190	&  320.78 		&  0.0497826	\\
\hline
\end{tabular}
\begin{tablenotes}
  \item[a] The velocity shifts have been calculated relative to the brightest component `D' ($J, F \rightarrow J^{\prime}, F^{\prime}$ = 5/2, 7/2 $\rightarrow$ 3/2, 5/2). 
  \item[b] The intensity of each component is normalized to the sum of the intensities of all components.
\end{tablenotes}
\end{threeparttable}
\end{table*}

\begin{table}
\centering
\caption{Integrated intensity ratios for the CN (2-1) hyperfine components}
\label{cn-ratios}
\begin{threeparttable}
\begin{tabular}{c||ccc||cccc} 
\hline
Component & \multicolumn{3}{c}{Ratio\tnote{a}} 			  \\
		& J182844    &	J182959 	& J183002	& 	LTE\tnote{b}     \\     
\hline

A		&	0.34	    &  0.16		& --			&	0.06		\\
B		&	0.26	    &	0.18		& --			&	0.06		\\	
C+D+E	&	1.00	    &	1.00		& 1.00		&	1.00 		\\		
F		&	0.23	    &	0.13		& 0.09		&	0.11		\\	
G		&	0.21	    &	0.15		& --			&	0.09		\\	
H		&	0.37	    &	0.22		& 0.25		&	0.31		\\	
I		&	0.25	    &	0.46		& 0.23		&	0.09		\\

\hline
\end{tabular}
\begin{tablenotes}
\item[a] Ratio of the integrated intensity of each hyperfine component relative to the `C'+`D'+`E' components in the CN (2-1) hyperfine structure. 
\item[b] The ratios under optically thin LTE conditions have been calculated from the normalized intensities listed in Table~\ref{cn-hfs}.
\end{tablenotes}
\end{threeparttable}
\end{table}

\begin{table}
\centering
\caption{Excitation temperature and optical depth}
\label{tex-tau}
\begin{threeparttable}
\begin{tabular}{llllllllllllllllllll} 
\hline
Object &  \multicolumn{2}{c}{HCN (3-2)\tnote{a}}  & \multicolumn{2}{c}{HNC (3-2)} & \multicolumn{2}{c}{CN (2-1)\tnote{a}}  \\
\hline
	   & $T_{ex}$ (K) & $\tau$ & $T_{ex}$ (K) & $\tau$ & $T_{ex}$ (K) & $\tau$ \\
	  
\hline

J182854   & 5.0 & 0.15       & 4.8 & 0.8 & -- & --  	\\

J182844 	& 4.8 & 0.6  & 4.6 & 1.1 & 5.0	   & 1.3  \\

J183002   & 5.7 & 0.4       & 5.9 & 1.3 & 6.4 	   & 0.8     	\\

J182959  	&  5.0    & 0.2	       & 4.8 & 1.5 & 4.8    & 0.7  \\

J163143  	& 	  -- & -- 	    & 4.3	  & 0.3      & 4.1    & 1.1  \\

J182953   &	 -- & -- 	     & 4.2	  & 0.4      &  4.5	   &  0.9    \\

J163136   &	  -- & -- 	   & 4.7	  & 0.2      &   -- & --     \\

J182940 	&	  -- & -- 	      & 4.7	  & 0.2      &  -- & --    	\\

J182927   &	 -- & -- 	     & 4.7	  & 0.2      &  -- & --     	\\

J182952   &	 -- & -- 	   & 4.7	  & 0.2      & 4.8	   & 0.1     	 \\

\hline
\end{tabular}
\begin{tablenotes}
  \item[a] The $T_{ex}$ and $\tau$ values are estimated using the brightest component `A' in HCN (3-2) and `C'+`D'+`E' components in CN (2-1). We estimate an uncertainty of $\sim$20\% on $T_{ex}$ and $\tau$ estimates. 
\end{tablenotes}
\end{threeparttable}
\end{table}


\bsp	
\label{lastpage}
\end{document}